# Revolutionising Antibacterial Warfare: Machine Learning and Molecular Dynamics Unveiling Potential Gram-Negative Bacteria Inhibitors

BY

Pritish Ranjan Joshi
Abhishek Bera

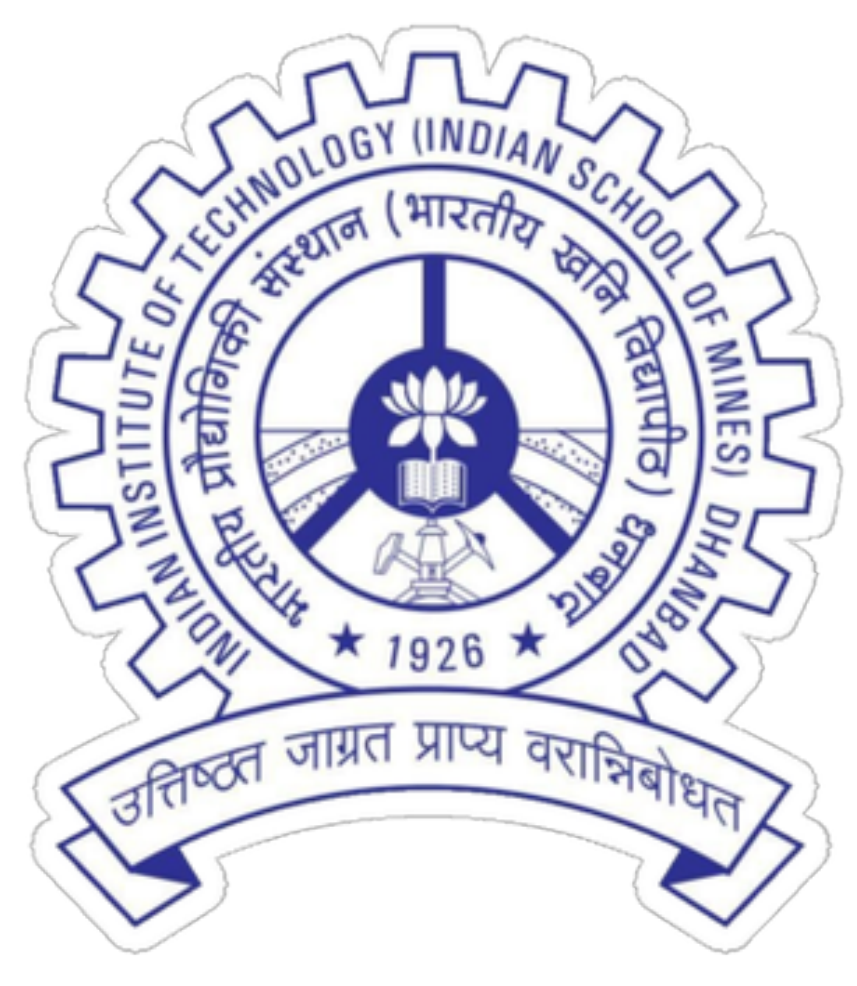

THESIS

SUBMITTED TO

INDIAN INSTITUTE OF TECHNOLOGY

(INDIAN SCHOOL OF MINES), DHANBAD

For the award of the degree of

MASTER OF SCIENCE

MAY 2024

# ACKNOWLEDGEMENT

On the very outset of this report. I want to extend my sincere obligation to everyone who has helped me in this endeavor without their active help, guidance, encouragement, and cooperation. I would not have made the headway in the project.

First and foremost, I would like to express my sincere gratitude to my project supervisor Dr. Niladri Patra, Assistant Professor, IIT(ISM) Dhanbad, for his invaluable guidance, encouragement, motivation, and inspiration throughout the curriculum and towards completing this project. I am very grateful to have a guide like him.

I want to express my sincere thanks to Dr. Parthasarathi Das, Head of the Chemistry and chemical biology department, IIT (ISM) Dhanbad, for allowing me to carry out this project.

I express my deep gratitude toward all the faculty members & research staff of Chemistry and chemical biology IIT(ISM) Dhanbad for their support and cooperation.

It deserves no less mention that this project work would have been unachievable without the relentless help and support of my seniors Abhishek Bera, Rakesh Kumar Roy, Diship Srivastava, Gourav Chakraborty, Kousik Kumar Bhanja, and Avigyan Naskar.

I sincerely thank my family and friends for their blessings, motivation, love, and constant support throughout my endeavors, without which I would not be where I am.

Lastly and most importantly, I am very thankful to Almighty Nature for all she has given me.

PRITISH RANJAN JOSHI
Admission No.: 22MS0105

# ABSTRACT

Diseases caused by bacteria have been a threat to human civilisation for centuries. Despite the availability of numerous antibacterial drugs today, bacterial diseases continue to pose life-threatening challenges. The credit for this goes to Gram-Negative bacteria, which have developed multi-drug resistant properties towards β-lactams, chloramphenicols, fluoroquinolones, tetracyclines, carbapenems, and macrolide antibiotics. Various mechanisms of bacterial defence contribute to drug resistance, with Multi-Drug Efflux Pumps and Enzymatic degradation being the major ones. An effective approach to cope with this resistance is to target and inhibit the activity of efflux pumps and esterases. Even though various Efflux Pump Inhibitors and Esterase resistant macrolide drugs have been proposed in the literature, none of them has achieved FDA approval due to several side effects. This research has provided valuable insights into the mechanism of drug resistance by RND efflux pump and Erythromycin esterase. A handful of potential efflux pump inhibitors have been predicted through machine learning and molecular dynamics.

# 1. Introduction

## 1.1 Bacterial stains

Diseases caused by bacteria, such as pneumonia, cholera, and tuberculosis, have been a threat to human civilization for centuries. Hans Christian Gram, a Danish bacteriologist, initially introduced a method to identify pneumonia-causing organisms. This method, called the Gram staining method, is used to classify bacteria into two broad classes [1]. The first step of this method involves drenching the bacteria in a primary color such as methylene blue or crystal violet. The second step uses iodine to prevent dye removal by forming a crystal violet-iodine complex. The third step involves decolorization using acetone or ethanol as solvents [2]. Finally, the bacteria are observed under a microscope using safranin. Bacteria that retain the crystal-violet coloration appear purple-brown under a microscope and are termed 'Gram-Positive'. In contrast, bacteria that do not retain the crystal violet and appear pink-red under a microscope due to safranin are termed 'Gram-Negative' [3]. The reason for this difference lies in the composition of the bacterial cell wall. The Gram-negative bacteria have a higher lipid content in their cell wall while the Gram-positive bacteria have a higher peptidoglycan content [4]. Initially, both types of bacteria take up the crystal-violet dye, however, during subsequent solvent washing, the lipid layer of Gram-negative bacteria dissolves, causing them to lose the purple coloration [5].

## 1.2 Antibiotics

Drugs which are used to treat bacterial infections are termed as antibiotics. The fortuitous discovery of the first antibiotic, penicillin (a β-lactam), by Alexander Fleming provided mankind with a significant advantage in the centuries-long fight against bacterial infections [6]. As of today, there are several advanced antibiotics available that operate based on the principles of minimum bactericidal concentration (MBC) and minimum inhibitory concentration (MIC). MBC refers to the antibiotic concentration that reduces bacterial density by 1000-fold within 24 hours, while MIC indicates the lowest concentration that inhibits visible bacterial growth within the same time frame. Antibiotics having an MBC to MIC ratio greater than 4 are classified as bacteriostatic, while those having an MBC to MIC ratio equal to or less than 4 are classified as bactericidal [7]. There are several modes of action of antibiotic drugs:

- Inhibition of cell wall synthesis

  Antibiotics such as β-lactams and cephalosporins disrupt the production of peptidoglycans, leading to the breakdown of the bacterial cell wall. Targeting the biosynthesis of peptidoglycan in the cell wall is a favored approach in the search for antibacterial agents, as mammalian cells lack this structure. Consequently, inhibiting peptidoglycan biosynthesis poses minimal risk to mammalian host cells [8].
- Inhibition of protein synthesis

  Antibiotics such as tetracyclines impede the binding of aminoacyl-tRNA by obstructing the A (aminoacyl) site of the 30S ribosome, thereby hindering protein synthesis in both 70S and 80S (eukaryotic) ribosomes [9].
- Inhibition of nucleic acid synthesis

  A group of antibiotics referred to as Quinolones have the capability to disrupt DNA synthesis by blocking topoisomerase, an enzyme crucial for DNA replication [9].
- Inhibition of membrane function

  Antibiotics such as Valinomycin disrupt various targets by interacting with a lipophilic component within the bacterial membrane, resulting in the deterioration of membrane structures and functional impairment [8].
- Inhibition of metabolic pathways

  Antibiotics belonging to the class Sulfonamides hinder bacterial metabolic pathways by disrupting the synthesis of bacterial Tetrahydro-folic Acid, a vital coenzyme used in the synthesis of nucleic acids and specific amino acids across all life forms [10].

### 1.3 ESKAPE Pathogens

Despite the availability of numerous antibacterial drugs, bacterial diseases continue to pose life-threatening challenges even today. The credit for this goes to the ESKAPE pathogens, namely Enterococcus spp., Staphylococcus aureus, Klebsiella pneumoniae, Acinetobacter baumannii, Pseudomonas aeruginosa, and Enterobacter spp.[11], which have developed multi-drug resistant properties towards β-lactams, chloramphenicols, fluoroquinolones, tetracyclines, carbapenems, and macrolide antibiotics. The last four belong to the Gram-negative class of bacteria. These pathogens exhibit various kinds of drug resistance:

i. Multidrug-resistant (MDR), resistant to three or more classes of structurally unrelated antibiotics.

ii. Extensively drug-resistant (XDR), resistant to all antimicrobials except a few classes.

iii. Pan drug-resistant (PDR), resistant to all available classes of antimicrobials [12].

## 1.4 Mechanisms of antimicrobial resistance

The biofilms formed by these bacteria are now recognized as crucial factors in the emergence of antimicrobial-resistant strains, presenting a significant global menace by greatly complicating the treatment of related infections. The severity of this problem has spurred both in-vitro and in-silico novel drug exploration efforts, resulting in advancements in antibiotic treatments. These Gram-negative bacteria exhibit a notable degree of antimicrobial resistance through the four fundamental mechanisms, which are efflux pumps, enzymatic degradation of antibiotics, antibiotic target alteration, and membrane permeability modification, the first two being the major ones [13] (Figure 1).

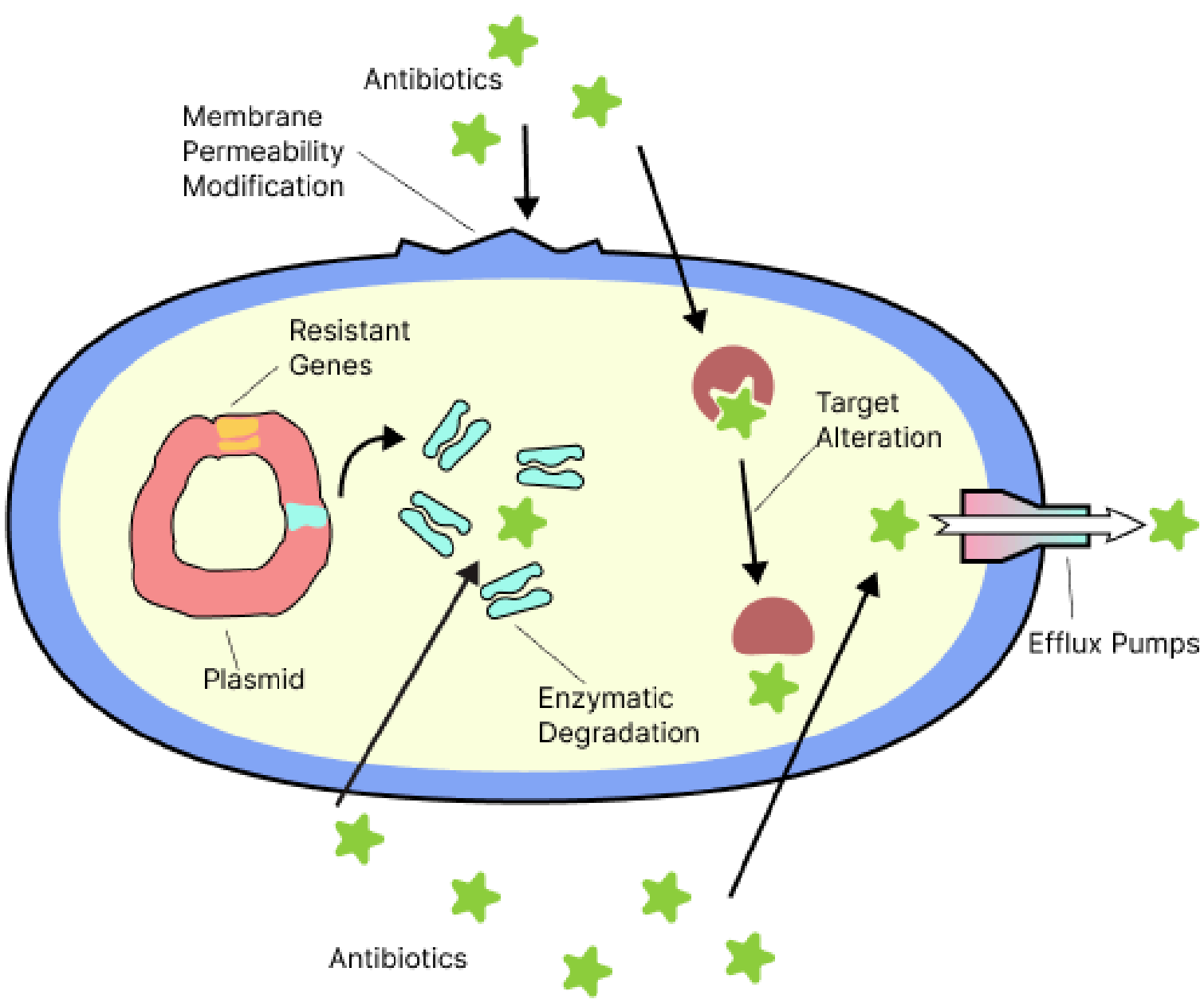

**Figure 1.** *Diagram showing various antimicrobial resistance mechanisms*

### 1.4.1 Efflux pumps

The Gram-negative bacteria is protected by a trilayer cell envelope: cytoplasmic inner membrane, peptidoglycan cell wall, and lipopolysaccharide outer membrane. In between these membranes, there is an aqueous cellular compartment called the periplasm [14]. In addition to giving resistance to the Gram-negative bacteria against the outer environment, the diderm cell envelope also presents a transmembrane transport mechanism known as tripartite efflux pumps. A majority of these pumps show specificity towards a large number of substrates, facilitating their efflux across the membranes [15]. On the basis of their structure and energy source, efflux pumps in bacteria can be broadly classified into five families (Figure 2) [16]:

i. ATP-binding cassette family (ABC)

They transport the substrates across the membrane by utilizing the ATP hydrolysis to. E.g. The MacAB-TolC pump in A. baumannii is responsible for toxin secretion and macrolide resistance [17].

ii. Multidrug and toxic compound extrusion family (MATE)

They can be further classified into NorM (drug/Na+ antiport) and DinF (drug/H+ antiport) subfamilies. They are capable of transporting cationic and poly-aromatic compounds, like tetraphenylphosphonium (TPP), rhodamine-6G, and ethidium bromide [18].

iii. Small multidrug resistance family (SMR)

They form parallel or antiparallel homo- or hetero-dimers comprising 4 transmembranes to transport substrates utilizing the proton motive force. E.g. EmrE in E. coli [19].

iv. Proteobacterial Antimicrobial Compound Efflux family (PACE)

These small membrane proteins consist of nearly 150 amino acids and are topologically close to the SMR family. E.g. AceI in A. baumannii [20].

v. Major facilitator superfamily (MFS)

They are antiporters, uniporters, and symporters consisting of approximately 500 amino acids and 12-14 transmembrane helices. E.g. EmrAB-TolC in E. coli [21].

vi. Resistance nodulation−cell division superfamily (RND)

These structures are defined by preserved transmembrane arrangements consisting of 12 trans-membrane (TM) helices organized into two pseudo-symmetric bundles. The TM regions of these secondary active transporters employ the H+ (or Na+) antiport mechanism to efflux substrates [22]. These RND transporters can further be categorized into distinct families based on their functions and phylogeny:

a. The hydrophobe amphiphile efflux exporter family (HAE)

   They are further classified into 3 subfamilies, namely, HAE-1 (secondary active drug efflux transporters of Gram-Negative bacteria. E.g. AcrAB-TolC in E. coli, MexAB-OprM in P. aeruginosa), HAE-2 (lipid exporters of Gram-positive bacteria. E.g. MmpL3 in M. smegmatis), HAE-3 (hopanoids exporters of archaea and bacteria. E.g. HpnN in B. multivorans) [23-25].

b. The heavy metal ion exporter family (HME)

   They are further classified into five subfamilies (HME1-5) and are involved in the efflux of heavy metal ions from the periplasm across the outer membrane. E.g. ZneCAB and CusCBA [26-29].

c. The nodulation factor exporter family (NFE)

   They act as a single component of nodulation factors. E.g. YerP in B. subtilis [26, 30].

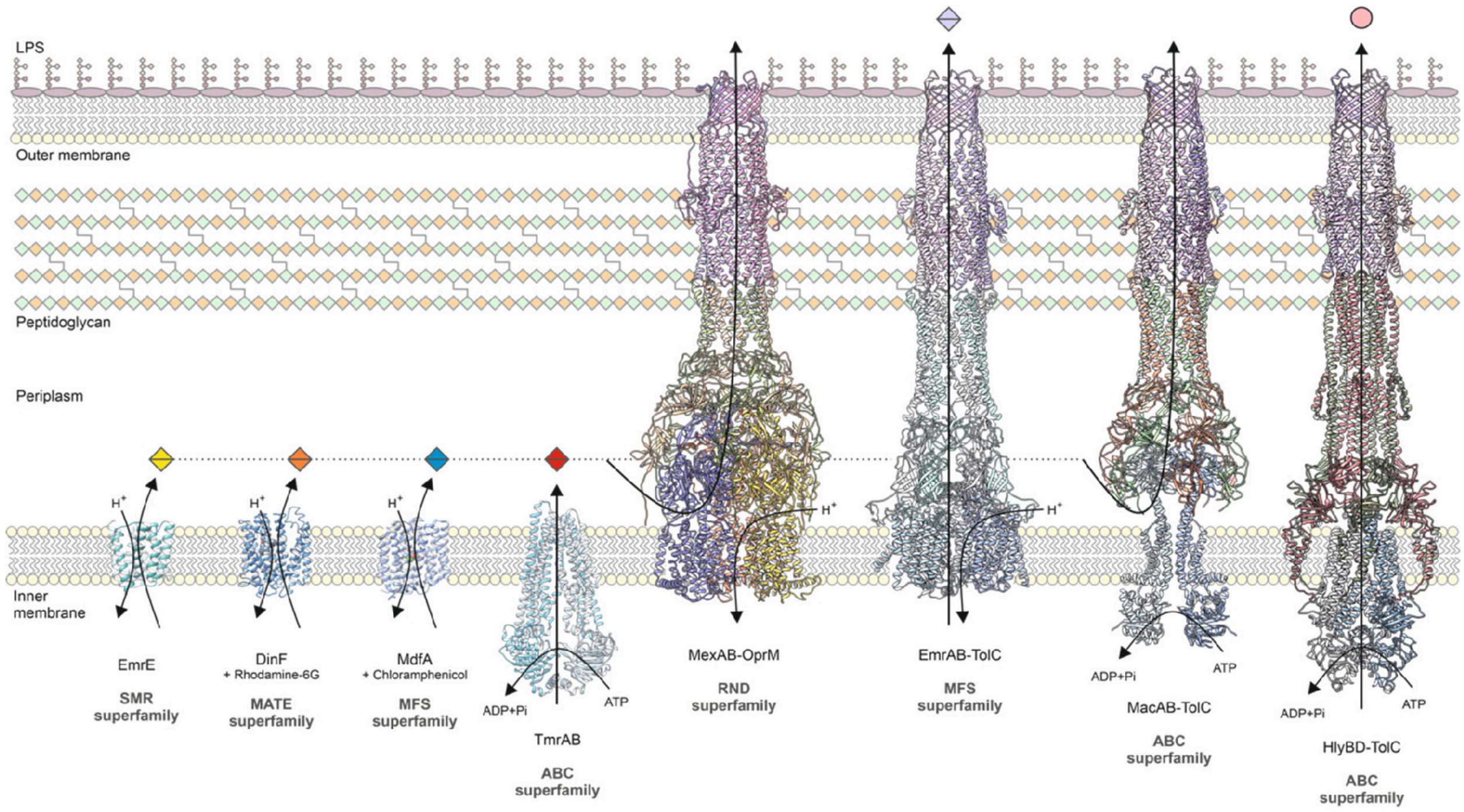

**Figure 2.** *Structures of various multidrug transporters and tripartite assemblies. Alav et al. [16]*

### 1.4.2 Enzymatic degradation of antibiotics

Drugs such as β-lactams and macrolides are modified and degraded by bacteria with the help of specific enzymes which catalyze these processes. For macrolide resistance three classes of enzymes are known:

i. macrolide phosphotransferases (MPHs):
They facilitate the transfer of the γ-phosphate group from Guanosine-5'-triphosphate (GTP) to macrolide substrates, thereby imparting resistance to bacteria including E. coli, Pseudomonas, and Klebsiella [31, 32].

ii. Macrolide Esterases (Eres):
Macrolide drugs like erythromycin, azithromycin, and clarithromycin can be catalytically hydrolysed by a family of enzymes known as esterases. These esterases catalyze the cleavage of the macrolactone ring of the macrolides, rendering them inactive. Among the various esterases, the Erythromycin esterases (Ere) have been extensively studied. Ere can be further subdivided into various classes such as EreA, B, C, and D. Sequence analysis shows that EreC and EreA are closely related with nearly 94% similarity and 90% sequence identity, while EreD and EreB are related somewhat closely with nearly 45% sequence identity. Even though both of these groups belong to the same Ere family, they are highly unrelated, sharing less than 25.4% of the sequence identity [33, 34].

iii. Macrolide glycosyltransferases (MGTs):

Glycosylation is a more recently discovered method to inactivate macrolide molecules by bacteria such as Streptomyces. This is catalyzed by glycosyltransferases [35, 36].

Another group of enzymes called β-lactamases are responsible for the catalytic degradation of β-lactam antibiotics. Based on the homology between amino acid sequences, they are classified as Ambler class A, B, C, and D, with class A, C, and D being structurally similar serine β-lactamases and class B being metallo-β-lactamases. Alternatively, based on the profile of substrate and inhibitor, they are classified as Bush-Jacoby-Medeiros groups 1, 2, 3, and 4 [37-39].

Despite the advancements in antibiotics, the enzymatic degradation of antibiotics by bacteria remains a formidable challenge in the treatment of bacterial infections.

# 2. AcrAB-TolC and MexAB-OprM Efflux Pumps

## 2.1 Literature Survey

### 2.1.1 Structure and Mechanism of action

MexAB-OprM and AcrAB-TolC are secondary active drug efflux transporters of Gram-Negative bacteria belonging to HAE-1 type of RND efflux pumps [23]. The transporter parts (AcrB and MexB) facilitate the efflux of substrates through a substrate/H+ antiport mechanism (proton motive force). These transporters work in coordination with a membrane fusion protein (MFP), also referred to as the periplasmic adaptor protein (PAP), and an outer membrane protein (OMP). Together, they enable the expulsion of substrates, including but not limited to antibiotics, detergents, dyes, heavy metals, and solvents, across the cell membrane [40] (Figure 3). The terminologies "AcrAB" refers to Acriflavine-resistant proteins A & B [41], "Tol" refers to tolerant to colicin [42], "MexAB" refers to Multiple-efflux proteins A & B [43], and "Opr" refers to Outer-membrane protein [44]. Both share a high degree of structural and functional homogeneity and confer critical resistance to chloramphenicol, fluoroquinolones, β-lactams, tetracyclines, and carbapenem-mediated antibiotic activities toward Gram-negative bacteria.

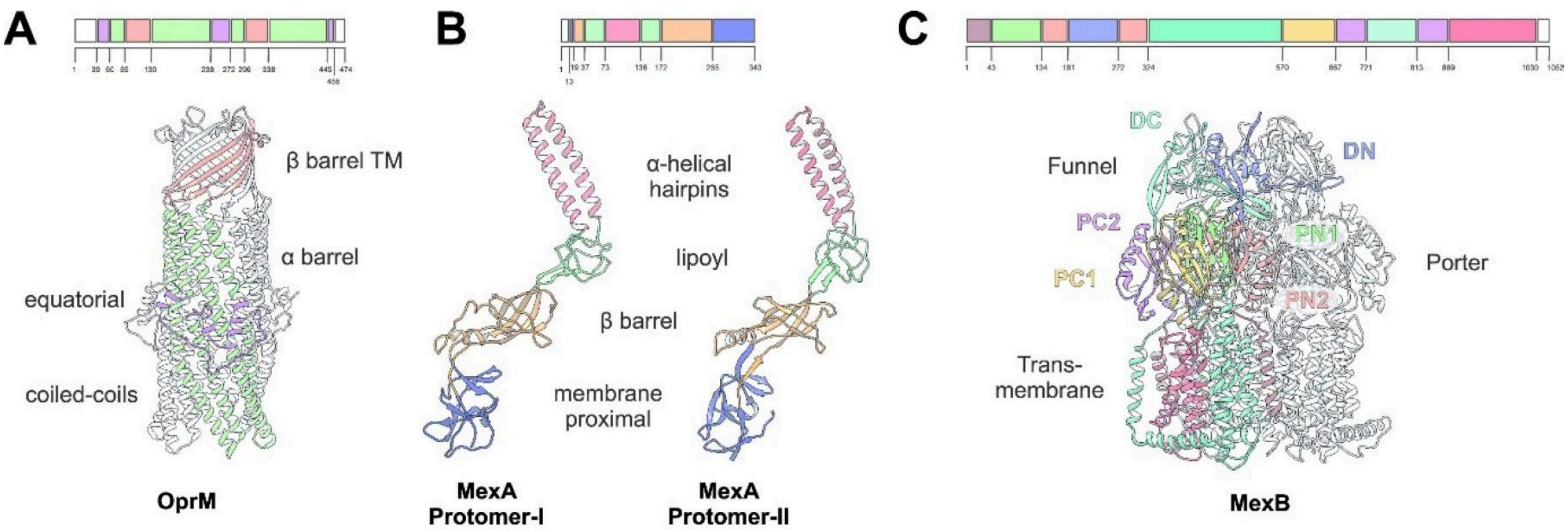

**Figure 3.** *Structure of MexAB-OprM efflux pump. (A) Outer membrane protein (OprM), (B) Periplasmic adaptor protein (MexA), (C) Inner membrane transporter (MexB). [16]*

The inner membrane transporter proteins, AcrB and MexB, exist as homotrimeric structures which are further divided into three domains, the lower 'Transmembrane domain', the middle 'Porter domain', and the top 'Funnel domain'. In each protomer the porter domain is further subdivided into four regions, the periplasmic N-terminal subdomains PN1 and PN2 that protrude from TM1 and TM2 of the transmembrane domain and the C-terminal subdomains PC1 and PC2 that protrude from the TM7 and TM8. The region between the PC1 and PC2 is known as the 'Access pocket' (AP) and the region between PN2 and PC1 is known as the 'Deep binding pocket' (DBP) [45] (Figure 4). A glycine-rich switch loop exists between the two distinct binding pockets, which functions to translocate different kinds of substrates [46, 47]. Two phenylalanine residues within this loop provide it with functional rigidity, which in turn affects the drug translocation activity [48].

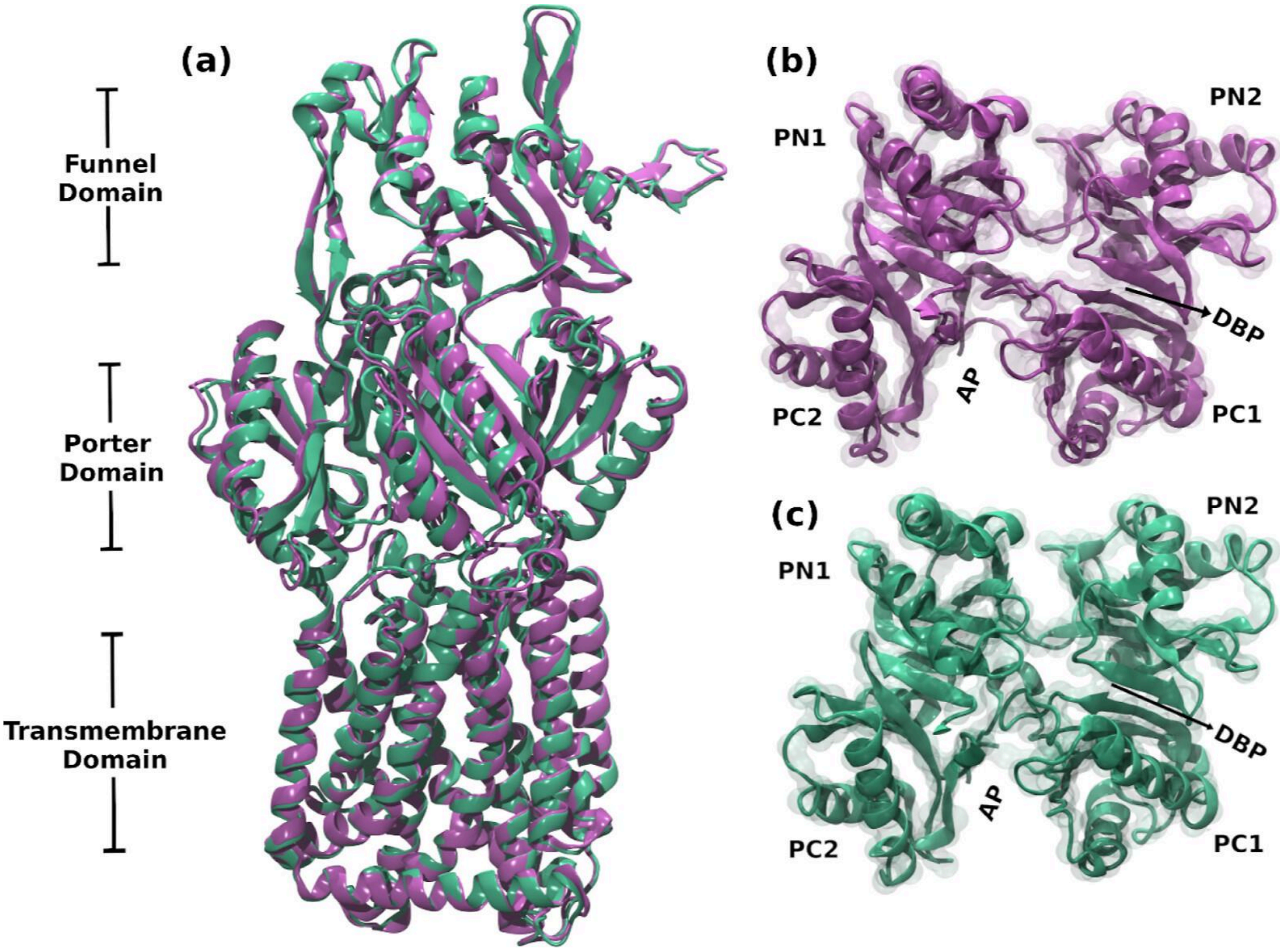

**Figure 4.** *Graphical representation of Tight protomers of AcrB and MexB. (a) Aligned T protomer of MexB (Green) and AcrB (magenta). (b) Porter domain of AcrB. (c) Porter domain of MexB. [49]*

Due to their high structural similarity, the drug efflux mechanism of both AcrB and MexB is almost similar. They actively transport drug-like molecules from the periplasm into the OMP. The mechanism involves the utilization of the proton motive force for the functional rotation of each protomer among three states: loose(L), tight(T), and open(O) (Figure 5). In the L state, the AP is open, while the DBP is closed. At first, the substrate binds to the AP in the L state, triggering a conformational shift to the T state, which opens both AP and DBP. In this state, the volume of AP decreases, causing the substrate to slide beneath the switch loop and enter the DBP (Figure 6). Finally, another conformational shift occurs from the T to O state, where the sole available space is within the funnel-shaped exit tunnel, enabling substrate expulsion [46].

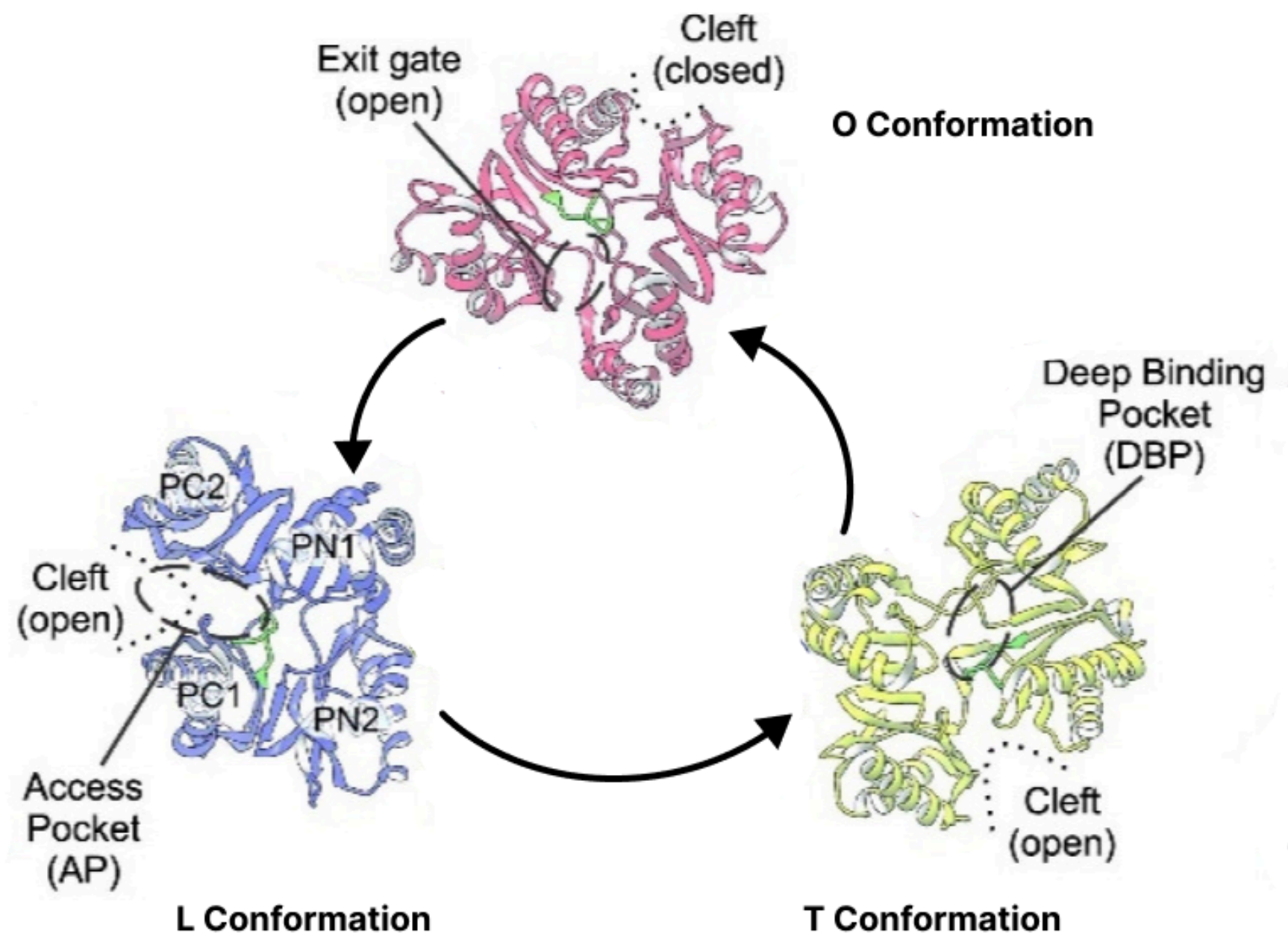

**Figure 5.** *Representative image showing functional rotational mechanism involving conformational changes among Loose(L), Tight(T), and Open(O) states of a protomer of AcrB. [16]*

Followed by this, the PAP (AcrA or MexA) acts as a bridge to enable the transfer of substrates from the inner membrane transporter to the OMP (TolC or OprM). The OMP provides a pathway for the substrates to exit the cell by forming a channel within the bacterial outer membrane [15, 50].

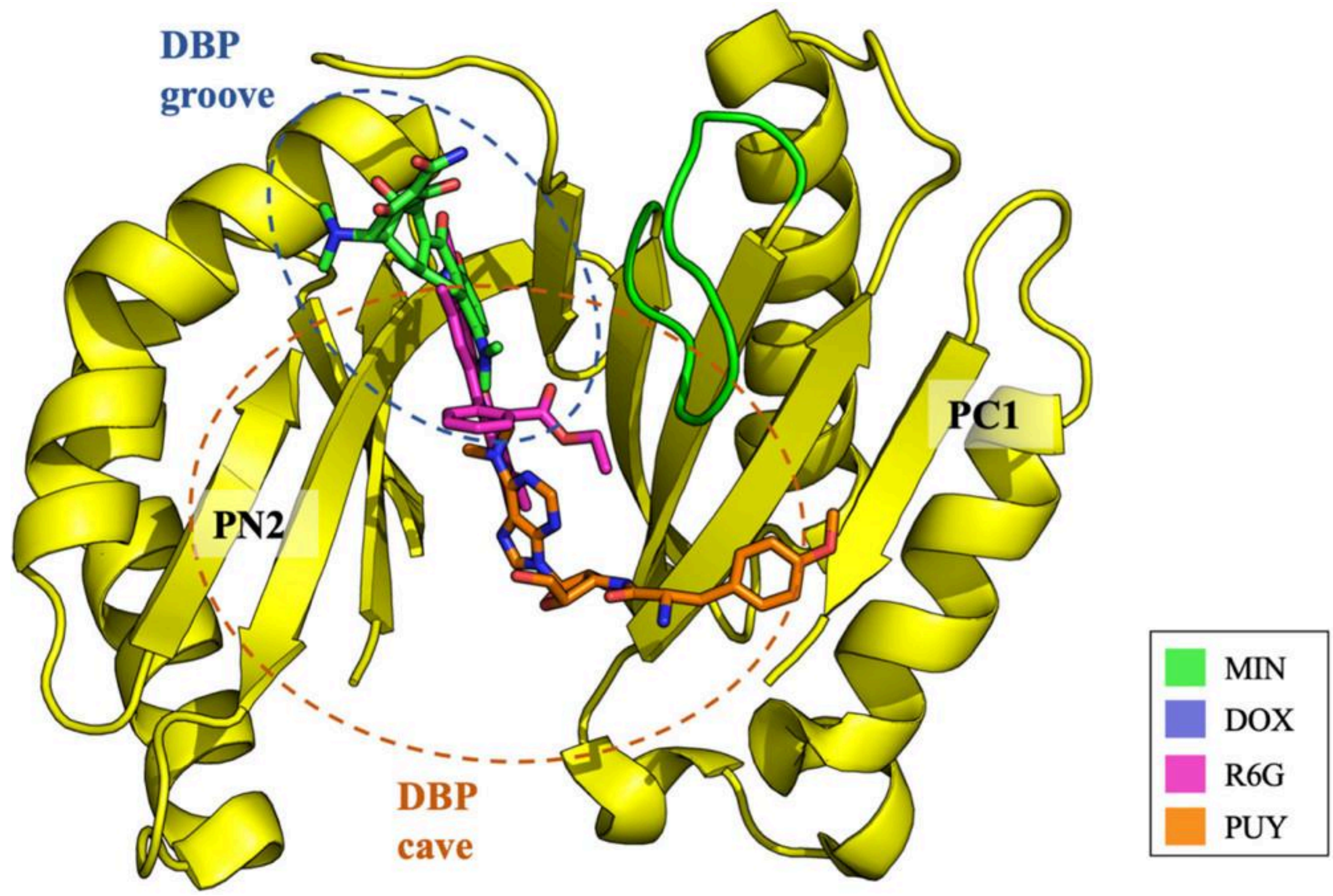

**Figure 6.** *The DBP cave of the AcrB T conformer. The switch-loop is highlighted in green. [16]*

### 2.1.2 Inhibition of Efflux pumps

Targeting and inhibiting the activity of efflux pumps is one of the most effective approaches to address this antimicrobial resistance. A class of chemical compounds called Efflux pump inhibitors (EPI) can impede the functioning of efflux pumps through a diverse set of mechanisms, which include disruption of the proton motive force, inhibition of gene expression, and competitive binding [51]. The most prevalent among these mechanisms is the competitive binding of inhibitors. The competitive-binding EPI binds either to the same binding site as the substrates or to a significantly larger region within the DBP. In addition to this, these inhibitors can arrest the protomer in the T conformation, disrupting the LTO cycle and rendering the efflux mechanism inactive. As a consequence, the presence of these inhibitors enhances the antimicrobial activities [52]. In the literature, many such EPIs have been reported (Table 1). However, D13-9001 and related inhibitors are only reported to have low cytotoxicity. Traditional 'Trial & Error' methods for the discovery and synthesis of novel and non-carcinogenic EPIs are pretty challenging and time-consuming since extensive experimentation and virtual Screenings are necessary.

**Table 1.** *Structure and MIC values of some of the reported EPIs for MexB and AcrB.*

| CLASS | NAME | STRUCTURE | MIC | Ref. |
|---|---|---|---|---|
| Quinoline derivatives | PAβN (MC-207,110) | | 512 | [53] |
| | MC-002,595 | | >512 | [53] |
| Pyranopyridines | MBX2319 | | >100 | [54] |
| | MBX3132 | | >100 | [54] |
| Benzothiazole derivatives | BSN-004 | | 256 | [55] |
| | BSN-006 | | 512 | [55] |
| Pyridopyrimidone | D13-9001 | | >64 | [56] |

## 2.2 Computational Methods

Computational methods can marginally boost up the drug discovery process. An extensive library of drug-like molecules can be screened using Docking and modern Machine Learning techniques to filter out potential drug candidates, which can be further analyzed by molecular dynamics (MD) simulations. Machine learning (ML) is used to predict various properties of data and acquire knowledge about it by using statistical methodologies [57]. Machine learning can further be divided into two broad groups:

I. Supervised Learning

It uses the training dataset which has labels or output variables (human-labeled supervisory signal) defined. It can further be classified as:

a. Regression: For continuous output variables.
b. Classification: For discrete output variables.

II. Unsupervised Learning

It uses the training data which do not have any labels or output variables, it groups the data points using the similarities and dissimilarities among the input variables [58].

To perform a supervised machine learning operation we need a known dataset to train the model and predict the output variable for the unknown data. To get the training dataset objectives a number of proposed EPIs were collected from the literature, along with their intrinsic minimum
inhibitory concentration (MIC) values, which served as targets or output variables. These MIC values are not precisely mentioned in the literature, rather mentioned with a '>' sign. Since the choice for the construction of the model, 'Regression' or 'Classification' is based on the nature of the output variables, we decided to build classification models for this study. The models were trained upon various molecular descriptors used as features for the dataset. These trained models were then employed for the in-silico prediction of potential novel EPIs using an unknown set of generated molecules. The predicted compounds then underwent rigorous filtering based on various criteria to obtain a final selection of highly effective potential EPIs (Figure 7).

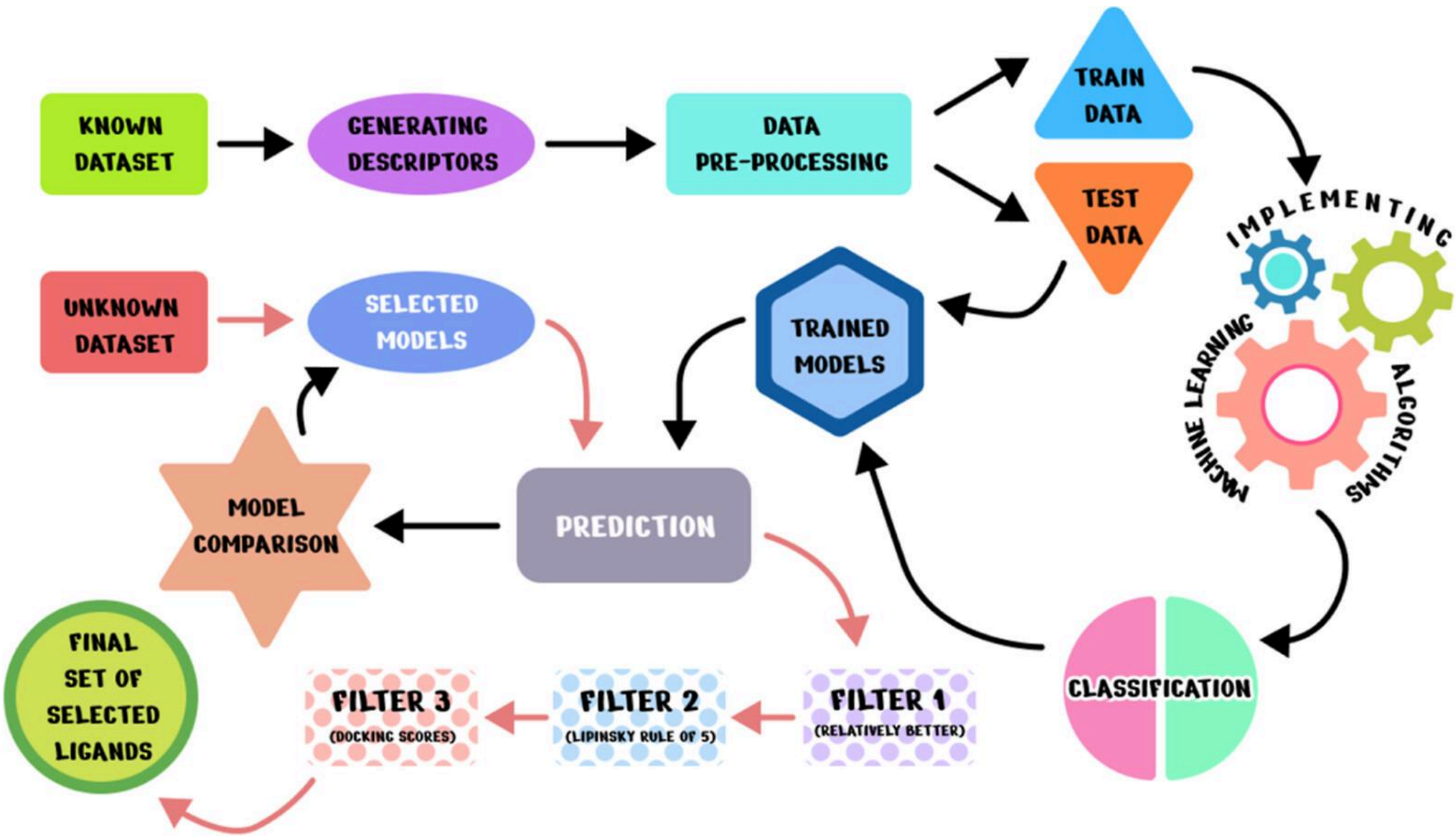

**Figure 7.** *Workflow of the implemented machine learning model in this study. [49]*

### 2.2.1 Data curation

To get the known dataset, a total of 53 effective AcrB and MexB efflux pump inhibitors were collected from the literature (Table 2), while the unknown dataset consisting of 5043 molecules was randomly generated, using a diverse set of organic moieties and functional groups attached to five selected pharmacophores (Figure 8). These pharmacophores are so chosen because they appear frequently in the molecules of the known dataset, and their structure‒activity relationships (SAR), synthetic routes, pharmacological properties, and in vitro and in vivo activities have been thoroughly established [59].

**Table 2.** *Effective AcrB and MexB efflux pump inhibitors collected from literature and used as the known dataset for the machine learning models. [49]*

| Sl. No. | IUPAC Name | SMILES | MIC (µg/mL) |
|---|---|---|---|
| 1 | (2S)-2-[[(2S)-2-amino-3-phenylpropanoyl]amino]-5-carbamimidamido-N-naphthalen-2-ylpentanamide | c1ccc(cc1)C[C@@H](C(=O)N[C@@H](CCCNC(=N)N)C(=O)Nc1cc2ccccc2cc1)N | 512 |
| 2 | (2S)-6-amino-2-[[(2S)-2-amino-3-phenylpropanoyl]amino]-N-naphthalen-2-ylhexanamide | c1ccc(cc1)C[C@@H](C(=O)N[C@@H](CCCCN)C(=O)Nc1cc2ccccc2cc1)N | 512 |
| 3 | (2S)-5-amino-2-[[(2S)-2-amino-3-phenylpropanoyl]amino]-N-naphthalen-2-ylpentanamide | c1ccc(cc1)C[C@@H](C(=O)N[C@@H](CCCN)C(=O)Nc1cc2ccccc2cc1)N | 512 |
| 4 | (2S)-2-amino-N-[(2S)-3-(1H-imidazol-4-yl)-1-(naphthalen-2-ylamino)-1-oxopropan-2-yl]-3-phenylpropanamide | c1ccc(cc1)C[C@@H](C(=O)N[C@@H](Cc1c[nH]cn1)C(=O)Nc1cc2ccccc2cc1)N | 512 |
| 5 | (2S)-5-amino-2-[[(2S)-2-amino-3-phenylpropanoyl]-methylamino]-N-naphthalen-2-ylpentanamide | c1ccc(cc1)C[C@@H](C(=O)N([C@@H](CCCN)C(=O)Nc1cc2ccccc2cc1)C)N | 256 |
| 6 | (2S)-5-amino-2-[[(2S)-2-amino-3-(4-fluorophenyl)propanoyl]amino]-N-naphthalen-2-ylpentanamide | c1(ccc(cc1)C[C@@H](C(=O)N[C@@H](CCCN)C(=O)Nc1cc2ccccc2cc1)N)F | 512 |
| 7 | (2S)-5-amino-2-[[(2S)-2-aminopropanoyl]amino]-N-naphthalen-2-ylpentanamide | C[C@@H](C(=O)N[C@@H](CCCN)C(=O)Nc1cc2ccccc2cc1)N | 512 |
| 8 | (2S)-2,5-diamino-N-naphthalen-2-ylpentanamide | N[C@@H](CCCN)C(=O)Nc1cc2ccccc2cc1 | 512 |
| 9 | (2S)-5-amino-2-[[(2S)-2-amino-4-phenylbutanoyl]amino]-N-naphthalen-2-ylpentanamide, | c1ccc(cc1)CC[C@@H](C(=O)N[C@@H](CCCN)C(=O)Nc1cc2ccccc2cc1)N | 128 |
| 10 | (2S)-5-amino-2-[[(2S)-2-amino-4-phenylbutanoyl]amino]-N-(2,3-dihydro-1H-inden-5-yl)pentanamide | c1ccc(cc1)CC[C@@H](C(=O)N[C@@H](CCCN)C(=O)Nc1cc2c(cc1)CCC2)N | 256 |
| 11 | (2S)-5-amino-2-[[(2S)-2-amino-4-phenylbutanoyl]amino]-N-phenylpentanamide | c1ccc(cc1)CC[C@@H](C(=O)N[C@@H](CCCN)C(=O)Nc1ccccc1)N | 512 |
| 12 | (2S)-5-amino-2-[[(2S)-2-amino-4-phenylbutanoyl]amino]-N-quinolin-6-ylpentanamide | c1ccc(cc1)CC[C@@H](C(=O)N[C@@H](CCCN)C(=O)Nc1cc2cccnc2cc1)N | 512 |
| 13 | (2S)-5-amino-2-[[(2S)-2-amino-4-phenylbutanoyl]amino]-N-quinolin-3-ylpentanamide | c1ccc(cc1)CC[C@@H](C(=O)N[C@@H](CCCN)C(=O)Nc1cc2ccccc2nc1)N | 512 |
| 14 | (2S)-2-amino-N-[(2S)-5-amino-1-(1,3-benzothiazol-2-ylsulfanyl)pentan-2-yl]-N-methyl-4-phenylbutanamide | c1ccc(cc1)CC[C@@H](C(=O)N([C@@H](CCCN)CSc1sc2ccccc2n1)C)N | 128 |
| 15 | (2R,4R)-4-(aminomethyl)-N-[(2R)-1-oxo-4-phenyl-1-(quinolin-6-ylamino)butan-2-yl]pyrrolidine-2-carboxamide | O=C(N[C@H](CCc1ccccc1)C(=O)Nc1cc2c(nccc2)cc1)[C@@H]1NC[C@H](C1)CN | 512 |
| 16 | (2R,4R)-4-(2-aminoethoxy)-N-[(2R)-1-oxo-4-phenyl-1-(quinolin-6-ylamino)butan-2-yl]pyrrolidine-2-carboxamide | O=C(N[C@H](CCc1ccccc1)C(=O)Nc1cc2c(nccc2)cc1)[C@@H]1NC[C@@H](C1)OCCN | 512 |
| 17 | (2R,4R)-4-(aminomethyl)-N-(6-tert-butylquinolin-3-yl)pyrrolidine-2-carboxamide | O=C(Nc1cc2c(ccc(C(C)(C)C)c2)nc1)[C@@H]1NC[C@H](C1)CN | 128 |
| 18 | (2R)-2,5-diamino-N-[2-[(6-tert-butylquinolin-3-yl)amino]-2-oxoethyl]pentanamide | O=C(Nc1cc2c(ccc(C(C)(C)C)c2)nc1)CNC(=O)[C@@H](CCCN)N | 64 |
| 19 | 5-hydroxy-2-methylnaphthalene-1,4-dione | CC1=CC(=O)c2c(C1=O)cccc2O | 128 |
| 20 | 4-[4-(3,4-dihydroxyphenyl)-2,3-dimethylbutyl]benzene-1,2-diol | C[C@@H](Cc1cc(c(cc1)O)O)[C@@H](C)Cc1cc(c(cc1)O)O | 512 |
| 21 | 4-[5-(3,4-dihydroxyphenyl)pentyl]benzene-1,2-diol | C(CCCCc1cc(c(cc1)O)O)c1cc(c(cc1)O)O | 512 |
| 22 | 5-[4-(3,4-dihydroxyphenyl)butyl]benzene-1,2,3-triol | Oc1cc(ccc1O)CCCCc1cc(c(c(c1)O)O)O | 512 |
| 23 | 4-(4-pyren-4-ylbutyl)benzene-1,2-diol | c1ccc2c(cc3cccc4ccc1c2c34)CCCCc1cc(c(cc1)O)O | 512 |
| 24 | 4-[8-(3,4-dihydroxyphenyl)octyl]benzene-1,2-diol | C(CCCCCc1cc(c(cc1)O)O)CCc1cc(c(cc1)O)O | 512 |
| 25 | 3,5-diamino-N-[2-(3,4-dihydroxyphenyl)ethyl]benzamide | Nc1cc(C(=O)NCCc2cc(c(cc2)O)O)cc(c1)N | 512 |
| 26 | N-[2-(3,4-dihydroxyphenyl)ethyl]-3,5-dihydroxybenzamide | Oc1cc(CCNC(=O)c2cc(cc(c2)O)O)ccc1O | 512 |
| 27 | N-[2-(3,4-dihydroxyphenyl)ethyl]-3,4,5-trihydroxybenzamide | Oc1cc(CCNC(=O)c2cc(c(c(c2)O)O)O)ccc1O | 512 |
| 28 | 4-amino-N-[2-(3,4-dihydroxyphenyl)ethyl]benzamide | Nc1ccc(C(=O)NCCc2cc(c(cc2)O)O)cc1 | 512 |

| 29 | 2-amino-N-[2-(3,4-dihydroxyphenyl)ethyl]benzamide | Nc1c(C(=O)NCCc2cc(c(cc2)O)O)cccc1 | 512 |
|---|---|---|---|
| 30 | 2-(3,4-dihydroxyphenyl)-3,5,7-trihydroxychromen-4-one | c1cc(c(cc1c1c(c(=O)c2c(cc(cc2o1)O)O)O)O)O | 1024 |
| 31 | 1,3,6,7-tetrahydroxy-2-[(2S,3R,4R,5S,6R)-3,4,5-trihydroxy-6-(hydroxymethyl)oxan-2-yl]xanthen-9-one | c1c2c(cc(c1O)O)oc1c(c2=O)c(c(c(c1)O)[C@H]1[C@@H]([C@H]([C@@H]([C@H](O1)CO)O)O)O)O | 512 |
| 32 | 5,8-dihydroxy-2-[(1S)-1-hydroxy-4-methylpent-3-enyl]naphthalene-1,4-dione | CC(=CC[C@@H](C1=CC(=O)c2c(ccc(c2C1=O)O)O)O)C | 256 |
| 33 | EA-371α | CCCCCc1cc2CCc3c(O)c4C(=O)c5c(O)cc(OS(=O)(=O)O)cc5C(C)(C)c4cc3-c2c(O)c1C(=O)O | 512 |
| 34 | EA-371δ | CCCCCc1cc2ccc3c(O)c4C(=O)c5c(O)cc(OS(=O)(=O)O)cc5C(C)(C)c4cc3c2c(O)c1C(=O)O | 512 |
| 35 | 2-[2-[[(3R)-1-[8-[(4-tert-butyl-1,3-thiazol-2-yl)carbamoyl]-4-oxo-3-[(E)-2-(2H-tetrazol-5-yl)ethenyl]pyrido[1,2-a]pyrimidin-2-yl]piperidin-3-yl]oxycarbonylamino]ethyl-dimethylazaniumyl]acetate | CC(C)(C)c1csc(n1)NC(=O)c1cc2nc(c(c(=O)n2cc1)/C=C/c1n[nH]nn1)N1CCC[C@H](C1)OC(=O)NCC[N+](C)(C)CC(=O)[O-] | 64 |
| 36 | 3,3-dimethyl-8-morpholin-4-yl-6-(2-phenylethylsulfanyl)-1,4-dihydropyrano[3,4-c]pyridine-5-carbonitrile | CC1(Cc2c(CO1)c(nc(c2C#N)SCCc1ccccc1)N1CCOCC1)C | 41 |
| 37 | N-[4-[2-[[5-cyano-8-[(2S,6R)-2,6-dimethylmorpholin-4-yl]-3,3-dimethyl-1,4-dihydropyrano[3,4-c]pyridin-6-yl]sulfanyl]ethyl]phenyl]acetamide | CC1(Cc2c(CO1)c(nc(c2C#N)SCCc1ccc(cc1)NC(=O)C)N1C[C@H](O[C@H](C1)C)C)C | 49 |
| 38 | N-[4-[2-[[5-cyano-8-[(2S,6R)-2,6-dimethylmorpholin-4-yl]-3,3-dimethyl-1,4-dihydropyrano[3,4-c]pyridin-6-yl]sulfanyl]ethyl]phenyl]prop-2-enamide | CC1(Cc2c(CO1)c(nc(c2C#N)SCCc1ccc(cc1)NC(=O)C=C)N1C[C@H](O[C@H](C1)C)C)C | 51 |
| 39 | 3-O-acetyl-urs-12-en-28-isopropyl ester | CC(OC(=O)[C@]12CC[C@H]([C@@H]([C@@H]1C1=CC[C@@H]3[C@@]([C@@]1(CC2)C)(C)CC[C@@H]1[C@]3(C)CC[C@@H](C1(C)C)OC(=O)C)C)C)C | 500 |
| 40 | 3-O-acetyl-urs-12-en-28-n-butyl ester | CCCCOC(=O)[C@]12CC[C@H]([C@@H]([C@@H]1C1=CC[C@@H]3[C@@]([C@@]1(CC2)C)(C)CC[C@@H]1[C@]3(C)CC[C@@H](C1(C)C | 1000 |

| | | )OC(=O)C)C)C | |
|---|---|---|---|
| 41 | N-[4-(1,3-benzothiazol-2-yl)phenyl]benzamide | c1ccc(cc1)C(=O)Nc1ccc(cc1)c1nc2c(s1)cccc2 | 256 |
| 42 | N-[4-(1,3-benzothiazol-2-yl)phenyl]-4-ethylbenzamide | c1(ccc(cc1)C(=O)Nc1ccc(cc1)c1nc2c(s1)cccc2)CC | 128 |
| 43 | N-[4-(1,3-benzothiazol-2-yl)phenyl]-2-(4-methoxyphenyl)acetamide | COc1ccc(cc1)CC(=O)Nc1ccc(cc1)c1nc2c(s1)cccc2 | 256 |
| 44 | N-[4-(1,3-benzothiazol-2-yl)phenyl]-2-(4-fluorophenyl)acetamide | c1(ccc(cc1)CC(=O)Nc1ccc(cc1)c1nc2c(s1)cccc2)F | 128 |
| 45 | N-[4-(1,3-benzothiazol-2-yl)phenyl]-2-(4-methylphenyl)acetamide | O=C(Cc1ccc(cc1)C)Nc1ccc(cc1)c1nc2c(s1)cccc2 | 512 |
| 46 | N-[4-(1,3-benzothiazol-2-ylmethyl)phenyl]-4-ethylbenzamide | O=C(c1ccc(cc1)CC)Nc1ccc(Cc2nc3c(s2)cccc3)cc1 | 64 |
| 47 | N-[4-(1,3-benzothiazol-2-yl)phenyl]-3-phenylpropanamide | O=C(Nc1ccc(cc1)c1nc2c(s1)cccc2)CCc1ccccc1 | 512 |
| 48 | (S)-[2,8-bis(trifluoromethyl)quinolin-4-yl]-[(2S)-piperidin-2-yl]methanol | C1CCN[C@@H](C1)[C@H](c1cc(nc2c1cccc2C(F)(F)F)C(F)(F)F)O | 32 |
| 49 | (3S)-4,5,6-trimethoxy-3-(3,4,5-trimethoxyphenyl)-2,3-dihydroinden-1-one | COc1c(OC)cc2c(c1OC)[C@@H](CC2=O)c1cc(OC)c(c(c1)OC)OC | 500 |
| 50 | 16,17-dimethoxy-5,7-dioxa-13-azoniapentacyclo[11800^2,100^4,80^15,20]henicosa-1(13),2,4(8),9,14,16,18,20-octaene | COc1c(c2c[n+]3c(-c4cc5c(cc4CC3)OCO5)cc2cc1)OC | 512 |
| 51 | 16,17-dimethoxy-21-[(2-methylphenyl)methyl]-5,7-dioxa-13-azoniapentacyclo[11800^2,100^4,80^15,20]henicosa-1(13),2,4(8),9,14,16,18,20-octaene | COc1c(c2c[n+]3c(-c4cc5c(cc4CC3)OCO5)c(c2cc1)Cc1ccccc1C)OC | 512 |
| 52 | 5,7,17,19-tetraoxa-13-azoniahexacyclo[111100^2,100^4,80^15,230^16,20]tetracosa-1(24),2,4(8),9,13,15(23),16(20),21-octaene | C1C[n+]2c(-c3cc4c(cc13)OCO4)cc1ccc3c(c1c2)OCO3 | 512 |
| 53 | 1-(naphthalen-1-ylmethyl)piperazine | C1CN(CCN1)Cc1cccc2ccccc12 | 400 |

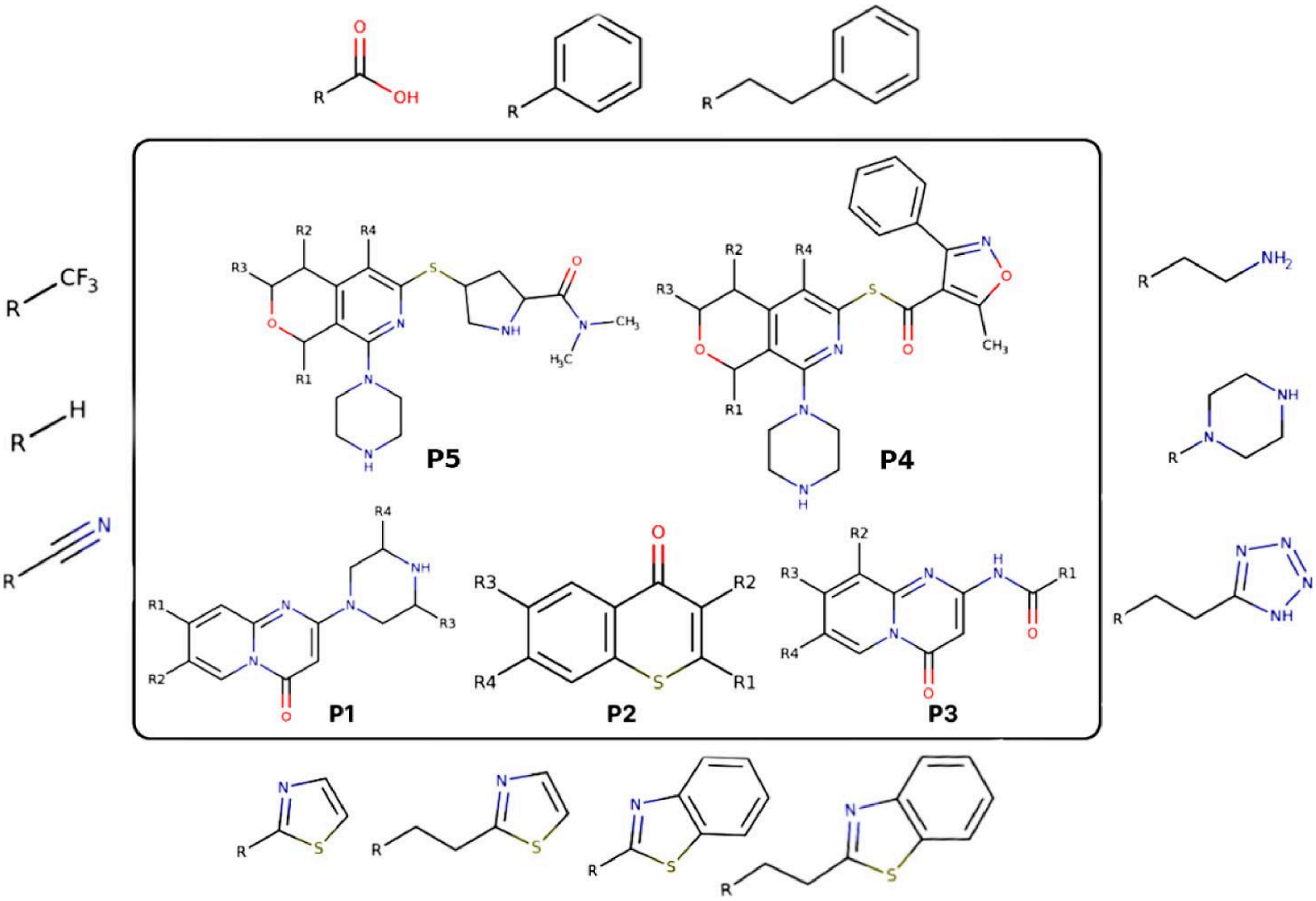

**Figure 8.** *5 Selected pharmacophores (P1 to P5 inside the box) and 13 molecular fragments (outside the box) used to generate the unknown dataset of 5043 molecules. [49]*

### 2.2.2 Calculation of descriptors and molecular properties

BIOVIA Discovery Studios 2022 was used to calculate more than 200 descriptors for the molecules of known and unknown datasets [60]. These descriptors include various molecular properties, as well as ADMET and TOPKAT properties (Table 3). The crystal structures for both MexB (PDB ID: 3W9J) [52] and AcrB (PDB ID: 4DX5) [46] were obtained from the Protein Data Bank. Subsequently, known molecules were docked into these structures using AutoDock Vina [61, 62] to get the docking scores, which were then treated as a descriptor. The descriptors with irrelevant and incomplete data were removed, which reduced the dimensionality of our features to 162. The correlation between these features is shown using Pearson's correlation coefficients [63] (Figure 9). The "OneHotEncoder" function of the Scikit-Learn Python library [64] was used to encode all the non-numerical data before further preprocessing.

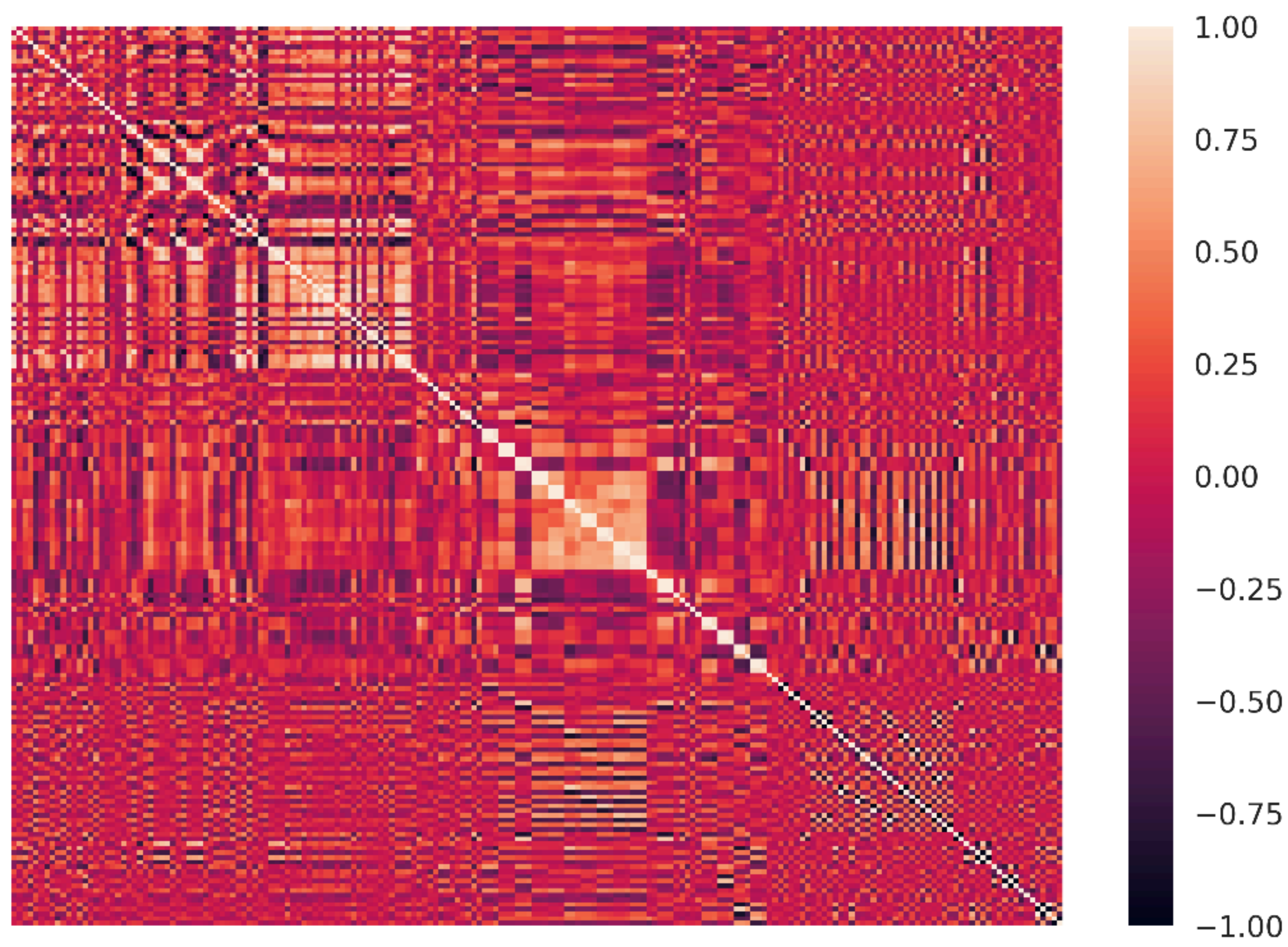

**Figure 9.** *Pearson's Correlation coefficients for all the features/descriptors used for the machine learning models. [49]*

**Table 3.** *List of features/descriptors generated using BIOVIA Discovery Studios 2022 for the machine learning models. [49]*

| **Descriptors List** | | | |
|---|---|---|---|
| Molecular Weight | Jurs_RPCS | Jurs_RPCG | ADMET_EXT_Hepatotoxic#Prediction |
| Exact Mol. Weight | Jurs_RPSA | ADMET_EXT_Hepatotoxic_Applicability#MD | TOPKAT_Rat_Female_FDA_None_vs_Carcinogen_Score |
| Net Formal Charge | Jurs_SASA | ADMET_EXT_Hepatotoxic_Applicability#MDpvalue | TOPKAT_Rat_Male_FDA_None_vs_Carcinogen_Prediction |
| Clean Energy | Jurs_TASA | ADMET_EXT_PPB | TOPKAT_Rat_Male_FDA_None_vs_Carcinogen_Probability |
| ALogP | Jurs_TPSA | ADMET_EXT_PPB#Prediction | TOPKAT_Rat_Male_FDA_None_vs_Carcinogen_Enrichment |
| Molecular_Weight | Jurs_WNSA_1 | ADMET_EXT_PPB_Applicability#MD | TOPKAT_Rat_Male_FDA_None_vs_Carcinogen_Score |
| Num_H_Donors | Jurs_WNSA_2 | ADMET_EXT_PPB_Applicability#MDpvalue | TOPKAT_WOE_Prediction |
| Num_H_Acceptors | Jurs_WNSA_3 | TOPKAT_Ames_Prediction | TOPKAT_WOE_Probability |

| Num_RotatableBonds | Jurs_WPSA_1 | TOPKAT_Ames_Probability | TOPKAT_WOE_Enrichment |
|---|---|---|---|
| Num_Rings | Jurs_WPSA_2 | TOPKAT_Ames_Enrichment | TOPKAT_WOE_Score |
| Num_AromaticRings | Jurs_WPSA_3 | TOPKAT_Ames_Score | TOPKAT_Carcinogenic_Potency_TD50_Mouse |
| Molecular_FractionalPolarSurfaceArea | AverageBondLength | TOPKAT_Mouse_Female_NTP_Prediction | TOPKAT_Carcinogenic_Potency_TD50_Rat |
| C_Count | Energy | TOPKAT_Mouse_Female_NTP_Probability | TOPKAT_DTP_Prediction |
| F_Count | Minimized_Energy | TOPKAT_Mouse_Female_NTP_Enrichment | TOPKAT_DTP_Probability |
| H_Count | RadOfGyration | TOPKAT_Mouse_Female_NTP_Score | TOPKAT_DTP_Enrichment |
| N_Count | Strain_Energy | TOPKAT_Mouse_Male_NTP_Prediction | TOPKAT_DTP_Score |
| O_Count | PMI_mag | TOPKAT_Mouse_Male_NTP_Probability | TOPKAT_Rat_Oral_LD50 |
| S_Count | PMI_X | TOPKAT_Mouse_Male_NTP_Enrichment | TOPKAT_Rat_Maximum_Tolerated_Dose_Feed |
| Dipole_mag | PMI_Y | TOPKAT_Mouse_Male_NTP_Score | TOPKAT_Rat_Maximum_Tolerated_Dose_Gavage |
| Dipole_X | PMI_Z | TOPKAT_Rat_Female_NTP_Prediction | TOPKAT_Rat_Inhalational_LC50 |
| Dipole_Y | Shadow_nu | TOPKAT_Rat_Female_NTP_Probability | TOPKAT_Chronic_LOAEL |
| Dipole_Z | Shadow_Xlength | TOPKAT_Rat_Female_NTP_Enrichment | TOPKAT_Skin_Irritancy |
| Jurs_DPSA_1 | Shadow_XY | TOPKAT_Rat_Female_NTP_Score | TOPKAT_Skin_Irritancy_None_vs_Irritant_Prediction |
| Jurs_DPSA_2 | Shadow_XYfrac | TOPKAT_Rat_Male_NTP_Prediction | TOPKAT_Skin_Irritancy_None_vs_Irritant_Probability |
| Jurs_DPSA_3 | Shadow_XZ | TOPKAT_Rat_Male_NTP_Probability | TOPKAT_Skin_Irritancy_None_vs_Irritant_Enrichment |
| Jurs_FNSA_1 | Shadow_XZfrac | TOPKAT_Rat_Male_NTP_Enrichment | TOPKAT_Skin_Irritancy_None_vs_Irritant_Score |
| Jurs_FNSA_2 | Shadow_Ylength | TOPKAT_Rat_Male_NTP_Score | TOPKAT_Skin_Sensitization |
| Jurs_FNSA_3 | Shadow_YZ | TOPKAT_Mouse_Female_FDA | TOPKAT_Skin_Sensitization_None_vs_Sensitizer_Prediction |
| Jurs_FPSA_1 | Shadow_YZfrac | TOPKAT_Mouse_Male_FDA | TOPKAT_Skin_Sensitization_None_vs_Sensitizer_Probability |
| Jurs_FPSA_2 | Shadow_Zlength | TOPKAT_Rat_Female_FDA | TOPKAT_Skin_Sensitization_None_vs_Sensitizer_Enrichment |
| Jurs_FPSA_3 | Molecular_3D_PolarSASA | TOPKAT_Rat_Male_FDA | TOPKAT_Skin_Sensitization_None_vs_Sensitizer_Score |
| Jurs_PNSA_1 | Molecular_3D_SASA | TOPKAT_Mouse_Female_FDA_None_vs_Carcinogen_Prediction | TOPKAT_Ocular_Irritancy |

| Jurs_PNSA_2 | Molecular_3D_SAVol | TOPKAT_Mouse_Female_FDA_None_vs_Carcinogen_Probability | TOPKAT_Ocular_Irritancy_None_vs_Irritant_Prediction |
|---|---|---|---|
| Jurs_PNSA_3 | Molecular_Volume | TOPKAT_Mouse_Female_FDA_None_vs_Carcinogen_Enrichment | TOPKAT_Ocular_Irritancy_None_vs_Irritant_Probability |
| Jurs_PPSA_1 | RandomNumber | TOPKAT_Mouse_Female_FDA_None_vs_Carcinogen_Score | TOPKAT_Ocular_Irritancy_None_vs_Irritant_Enrichment |
| Jurs_PPSA_2 | ADMET_EXT_CYP2D6 | TOPKAT_Mouse_Male_FDA_None_vs_Carcinogen_Prediction | TOPKAT_Ocular_Irritancy_None_vs_Irritant_Score |
| Jurs_PPSA_3 | ADMET_EXT_CYP2D6#Prediction | TOPKAT_Mouse_Male_FDA_None_vs_Carcinogen_Probability | TOPKAT_Aerobic_Biodegradability_Prediction |
| Jurs_RASA | ADMET_EXT_CYP2D6_Applicability#MD | TOPKAT_Mouse_Male_FDA_None_vs_Carcinogen_Enrichment | TOPKAT_Aerobic_Biodegradability_Probability |
| Jurs_RNCG | ADMET_EXT_CYP2D6_Applicability#MDpvalue | TOPKAT_Mouse_Male_FDA_None_vs_Carcinogen_Score | TOPKAT_Aerobic_Biodegradability_Enrichment |
| Jurs_RNCS | ADMET_EXT_Hepatotoxic | TOPKAT_Rat_Female_FDA_None_vs_Carcinogen_Prediction | TOPKAT_Aerobic_Biodegradability_Score |
| TOPKAT_Daphnia_EC50 | TOPKAT_Fathead_Minnow_LC50 | TOPKAT_Rat_Female_FDA_None_vs_Carcinogen_Probability | TOPKAT_Rat_Female_FDA_None_vs_Carcinogen_Enrichment |

### 2.2.3 Docking Studies

Docking of molecules from both known and unknown datasets were performed on T state protomers of AcrB and MexB, which are obtained from RCSB. All the structures of molecules (ligands) were constructed using Avogadro 1.97 [65], following this a UFF optimization was done [66]. Vina-compatible receptor and ligand 'PDBQTs' were generated using Auto Dock Tools [61,62]. For the grid box setup, the dimensions were set as 26 Å × 26 Å × 26 Å with a grid spacing of 1.0, keeping the rest of the parameters as Vina default. The grid box was built around the DBP of both AcrB and MexB including the hydrophobic residues such as PHE-136, PHE-178, PHE-281, PHE-573 in MexB, PHE-610, PHE-615, PHE-617, and PHE-628 [47, 52]. The docking scores of the known set of molecules is shown in Figure 10.

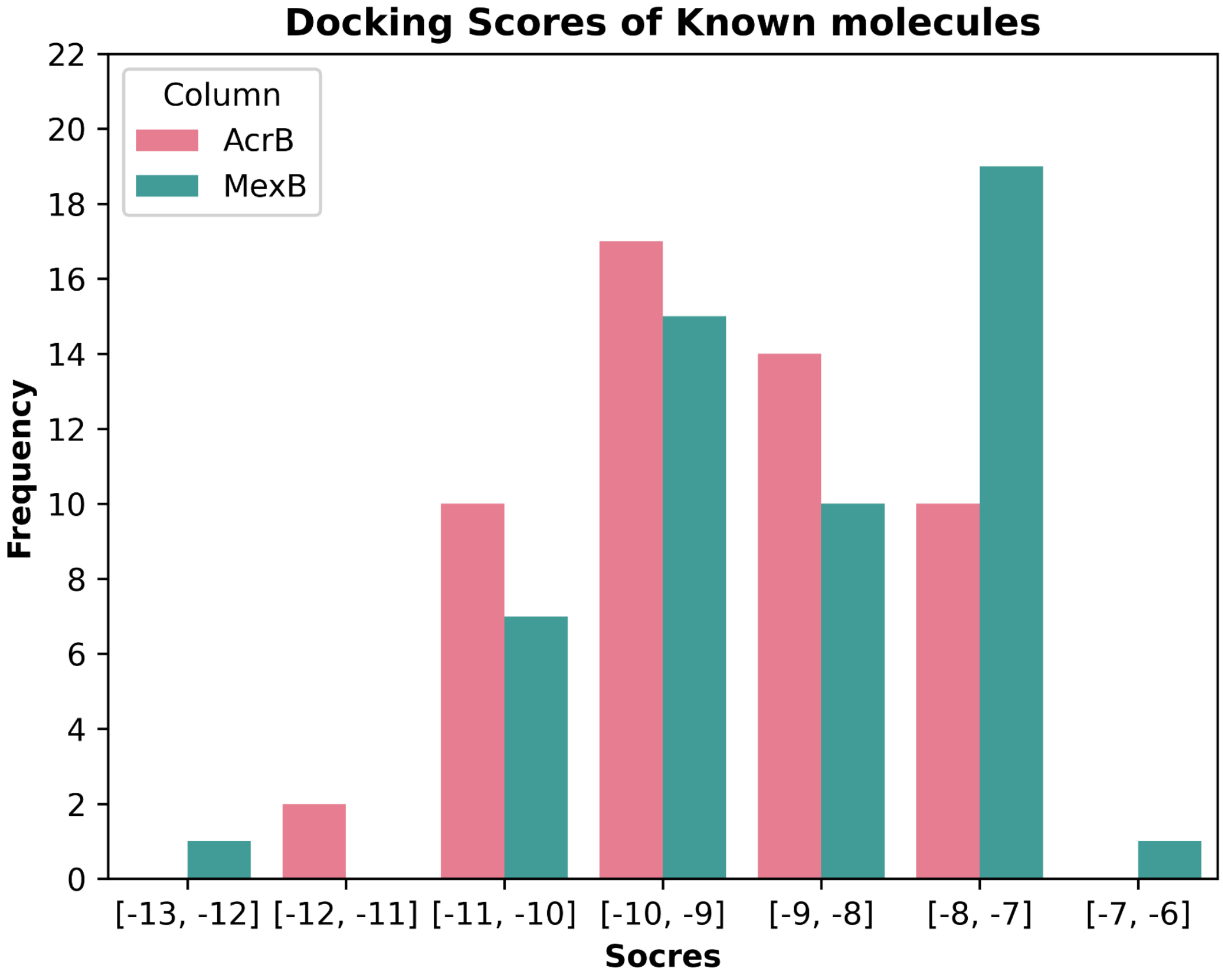

**Figure 10.** *Frequency distribution of 53 known molecules representing their range of docking scores in AcrB (Green) and MexB (Magenta).*

### 2.2.4 Machine Learning Models

2.2.4.1 Model Setup

The data was preprocessed using the Python libraries NumPy and Pandas in Jupyter Notebook [67, 68]. Following this various Supervised classification and Unsupervised clustering machine learning algorithms were employed to build a robust pipeline inorder to obtain predictions as accurate as possible:

I. Unsupervised Learning

a. K-Means Clustering

It is an unsupervised ML algorithm that allocates data points to one of the "K" groups called clusters, based on their distance from the centroids of the clusters. At first the centroids are randomly assigned in space. Subsequently, each data point is allocated to the nearest cluster centroid. Following this assignment, new centroids are recalculated for the clusters. This iterative process continues until an optimal clustering arrangement is attained [69].

b. Gaussian Mixture model

It's a statistical model that posits the acquisition of all data points from a mixture of a finite set of Gaussian distributions with unspecified parameters. Essentially, it extends the concept of k-means clustering by incorporating supplementary data insights, including the covariance structure and the centroids of the latent Gaussians [70].

II. Supervised Learning

a. Logistic Regression

It's a supervised machine learning method primarily applied in classification scenarios. It utilizes the output from the linear regression function and employs a sigmoid function($\sigma(z) = \frac{1}{1-e^{z}}$ ) to assess the probability of the given class. As z approaches ∞, σ(z) approaches 1, and as z approaches -∞, σ(z) approaches 0, ensuring that σ(z) always falls within the range of 0 to 1 [71].

b. K Nearest Neighbours

It is a non-parametric supervised learning algorithm which can be used for regressions as well as classification tasks. It uses proximal distance-based voting (in case of classification) or averaging (in case of regression) to make predictions. The distance computation can be done using Euclidean distance (c=2), Manhattan distance (c=1), and Minkowski distance (Eq. 1) [72].

$$d(x, y) = \sqrt[c]{\sum_{i} |x_i - y_i|^c} \qquad (1)$$

c. Support Vector Machine

This algorithm creates the best line or decision boundary, called a hyperplane, that can segregate n-dimensional space into classes so that new data points can be put in the correct category in the future. The features present in the dataset govern the dimensions of the hyperplane. Support Vectors are the data points that are closest to the hyperplane, and they also affect its position [73].

d. Decision Tree

It is a versatile supervised ML algorithm that can be used for regressions as well as classification tasks. The tree structure can be thought of as a flowchart consisting of the Root Node (representing the complete dataset), internal nodes (representing features), branches (representing rules), and leaf nodes (representing the results of the algorithm). An internal node represents a decision concerning an input feature, with branching linking them to leaf nodes or other internal nodes. A leaf or terminal node does not have any child nodes that indicate a class label or a numerical value. A branch is a subsection of the decision tree that starts at an internal node and ends at the leaf nodes [74].

e. Gaussian Naive Bayes

These classification techniques rely on Bayes' Theorem, which calculates the probability of occurrence of an event (y) given the probability of another event (X) that has already happened (Eq. 2). In Gaussian Naive Bayes, it is assumed that the continuous values linked with each feature follow a normal distribution. Thus, the probability distribution of the features is assumed to be Gaussian (Eq. 3). [75].

$$P(y|X) = \frac{P(X|y) \cdot P(y)}{P(X)} \qquad (2)$$

$$P(x_i|y) = \frac{1}{\sqrt{2\pi\sigma y^2}} e^{-\frac{(x_i - \mu y)^2}{2\sigma y^2}} \qquad (3)$$

## III. Ensemble Learning

### a. Bagging algorithms

It is an ensemble learning algorithm, which stands for 'Bootstrap Aggregating'. It uses independent homogeneous weak learners in parallel and aggregates them to get a strong learner [76].

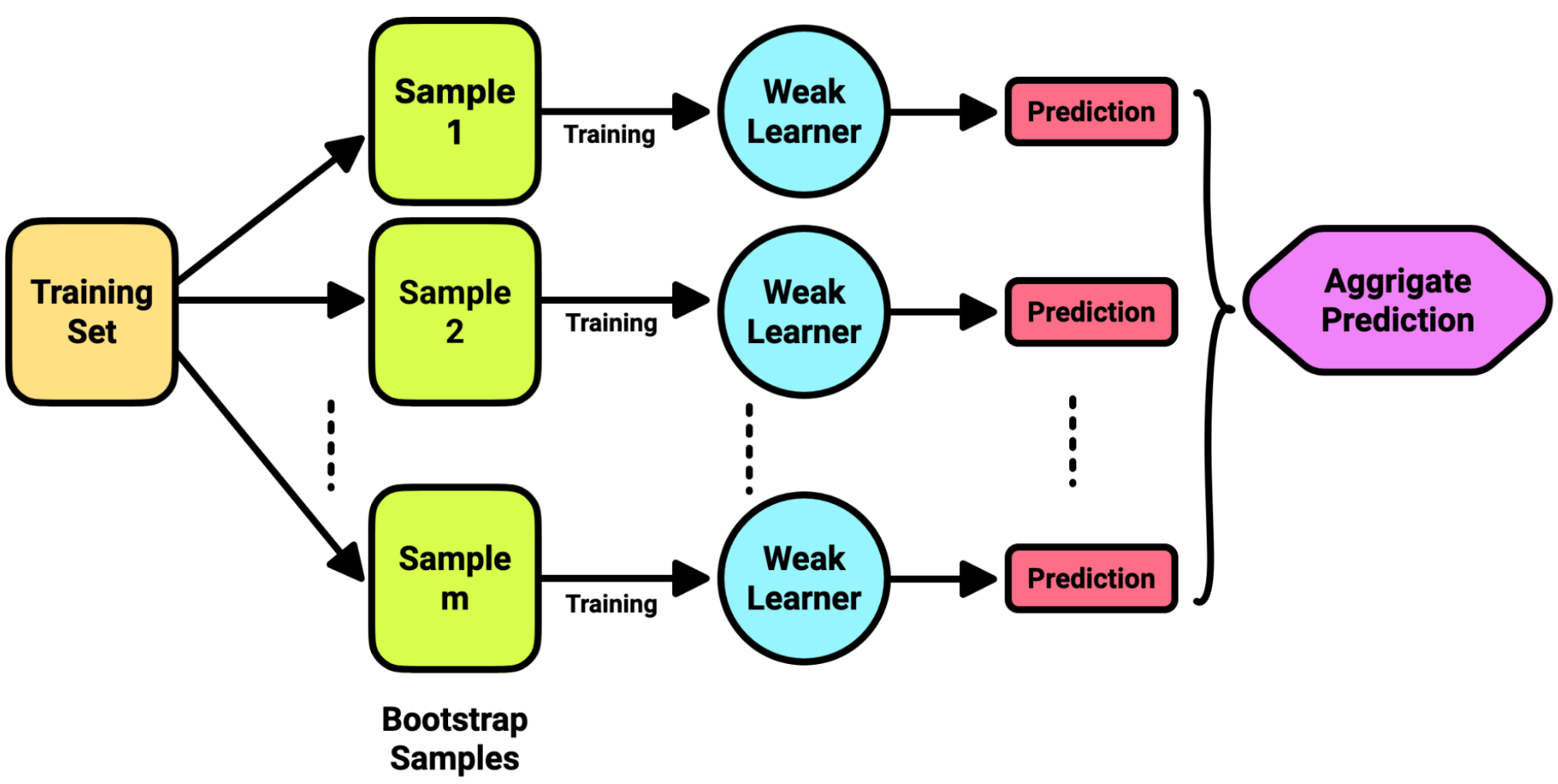

**Figure 11.** *Flow diagram of Bagging algorithm*

### b. Boosting algorithms

It is an ensemble learning algorithm that uses homogeneous weak learners in series adaptively to get a strong learner [77].

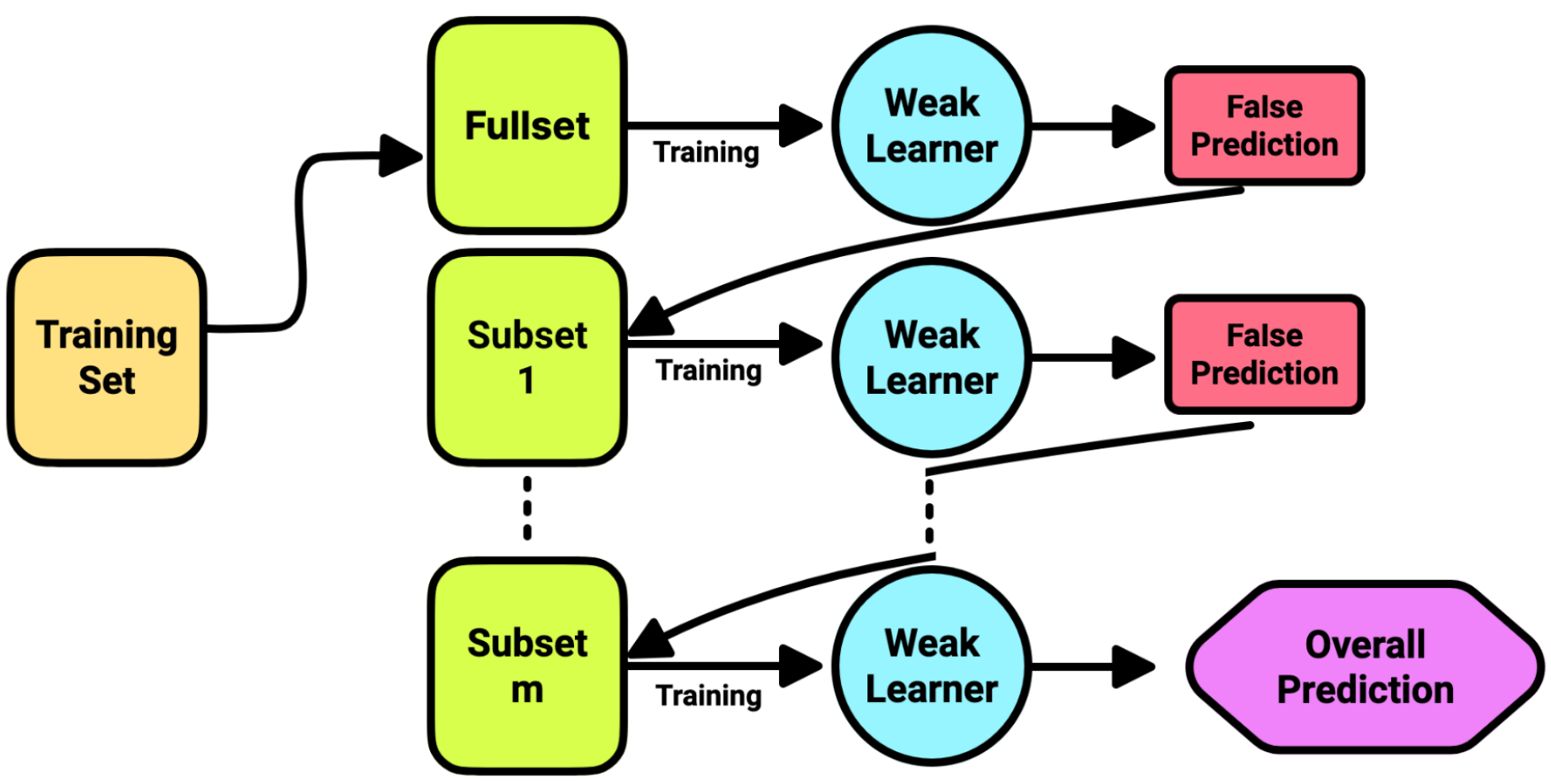

**Figure 12.** *Flow diagram of Boosting algorithm*

Given below are the Bagging and Boosting algorithms used in this study:

- Random Forest

  It is a bagging technique which uses multiple decision trees for weak learners. Since decision trees are prone to overfitting, they show high variance. Random forest improves accuracy by pruning, row sampling, and feature sampling. As a result, we get a lower variance model [78].

- AdaBoost

  It is a boosting algorithm, which stands for Adaptive Boosting. It uses 'Decision Stumps', which are one-level decision trees, as weak learners. It prioritizes models based on weights, i.e., wrong predictions will have more weight as compared to right predictions [79].

- XGBoost

  It is a gradient boosting algorithm, which stands for eXtreme Gradient Boosting. It uses decision trees with 8 to 32 leaf nodes for weak learners. It can handle mixed data types, and it is quite robust against outliers. It supports gradient boosting algorithms, including learning rate, stochastic gradient boosting, and regularizations [80].

- LightGBM

  It is an advanced gradient boosting framework, which stands for Light Gradient Boosting Machine. It leverages decision trees to enhance model efficiency and reduce memory consumption. Unlike other tree-based learning algorithms that grow trees horizontally (level-wise), LightGBM adopts a vertical growth approach (leaf-wise). By selecting splits based on their impact on the overall loss rather than solely on the loss along a specific branch, the leaf-wise strategy can often result in faster learning of lower-error trees. This makes it particularly advantageous when dealing with a limited number of nodes, where leaf-wise growth is likely to outperform level-wise growth. LightGBM's unique approach enables improved performance and resource utilization in gradient boosting models [81].

After preprocessing the data, K-Means clustering was carried out using 'Docking Scores AcrB', 'Docking Scores MexB', and MICs as three features, the value of "K" was determined by elbow method (Figure 13). The resulting three clusters were assigned as Cluster 1 (with MIC ≤256 μg/mL), Cluster 2 (400 ≤ MIC ≤ 512 μg/mL), and Cluster 3 (MIC ≥1000 μg/mL). Also, no direct relation between MIC values and Docking Scores was found. Therefore, they were removed from the descriptor set, and further analysis was done with the remaining 160 descriptors. Then, for the classification models, the MIC values were used as target variables, which were labeled as “Highly Potent” (MIC <512 µg/mL) and “Less Potent” (MIC ≥512 µg/mL). This decision was taken based on a recent study of the phytochemical activity of drug-resistant strains of P. aeruginosa [82]. The targets were then encoded as 1 and 0, respectively, “Highly Potent” and “Less Potent”, using the “LabelEncoder” function of the Scikit-Learn Python library. Taking all the features, additional clustering was done using K-Means and Gaussian mixture model with the number of clusters set to two. Both resulted in successfully grouping the data into two clusters that had considerable overlap with the classes defined above.

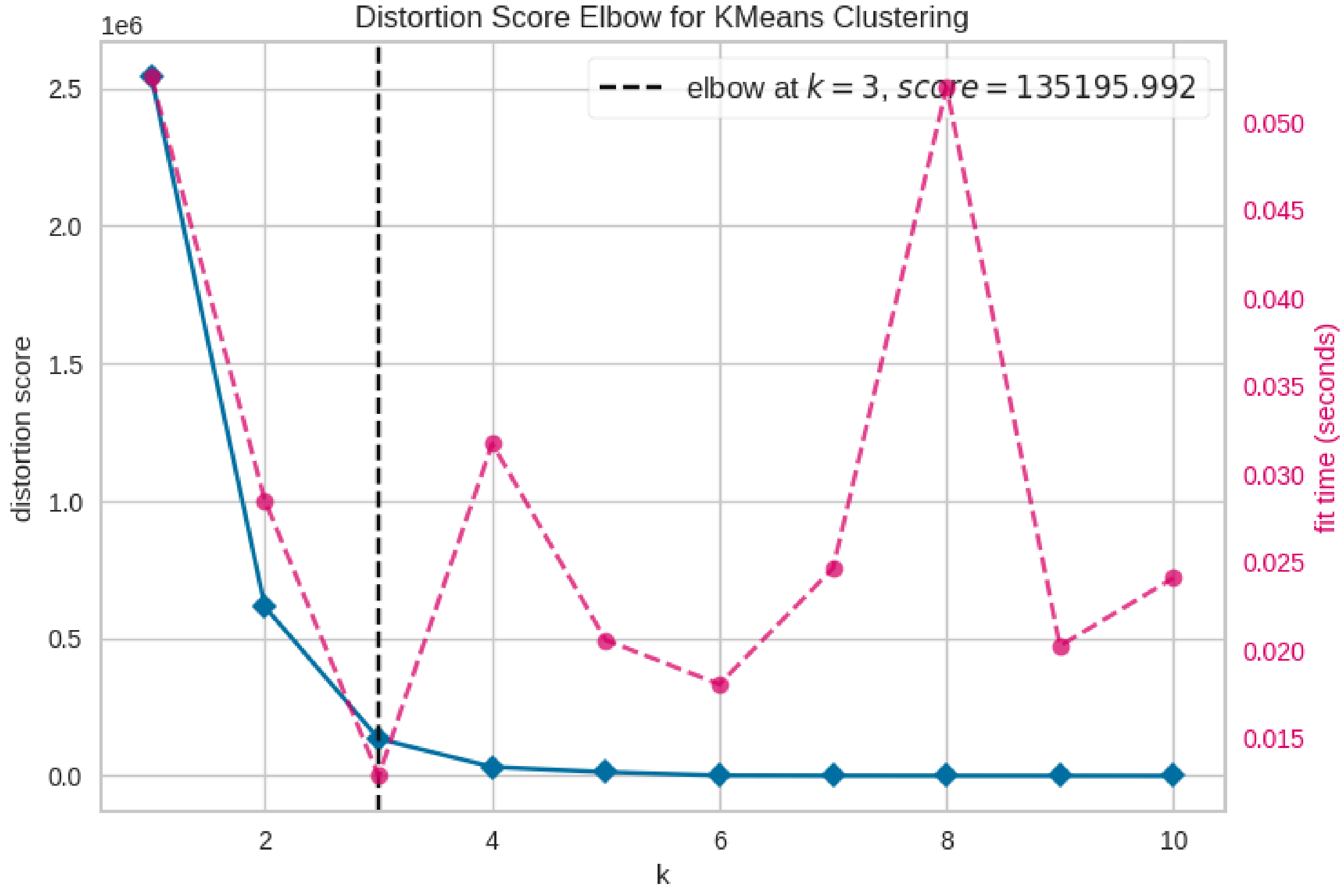

**Figure 13.** *Elbow plot for K-Means clustering of known dataset. [49]*

2.2.4.2 Model Performance Evaluation

The performance evaluation of the classification algorithms can be done using the Confusion matrix [83] and Receiver Operating Characteristic (ROC) plots [84] using the Scikit-learn library. The confusion matrix gives four terms, TN, TP, FP, and FN, which play a significant role in assessing the model's accuracy. TN (True Negative) represents the count of accurately classified negative examples. TP (True Positive) indicates the count of accurately classified positive examples. On the other hand, FP (False Positive) denotes the count of actual negative examples misclassified as positive. FN (False Negative) signifies the count of actual positive examples misclassified as negative. The various scoring parameters of a model can be calculated from its confusion matrix using the following formulas:

- Accuracy: Measure of the closeness of a given set of readings to their true value.

$$ACC = \frac{TN + TP}{TN + FP + FN + TP} \quad (4)$$

- Cohen Kappa: Measures the inter-annotator agreement [85].

$$CK = \frac{2(TP \cdot TN - FP \cdot FN)}{(TP + FP) \cdot (FP + TN) + (TP + FN) \cdot (FN + TN)} \quad (5)$$

- Matthews correlation coefficient: Measures the quality of binary classification [85].

$$MCC = \frac{TP \cdot TN - FP \cdot FN}{\sqrt{(TP + FP) \cdot (FP + TN) \cdot (TP + FN) \cdot (FN + TN)}} \quad (6)$$

- Precision: Measure of the closeness of the measurements to each other.

$$Pr = \frac{TP}{TP + FP} \quad (7)$$

- True Positive Rate: Fraction of actual positive cases that are correctly identified.

$$recall = \frac{TP}{TP + FN} \quad (8)$$

- False Positive Rate: Probability that a true positive is missed by the test.

$$FPR = \frac{FP}{FP + TN} \quad (9)$$

- Balanced Accuracy: Average of recall obtained for each class.

$$BA = \frac{1}{2}\left(\frac{TP}{TP + FN} + \frac{TN}{TN + FP}\right) \quad (10)$$

- F1 Score: Harmonic mean of precision and recall [86].

$$F1 = 2 \cdot \frac{(Pr)(recall)}{(Pr) + (recall)} \quad (11)$$

The ROC curve is a plot between the true positive rate (TPR) and false positive rate (FPR). The area under the curve (AUC) quantifies the effectiveness of the classification model; larger areas indicate superior performance. For a predictor f, AUC is defined as [87]:

$$AUC(f) = P(x) < f(y) \quad C(x) = 0, C(y) = 1 \quad (12)$$

The diagonal line crossing the origin is called the line of no- discrimination, it signifies the Random Classifier (50−50 chance). Points above this line denote favorable classification outcomes, while those below it indicate poorer results.

### 2.2.5 Classical MD Simulations

#### 2.2.5.1 Preparation of Protein-Ligand Systems

The ligand structure and combined parameter/ topology files (prmtop) were generated in the Antechamber [88] module provided by AmberTools 18. All the eight molecules were optimized with the 6-311+G**/ B3LYP level in TeraChem [89-91], followed by calculation of restrained electrostatic potential charge (RESP) [92] using 6-31G/HF method. The ligand parameters were generated using GAFF [93] and the RESP charges are provided as partial atomic charges. The protein structures of AcrB and MexB were modeled using the Amber ff19SB force field [94]. The different protonation states of histidine residues were obtained from the H++ server [95] at pH 7.0 at the time of input generation. The ligand-bound AcrB and MexB structure and prmtop were prepared using the LEaP module. Each component was then assembled and solvated with the TIP3P water model [96] with 14 Å. The molarity and neutrality of each system was maintained at 150 mM by adding an appropriate number of $Na^+$ and $Cl^−$ ions, for AcrB the number of $Na^+$ and $Cl^-$ ions were 197 and 189, whereas for MexB they were 187 and 183. Throughout the whole simulation process, the @ CA, CB atoms of TM region were subjected to positional restraints, thereby preserving the TM α-helices. For comparative trajectory analysis, solvated apo-AcrB, apo-MexB, and individual ligand systems were also prepared using the same procedure. Periodic boundary conditions were applied in each direction. A cutoff value of 10 Å was set for the calculation of van der Waals (vdW) and electrostatic interactions. Particle mesh Ewald (PME) [97] was used for electrostatics and SHAKE algorithm [98] for covalent interactions of heavy and hydrogen atoms.

#### 2.2.5.2 Molecular Dynamics simulation

All the MD simulations were done using the AMBER 18 package [99]. Energy minimization was performed in a sequence of five steps. In the first step, only the water molecules were minimized by imposing restraints into the complex. Then, gradually, restraints were decreased for the protein complex in the following three steps, and finally, the entire unrestrained system was energy-minimized. Successively, the heating was done from 0 to 310 K over a 3 ns period. followed by a total 4 ns simulation to equilibrate the system in six stages (restraints on the backbone with the restraint wt 200, 100, 50, 25, 10, and 5 kcal/Å-mol) in NPT ensemble at a constant temperature of 310 K and pressure of 1 atm. The production run was done for 200 ns, where the MD trajectory was produced in two 100 ns segments by imposing the TM region restraint weights at 1.0 and 0.25 kcal/Å-mol, respectively. Here, NPT ensemble was also used for the equilibration step. A 2 fs integration step was set for the equilibration run and long simulations. During this NPT simulation, the temperature was maintained using a Langevin thermostat [100] with a collision frequency of 1.0 $ps^{-1}$ and 1 atm pressure was maintained using a Berendsen barostat [101].

2.2.5.3 Structural analysis and Binding Free-Energy Estimation

From the last 100 ns trajectory data, the Root Mean Square Deviation (RMSD), Root Mean Square Fluctuations (RMSF), Radius of Gyration (RoG), H bonds, and Solvent Accessible Surface Area (SASA) analyses were performed using the CPPTRAJ module implemented in AmberTools 22. The DBP region was defined by the following residues: 135 to 144, 173 to 180, 278 to 292, 323 to 329, 571 to 578, and 606 to 632 for both proteins, which formed a nest-like cavity comprising the β-sheet region of PN2 and PC1 subdomains. Snapshots of structures were visualized and generated in VMD and Discovery Studio Visualizer. The DBP volumes of AcrB (T) and MexB (T) were calculated using the Epock plugin [102]. The H-bond distance and angle cutoff were set to 3.0 Å and 135°, respectively.

The binding energy of the molecules for both AcrB and MexB were calculated using the well-established implicit solvation model, Poisson− Boltzmann (PB) method [103]. The MMPBSA.py module [104] implemented in AmberTools 22 was invoked to calculate the MMPBSA binding free energy (Eq. 13). For these, a total of 51 frames were extracted from the last 50 ns of trajectories of each system. Entropy contribution to the overall free-energy change ($\Delta G_{MMPBSA}$) was not calculated due to the small contribution for protein−small-molecule systems. The dielectric constants for the solute and solvent were kept at 1.0 and 78.5, respectively.

$$\Delta G_{MMPBSA} = \Delta E_{MM} + \Delta G_{solvation} - T\Delta S_{config} \quad (13)$$

Where,

$$\Delta G_{solvation} = \Delta G_{complex} - \Delta G_{receptor} - \Delta G_{ligand} \quad (14)$$

## 2.3 Results and Discussion

### 2.3.1 Comparison of Machine Learning Models

From the results of the clustering algorithms we finalized our target variables for the classification models. The groups with targets set as 0 for 'Less Potent' (MIC ≥512 μg/mL) & 1 for 'Highly Potent' (MIC <512 μg/mL) and the two clusters formed as Cluster 1 (MIC ≥400 μg/mL) & Cluster 2 (MIC ≤256 μg/mL) showed a significant overlap of 64.15% (Figure 14). The known dataset was then split into "training" (80%) and "test" (20%) sets using the 'train_test_split' function of scikit-learn. The 10 classification models were trained using the "training set" and then tested using the "test set" of the known dataset.

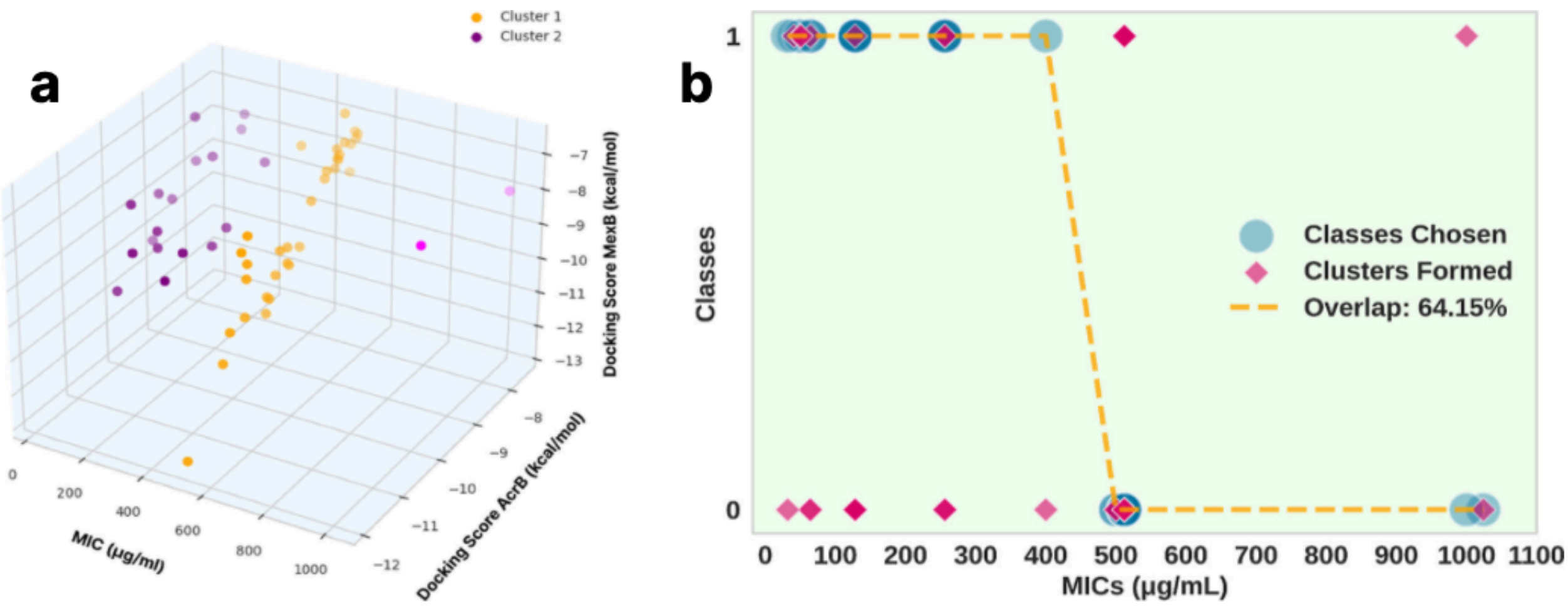

**Figure 14.** *(a) K-Means clustering of the known inhibitors using MIC values, the docking score in MexB, and the docking scores in AcrB. (b) Plot showing the distribution of the 53 known inhibitors in the two clusters formed by GMM unsupervised learning (pink diamonds) and in two classes chosen by us (blue circles). The orange line indicates the overlap of the data points (64.15%). [49]*

The quality of prediction of the models was quantitatively evaluated using scoring parameters, i.e., ACC, AUC, CK, MCC, Pr, Recall, BA, and F1 scores. It was observed that the top-scoring models were LightGBM, AdaBoost, Random Forest, and SVM. In terms of accuracy, LightGBM was 91% accurate, and the rest of the three top- scoring models were 82% accurate (Figure 15).

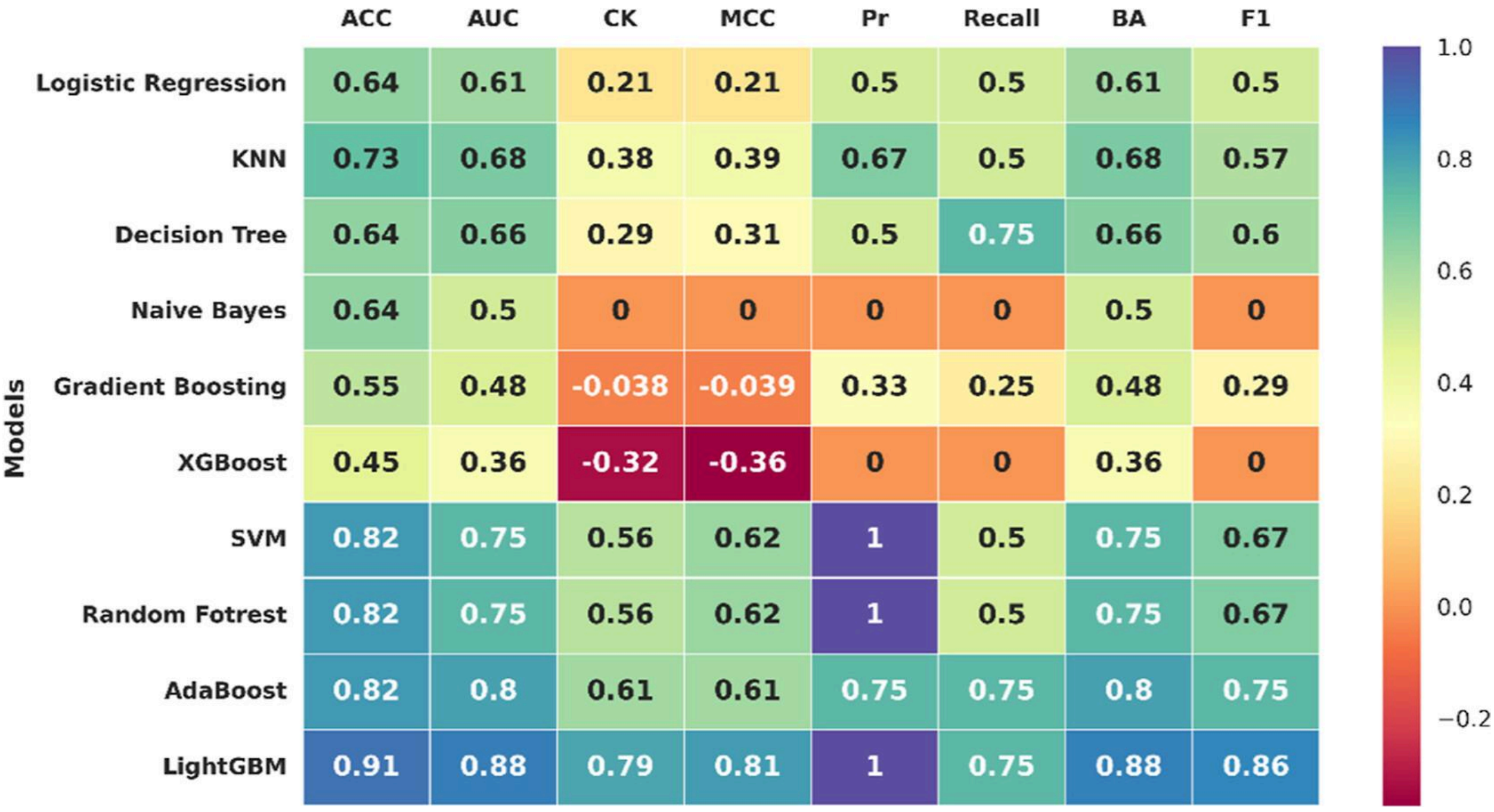

**Figure 15.** *Comparison of the 10 classification models using various scoring parameters. [49]*

As the "test set" included 20% of the 53 inhibitors (i.e. 11 molecules), from the confusion matrices of the top-scoring models (Figure 16a), it can be seen that LightGBM predicted TN 7 times, TP 3 times, FN 1 time, and FP 0 times. Thus, out of 11 predictions, only one of them was wrongly predicted. This means that for a set of 11 potential inhibitors, it is expected for this model to classify one of the "highly potent" inhibitors as "less potent". Similarly, AdaBoost is expected to classify one of the "highly potent" inhibitors as "less potent" and one of the "less potent" inhibitors as "highly potent", and Random Forest and SVM are expected to classify two of the "highly potent" inhibitors as "less potent". Also, from the ROC plots of the four top-scoring models (Figure 16b) it can be seen that all the points are well above the Random Classifier, and the AUCs are 0.75 or above. Among these, LightGBM has the highest AUC of 0.88.

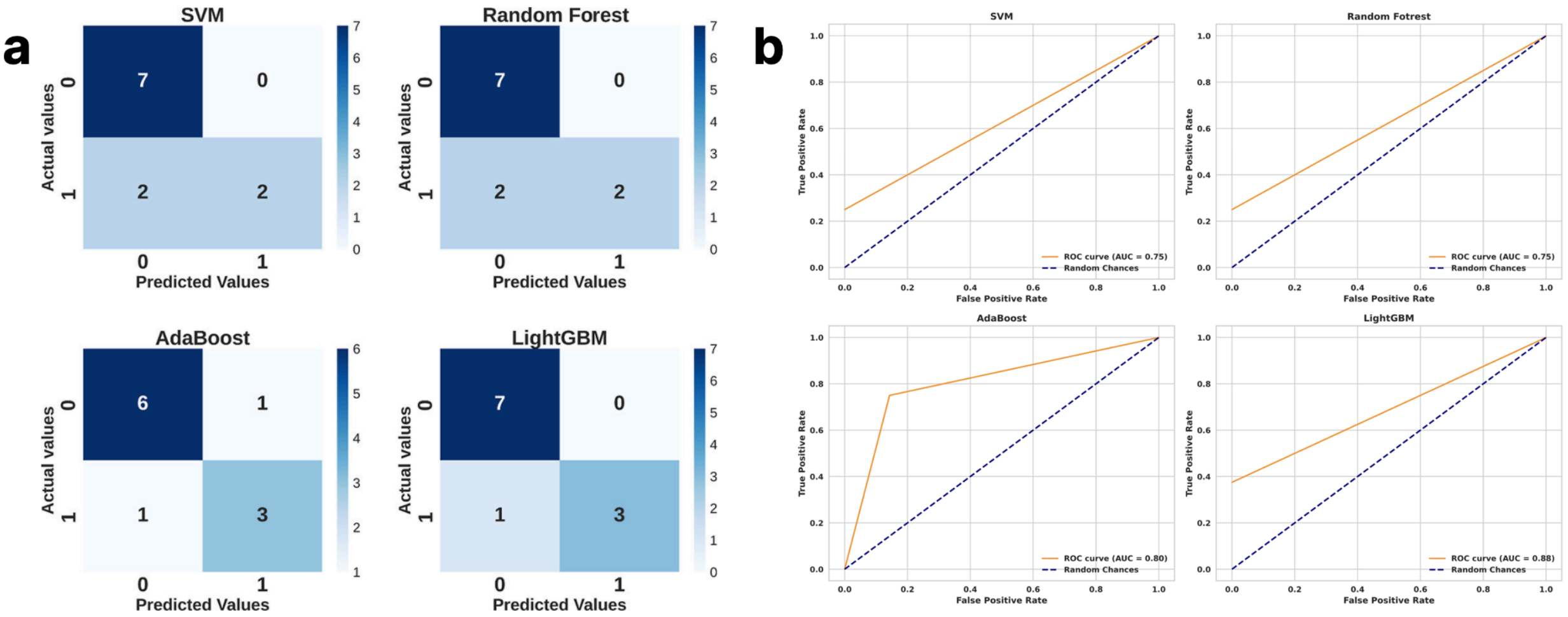

**Figure 16.** *(a) Confusion matrices for top four ML models. (b) ROC plots for top four ML models. [49]*

The superior performance of LightGBM can be understood in terms of its construction. Unlike other tree-based algorithms which grow a tree level-wise, LightGBM grows a tree leaf-wise. It intuitively selects the leaf that yields the largest decrease in loss. It also uses two advanced techniques called exclusive feature bundling (EFB), and gradient-based one side sampling (GOSS), which allow it to run efficiently with higher accuracy [79]. Also, from the "feature importance", it can be seen that LightGBM selected the features such as Shadow_Xlength, TOPKAT_Oral_LD50, N_Count, WPSA_1, and Num_H_Donors to train itself. These features are important structural and pharmacokinetic parameters that define a "highly potent" inhibitor [49].

### 2.3.2 Prediction of Inhibitor Molecules

After the model evaluation, the four top-scoring models mentioned above were trained with the known data set of 53 molecules. The unknown data set of 5043 molecules, which was generated using permutation and combination of the five selected pharmacophores and various side groups (Figure 8), was then fed into the trained models. After the ML models were run, the predicted targets were recorded. To filter the top hits from this predicted data set, we employed three filters. For the first filter, we sorted out the molecules that had a target value of "1" in LightGBM and at least in any two of the rest of the three models. This is done because the LightGBM model had the highest accuracy and precision among all of the other models. Consequently, a total of 493 molecules were separated out of the 5043 unknown molecules. For the second filter, we used Lipinski's rule of 5, which states that for a molecule to have more drug likeness, it must not violate more than one of the following criteria:

$number\ of\ H\ bond\ donors\ \leq\ 5$, $number\ of\ H\ bond\ acceptors\ \leq\ 5$, $log\ P\ \leq\ 5$, and $molecular\ mass\ <\ 500\ g/mol$. Molecules that violate more than one of the above criteria are likely to be toxic. By doing so, it was found that only 28 of 493 molecules ultimately passed through this filter. Aside from Lipinski's rule of 5, there are other rules to filter out molecules having more drug likeness, like Ghose filter and Veber’s rule, but we chose Lipinski’s rule because it is comparatively less strict than other filters. A strict rule would have rejected many potential inhibitors that might score well in future analyses. Now that we have molecules that are likely to be highly potent and less toxic, we decided to dock these molecules in both AcrB and MexB using AutoDock Vina. The docking scores were used as a third filter. An arbitrarily good docking score value of −10.5 kcal/mol in both AcrB and MexB was set as the threshold. This value was chosen because there was a dense collection of known inhibitors present around this value in our earlier clustering of known dataset. Only 8 of the 28 molecules could pass through this filter (Figure 17).

**Figure 17.** *Top eight predicted inhibitor-like molecules. Each consists of pyridopyrimidone as the core moiety. [49]*

It is noteworthy to state that all the eight molecules have pyridopyrimidone as their core pharmacophore, which means that it passed through all the filters we have applied on the various molecules. From the literature, we have seen that molecules with this pharmacophore have higher efflux pump inhibition activities. From the structure activity studies, it was found that the pyridopyrimidone core reduces plasma protein binding (PPB) [105]. Hence, these top eight molecules were taken to carry out MD simulations and the corresponding analyses thereafter.

### 2.3.3 MD Data Analyses

2.3.3.1 Structural Changes to Porter Domain

The RMSD data of the porter domain consisting of PN1, PN2, PC1, and PC2 subdomains over the last 100 ns trajectory were analyzed. The average RMSD values of the PN2 and PC1 subdomains were calculated for each ligand-bound AcrB and MexB system to gain insights about the structural changes of the DBP region (Figure 18). From the results (Table 4) it was evident that the AcrB PN2/PC1 domain experienced more structural changes than that of ligand-bound MexB systems. For Lig4-, 5-, and 6-bound AcrB, relatively larger fluctuations of the DBP region indicated the pocket- widening effect as well as orientational changes of the hydrophobic side chains comprising the interior of the deep binding pocket due to ligand accommodation. On the other hand, the PN2 subdomain in MexB almost preserved its native structure, and only the PC1 subdomain for Lig3 and 4 showed a slightly higher RMSD fluctuation.

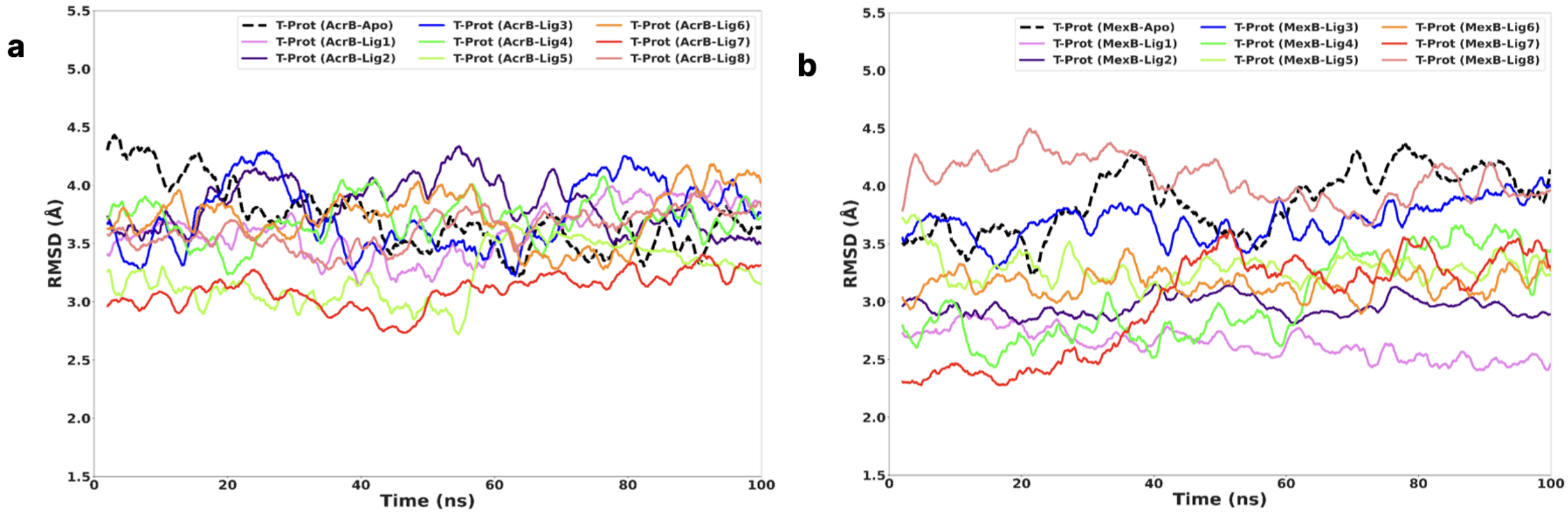

**Figure 18.** *Root Mean Squared Deviation of top eight Ligand-AcrB and -MexB systems. [49]*

**Table 4.** *Average RMSD values for PN2 and PC1 subdomains in AcrB and MexB. [49]*

| Protein-ligand systems | RMSD $_{AcrB}$ (Å) | | RMSD $_{MexB}$ (Å) | |
|---|---|---|---|---|
| | PN2 | PC1 | PN2 | PC1 |
| Lig1 | 1.05±0.08 | 1.11±0.09 | 1.01±0.10 | 0.96±0.10 |
| Lig2 | 1.01±0.10 | 1.06±0.09 | 1.06±0.11 | 1.05±0.11 |
| Lig3 | 1.16±0.13 | 1.08±0.09 | 1.14±0.11 | 1.28±0.10 |
| Lig4 | 1.37±0.10 | 1.25±0.12 | 0.95±0.08 | 1.47±0.19 |
| Lig5 | 1.32±0.09 | 1.06±0.09 | 0.92±0.09 | 1.20±0.12 |
| Lig6 | 1.19±0.16 | 1.32±0.11 | 0.99±0.13 | 1.05±0.11 |
| Lig7 | 1.03±0.12 | 1.03±0.12 | 1.08±0.10 | 1.22±0.09 |
| Lig8 | 1.06±0.12 | 1.44±0.16 | 0.97±0.10 | 0.90±0.09 |

The RoG of the β-sheet region of PN2-PC1 subdomains also provided useful insights against structural changes. The increased average RoG of the nest-like β-sheet groove from its apo state indicates the pocket-widening phenomena upon ligand binding (Figure 19). This effect was found to be more prominent in the case of AcrB with Lig2 and Lig6. For MexB, Lig6 binding slightly increased RoG with respect to its free state showed a pocket-widening effect.

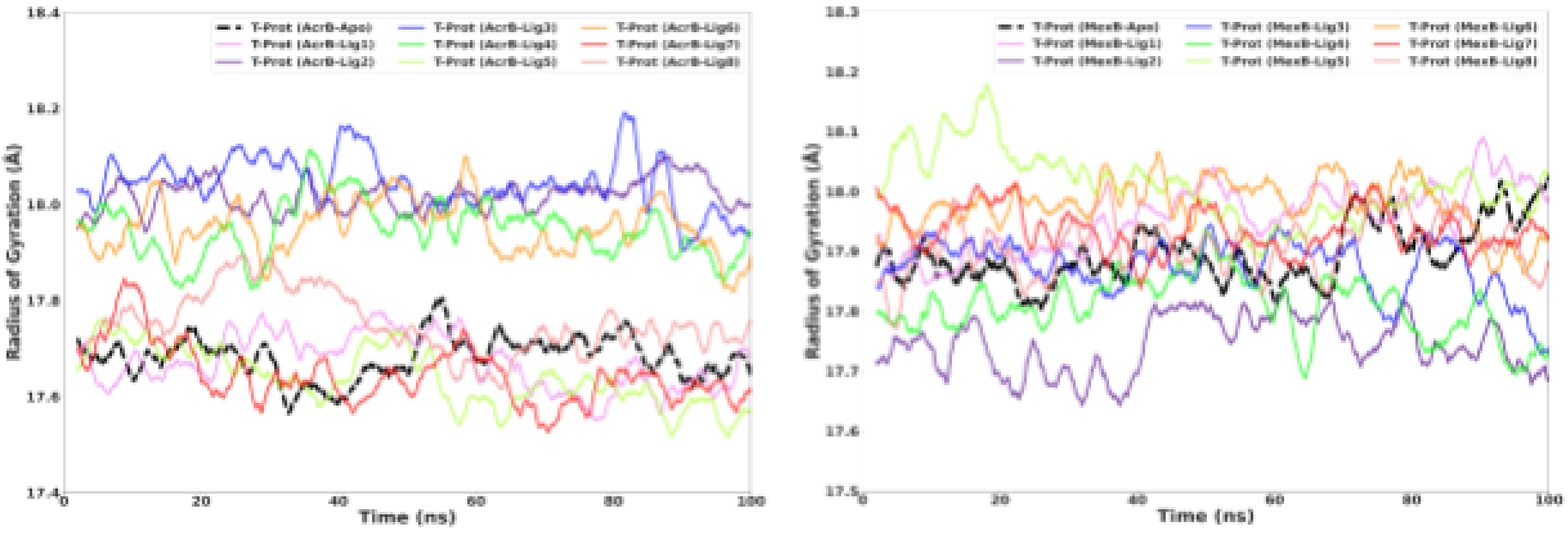

**Figure 19.** *Radius of Gyration of top eight Ligand bound AcrB and MexB. [49]*

2.3.3.2 MMPBSA and H-Bonding Analyses

All the selected molecules record a good binding affinity (ΔG) for both ligand- bound AcrB and MexB systems. Based on the ΔG score, we have selected the top three ligands and evaluated energy terms as shown in Table 5. In the case of AcrB, Lig6, Lig2, and Lig7 ranked as the top three, having ΔG values of −43.58 ± 3.91 kcal/mol, −42.45 ± 3.27 kcal/mol, and −39.33 ± 4.13 kcal/mol, respectively. For MexB, the ΔG values of Lig3, Lig6, and Lig4 were −45.91 ± 4.35 kcal/mol, −43.52 ± 4.57 kcal/mol, and −39.43 ± 4.24 kcal/mol, subsequently. It was observed that Lig6 can be found in the topmost scoring ligand both for AcrB and MexB systems, showing almost equal amounts of ΔG value and being comparable to the known inhibitors like, D13-9001, MBX3132, MC-207,110, MC-04,124, and Quercetin.

**Table 5.** *ΔG<sub>MMPBSA</sub> values in kcal/mol shown for top three ligands bound with (a) AcrB & (b) MexB. [49]*

(a)

| **Energy Terms** | **Lig2 (Mol. 2977)** | **Lig6 (Mol. 3488)** | **Lig7 (Mol. 3493)** |
|---|---|---|---|
| vdW | -67.09 ± 2.30 | -66.61 ± 2.59 | -69.94 ± 3.04 |
| $E_{EL}$ | -19.83 ± 3.95 | -7.23 ± 3.49 | -27.29 ± 4.48 |
| $E_{PB}$ | 49.33 ± 93.59 | 35.98 ± 3.21 | 63.01 ± 4.27 |
| $E_{NPOLAR}$ | -4.85 ± 0.10 | -5.71 ± 0.09 | -5.10 ± 0.12 |
| $\Delta G_{gas}$ | -86.93 ± 4.53 | -73.85 ± 3.91 | -97.24 ± 5.63 |
| $\Delta G_{solv}$ | 44.47 ± 3.55 | 30.26 ± 3.19 | 57.90 ± 4.25 |
| $\Delta G_{MMPBSA}$ | -42.45 ± 3.27 | -43.58 ± 3.91 | -39.33 ± 4.13 |

(b)

| **Energy Terms** | **Lig3 (Mol. 3032)** | **Lig4 (Mol. 3095)** | **Lig6 (Mol. 3488)** |
|---|---|---|---|
| vdW | -59.55 ± 2.78 | -57.79 ± 4.31 | -60.77 ± 3.61 |
| $E_{EL}$ | -16.79 ± 3.13 | -11.13 ± 4.68 | -6.87 ± 4.83 |
| $E_{PB}$ | 35.75 ± 2.95 | 34.62 ± 5.21 | 29.61 ± 4.39 |
| $E_{NPOLAR}$ | -5.32 ± 0.12 | -5.11 ± 0.15 | -5.49 ± 0.19 |
| $\Delta G_{gas}$ | -76.34 ± 3.98 | -68.93 ± 7.12 | -67.64 ± 6.00 |
| $\Delta G_{solv}$ | 30.43 ± 2.92 | 29.50 ± 5.13 | 24.12 ± 4.31 |
| $\Delta G_{MMPBSA}$ | -45.91 ± 4.35 | -39.4 ± 4.24 | -43.52 ± 4.57 |

After visually inspecting each protein−ligand system, we noticed that Lig6 had maintained almost a V-shaped structure and occupied the deep binding pocket of both proteins (Figure 20). As a result, the highest phenylalanine exposure was observed in the case of Lig6-bound AcrB and MexB. This fact also supported that Lig6 was deeply buried inside the hydrophobic pit region, thereby showing lot of π−π interactions, alkyl−π interactions, and π−sulfur interactions between the phenylalanine residues, PHE136, PHE178, PHE610, PHE615, and PHE628, and ligand π electron-rich moieties, e.g., pyridopyrimidone and benzothiazole moieties. In addition to that, a weak H bond was observed between pyridopyrimidone C=O and SER 134 side chain O−H for the Lig6-AcrB protein (Table 6).

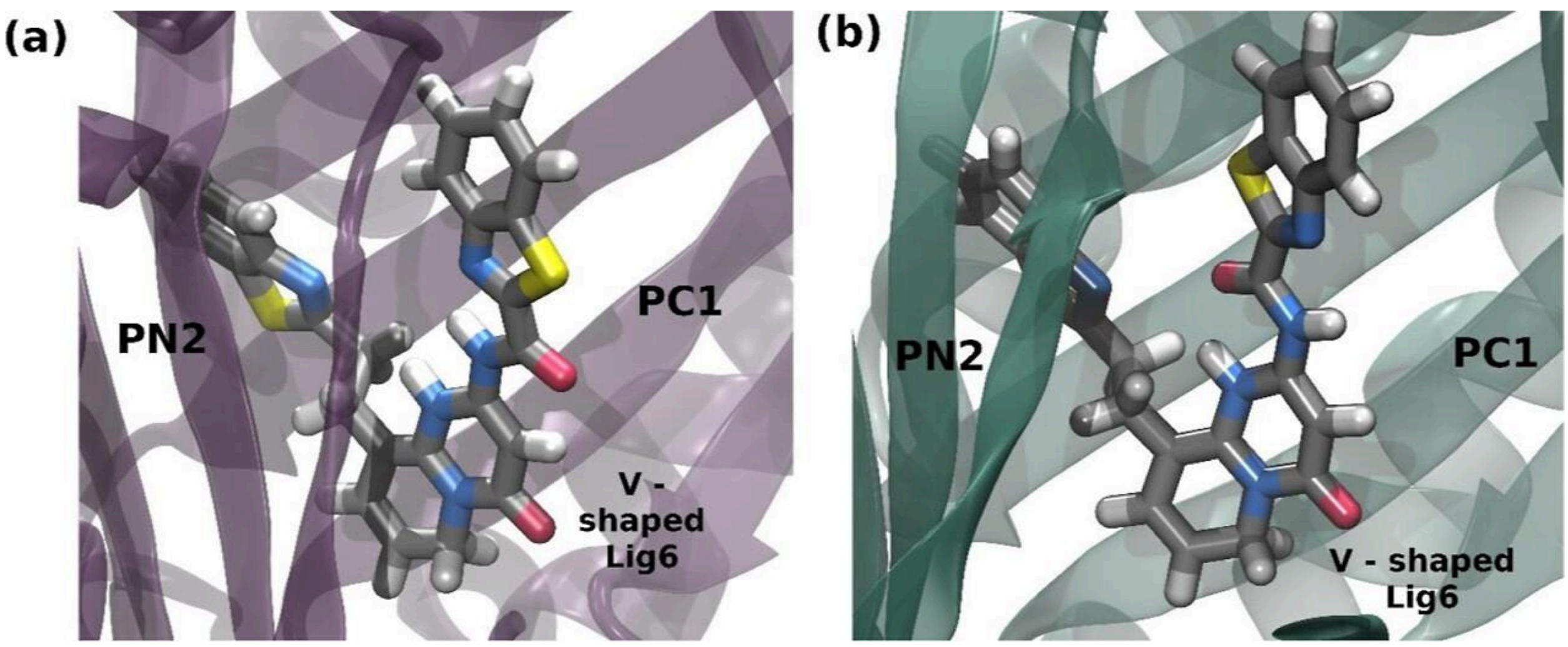

**Figure 20.** *Graphical representation of the V-shaped orientation of Lig6 inside the deep binding pocket of (a) AcrB and (b) MexB. [49]*

**Table 6.** *Hydrogen Bonding information for Lig6 with AcrB. [49]*

| #Acceptor | DonorH | Donor | Frames | Frac | AvgDist | AvgAng |
|---|---|---|---|---|---|---|
| LIG_1034@O1 | SER_134@HG | SER_134@OG | 611 | 0.2444 | 2.7784 | 159.7967 |
| LIG_1034@H15 | LIG_1034@H2 | LIG_1034@N2 | 39 | 0.0156 | 2.8188 | 140.0977 |
| LIG_1034@H17 | VAL_142@H | VAL_142@N | 17 | 0.0068 | 2.8981 | 142.1067 |
| LIG_1034@H13 | LIG_1034@H2 | LIG_1034@N2 | 13 | 0.0052 | 2.5533 | 137.9332 |

2.3.3.3 Ligand Binding and Switch Loop Dynamics

In both AcrB and MexB, a flexible loop named the “switch loop” lies between the access pocket and deep binding pocket. It contains four glycine residues in AcrB and three glycine residues in the case of MexB, which infer the intrinsic flexibility of the loop. It was also mentioned in literature that the inflexibility of the loop might lead to the loss of efflux function [48]. To get insights about the structural flexibility RMSF measurements were done. RMSF of the switch loop residues 613 to 623 revealed some drug-induced inflexibility of the loop (Figure 21).

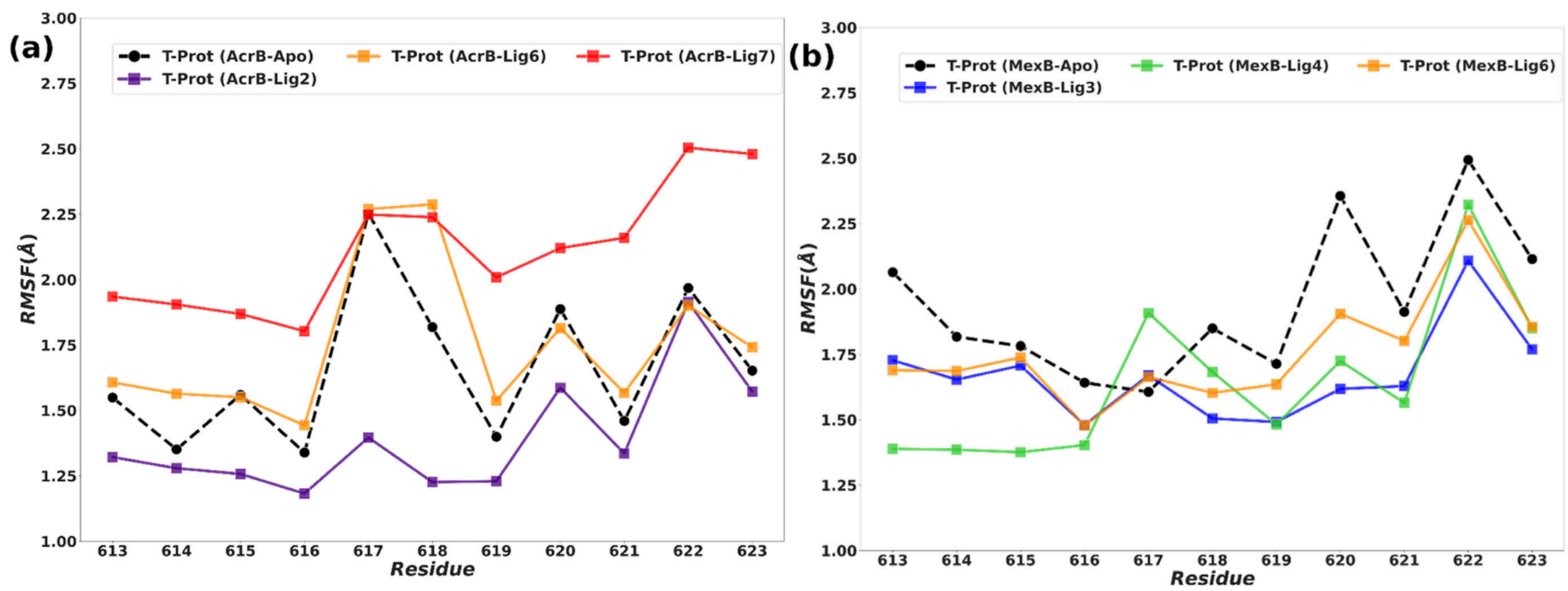

**Figure 21.** *Comparison of the RMSF of the switch loop of apo- and ligand-bound (a) AcrB and (b) MexB. [49]*

From this it is quite obvious that the average switch loop fluctuation of MexB apo-state is higher than that of the AcrB counterpart. The switch loop movement is less restricted in MexB because of its voluminous DBP region. However, our careful analyses indicate that the "switch loop" flexibility would be affected upon ligand binding inside the deep binding pocket. We identified two major factors for that: (i) steric crowding inside DBP and (ii) specific interaction of the three important switch loop residues: PHE 615, PHE 617, and ARG 620. The first factor is applicable for all the ligand-bound systems of AcrB and MexB. As the molecules occupy the DBP, switch loop motion becomes sterically restricted. The 2D protein residue-ligand interactions for the top three ligands are shown in Figure 22.

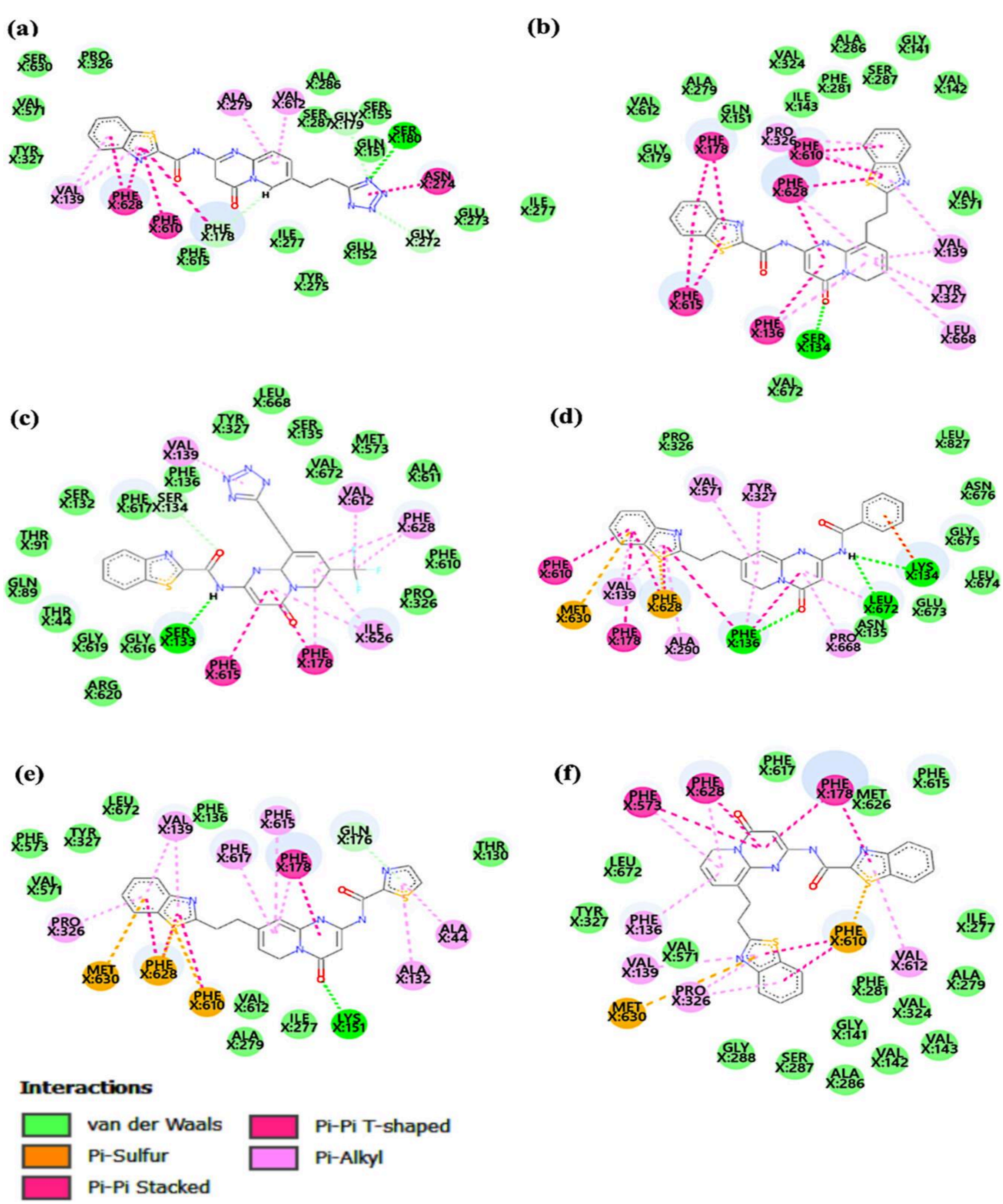

**Figure 22.** *Comprehensive graphical representation of the ligand−DBP residue interactions. (a) Lig2−AcrB, (b) Lig6−AcrB, (c) Lig7−AcrB, (d) Lig3−MexB, (e) Lig4−MexB, and (f) Lig6−MexB systems. [49]*

2.3.3.4 Solvent-Accessible Surface Area Analysis

In the protein−ligand interaction analysis, SASA is a widely used tool as a measure of protein folding, stability, and binding pocket alterations upon ligand accommodation. In this study, as the DBP interior is lined up mainly with hydrophobic residues, the binding of ligands is expected to lower the accessible surface area by the solvent. Also, the selected ligands have positive LogP values, which indicate their intrinsic hydrophobicities [49]. We calculated DBP-accessible surface areas for each of the AcrB and MexB systems after ligand binding. The average SASAprotein values for AcrB T-protomer with and w/o ligands is shown in Figure 23. The Lig3- and Lig4-bound structures indicated higher average $SASA_{protein}$ values than that of the apo form, which means the opening of DBP to host the molecules inside it. The other AcrB-ligand-bound structure shows almost similar SASA values to the apo-protein. However, in the case of MexB-bound structures, no such pocket-opening phenomenon was observed.

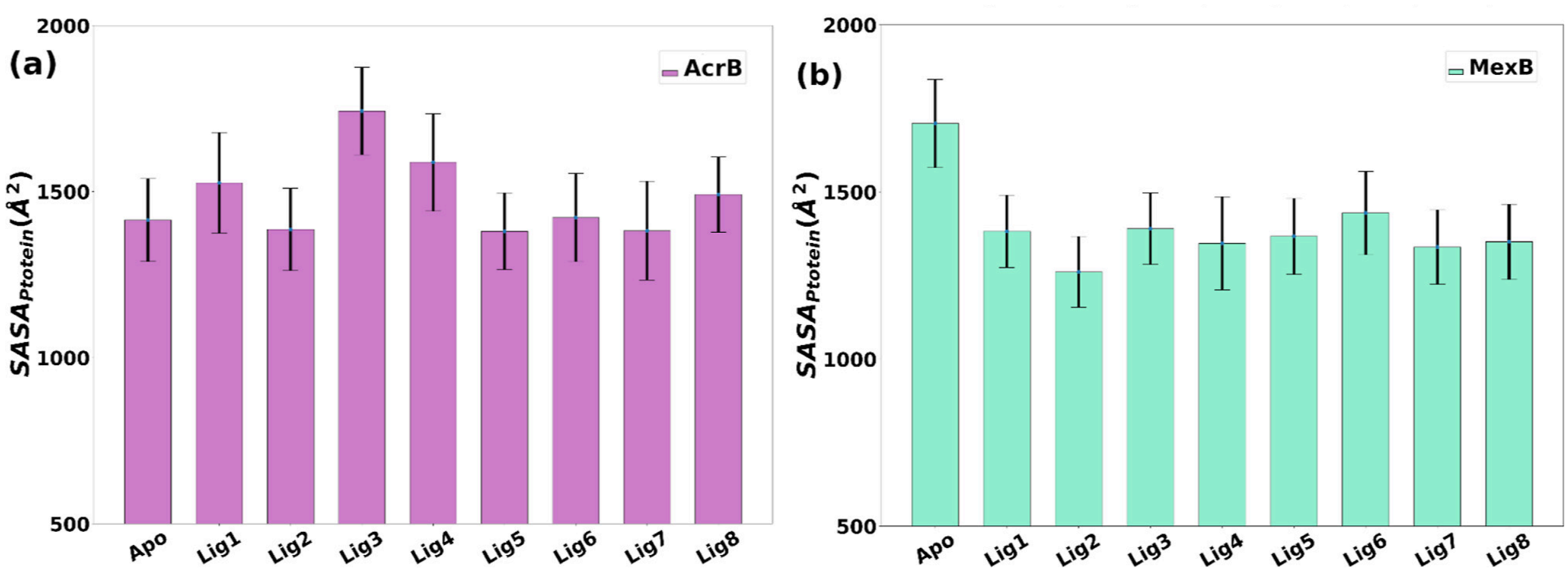

**Figure 23.** *Comparison of the surface-accessible surface area of apo and ligand-bound (a) AcrB and (b) MexB. [49]*

# 3. Macrolide Degradation by EreC

## 3.1 Literature Survey

### 3.1.1 Structure and Mechanism of action

Erythromycin esterases (Eres) deactivate the macrolides such as Erythromycin, Azithromycin, and Clarithromycin by enzymatically cleaving the macrolactone ring. The most frequently identified erythromycin esterase enzyme is the EreA. It confers resistance to most of the clinically used macrolides. Despite this, the research into the Ere family of enzymes has been sparse. An extremely close homologue of EreA, known as EreC, displays more than 90% sequence identity to EreA. Two distinct forms of crystal structures of EreC are available at RCSB with a resolution of 2.00 Å, $EreC^{Open}$(PDB ID: 6XCS) and $EreC^{Closed}$(PDB ID: 6XCQ). Both of these structures are effectively identical, except for a loop that lies from residue 295 to 303 (Figure 24). In the closed form, this loop is bent over the rest of the enzyme, while in the open form, it is turned away from the rest of the enzyme [34].

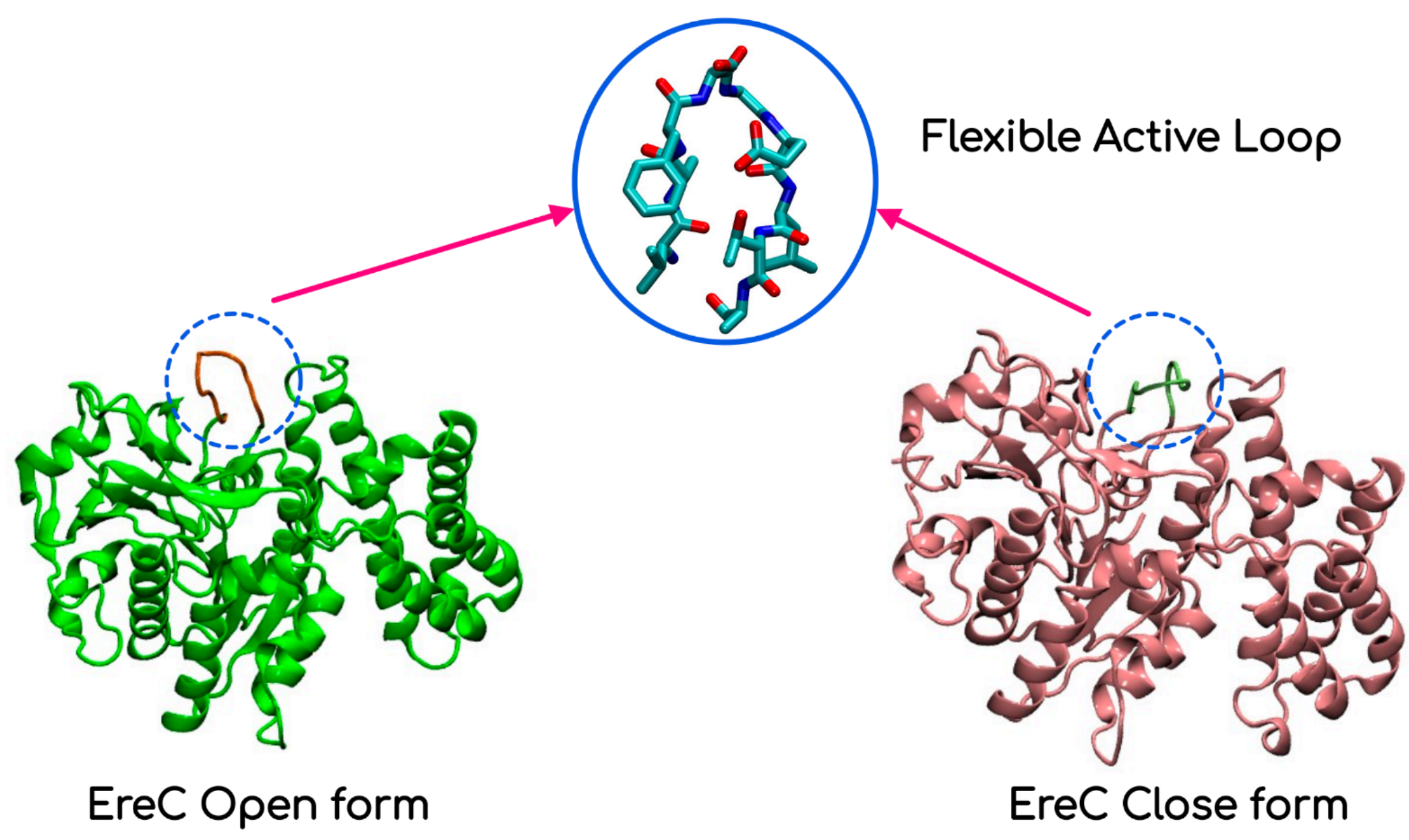

**Figure 24.** *Open and Closed conformations of EreC along with the flexible active loop.*

The protein in both conformations can be divided into two parts: a minor lobe and a major lobe. The major lobe, as the name suggests, represents the major portion of the protein and is composed of 8 β-strands forming a continuous sheet, surrounded by α- helices and loops. The remaining minor lobe contains a four-helix bundle motif (Figure 23).

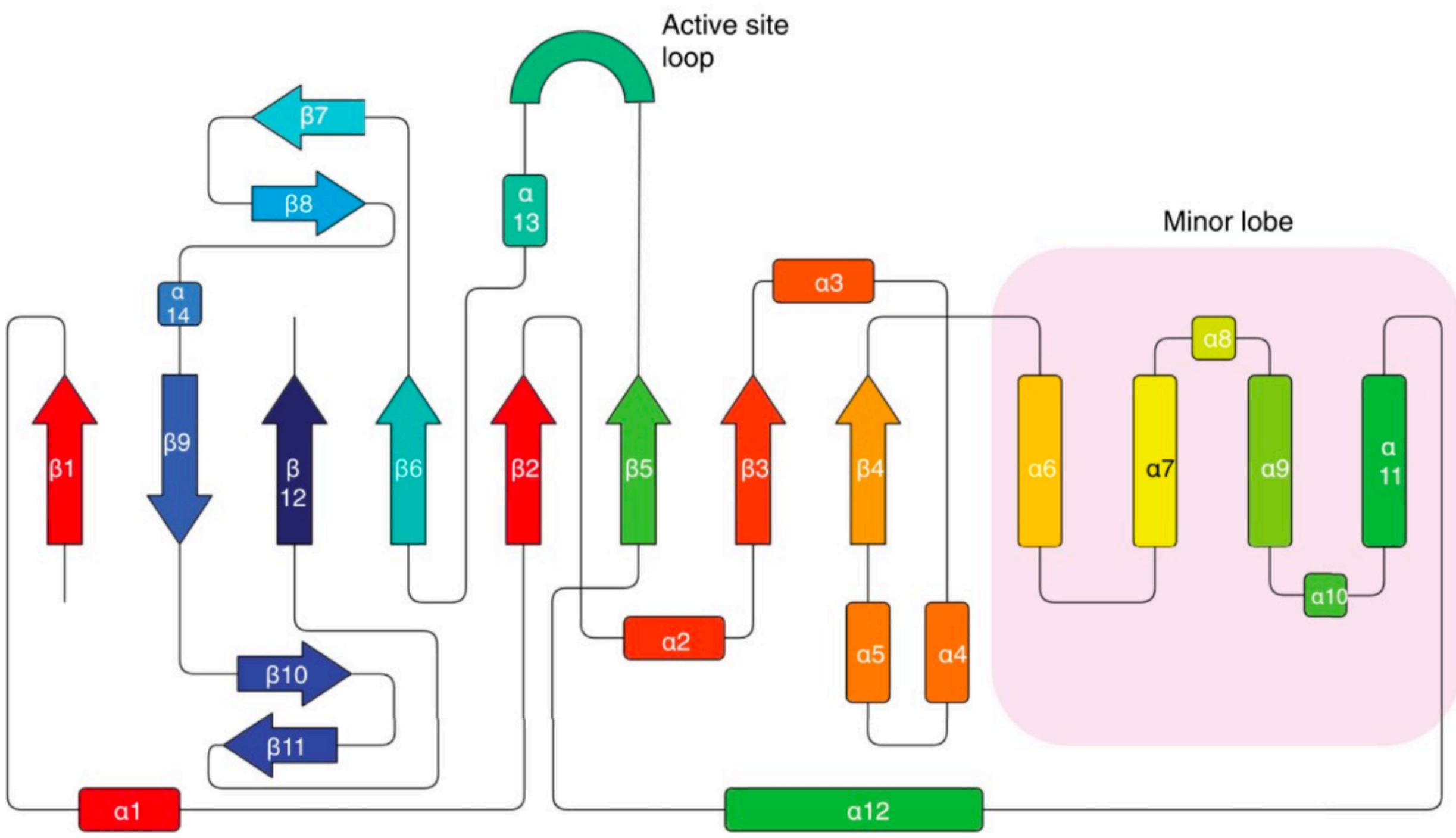

**Figure 25.** *Topological diagram of EreC with α-helices indicated by rectangles, β-sheets by arrows, and the active loop by a semi-circle. [34]*

The binding of the macrolide is facilitated by the closing of the active loop, which is rich in glycine and proline. Thereafter, the macrolide binds to the active pocket surrounded by the catalytic residues, His-50, Glu-78, His-289, Glu-47, Arg-261, and Asp-262 (Figure 24). The first three residues mentioned above align themselves and coordinate with a catalytic water molecule, preparing it for a nucleophilic attack. Thus, these residues play an important role in the catalytic ester bond cleavage of the macrolide ring. His-50 acts as the catalytic base, Glu-78 and His-289 are responsible for substrate specificity, while Glu-47 restricts the rotation of His-50 [34].

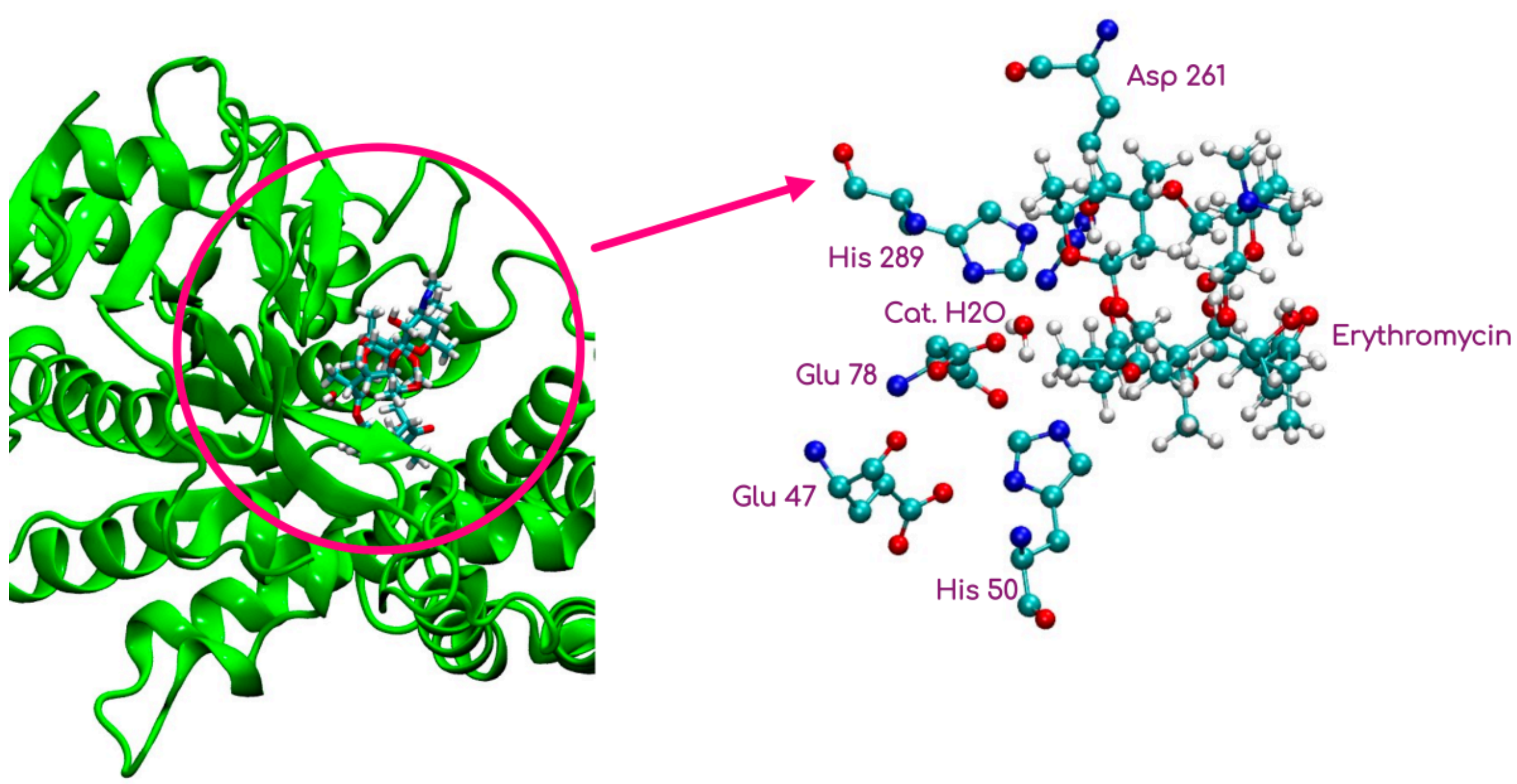

**Figure 26.** *Binding pocket of macrolide in EreC (Left), Catalytic residues in the vicinity of the macrolide (Right).*

The proposed mechanism of the catalytic cleavage of macrolide ring by the enzyme can be summarized by the following steps:

1. Erythromycin binds within the active site of EreC.
2. Upon binding, a flexible loop undergoes closure.
3. The ester bond is positioned proximally to His-50, Glu-78, and His-289.
4. Catalytic water coordinates to these residues.
5. Glu-47 stabilizes the position of His-50.
6. The pKa of His-50 decreases, increasing its basicity.
7. His-50 deprotonates water, yielding the hydroxide ion (OH-).
8. The hydroxide ion attacks the carbonyl group of the ester.
9. Arg-261 repositions to stabilize the resulting negative charge on the oxygen atom, forming a tetrahedral intermediate.
10. Cleavage of the C-O bond in the ester occurs, and the cleaved deprotonated alcohol abstracts a proton from His-50.
11. Following completion of the reaction, the flexible loop reopens to initiate the next cycle.

This mechanism was proposed by Zieliński et. al. [34], and is diagrammatically represented in Figure 25.

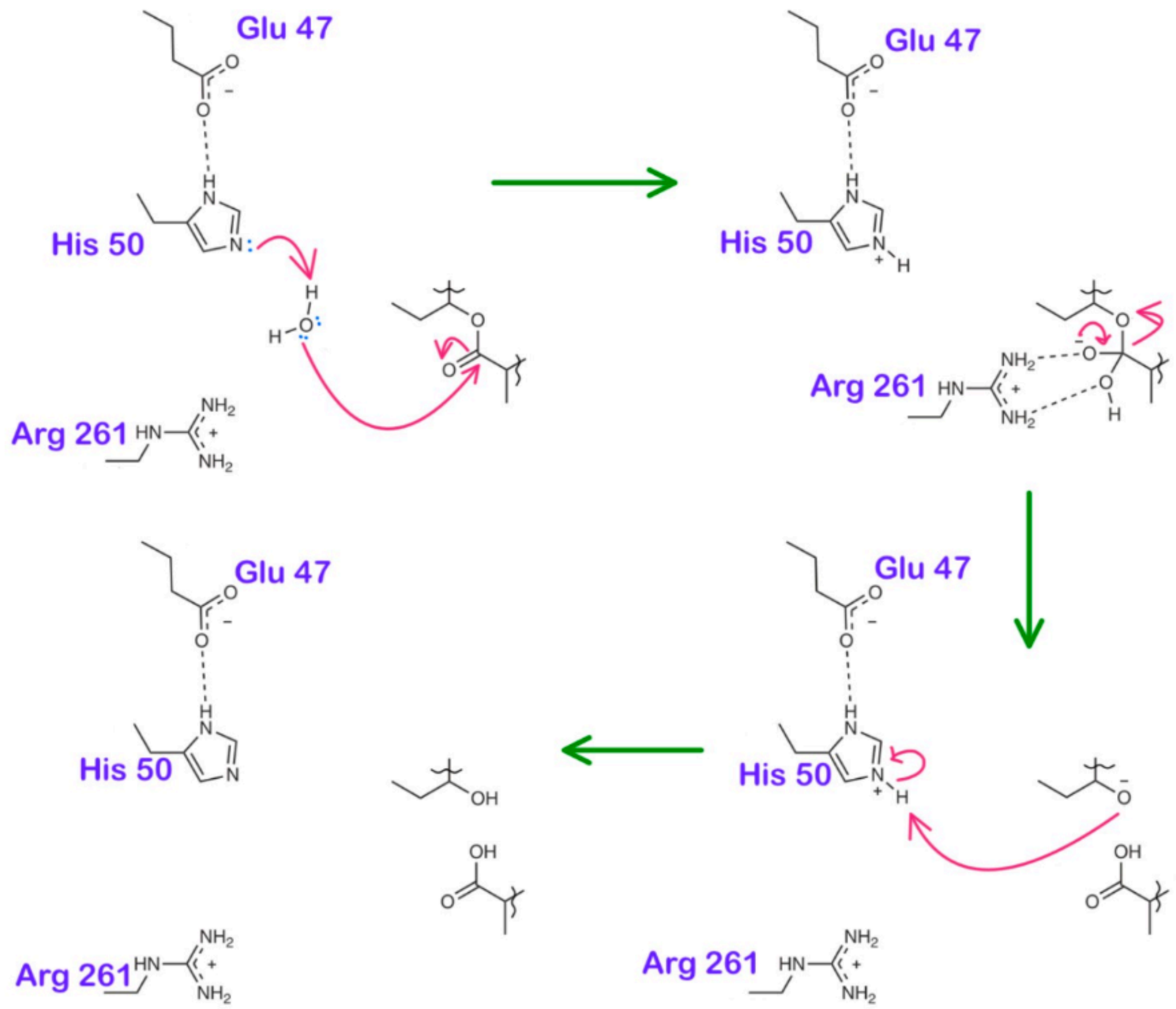

**Figure 27.** *Proposed mechanism of enzymatic macrolide cleavage by EreC. [34]*

### 3.1.2 Inhibition of Macrolide degradation

Previous studies established that macrolides affect bacterial protein synthesis by attaching to the 50S ribosome [106]. Specifically, the macrolide binds within the peptide exit tunnel of the large ribosomal subunit situated adjacent to the peptidyl transferase center. This action obstructs the tunnel's interior, hindering the passage of an elongating polypeptide chain and leading to either a bacteriostatic or bactericidal outcome, depending on the specific macrolide involved [107]. However, the precise mechanism of macrolide binding to the ribosome remains elusive. If this were elucidated and integrated into current research efforts, significant progress could be made in combating macrolide resistance mechanisms employed by Eres. Two rational approaches could be employed to inhibit macrolide resistance [32]:

1. Enhanced understanding of the features of macrolides recognized by resistance mechanisms, juxtaposed with their binding interactions with the ribosome, could inform the design of next-generation macrolides.
2. Development of inhibitors targeting the resistant proteins, which could subsequently serve as adjuvants to reinstate the efficacy of existing antibiotics.

## 3.2 Computational Methods

### 3.2.1 Protein and Ligand Systems

For this study, the two available crystal structures of EreC protein were collected from RCSB, $\mathrm{EreC}^{\mathrm{Open}}$(PDB ID: 6XCS) and $\mathrm{EreC}^{\mathrm{Closed}}$(PDB ID: 6XCQ). For the ligands Erythromycin and Azithromycin (Figure 26) were chosen, since both of them are good substrates of EreC [34]. For the molecular dynamics simulations the apo- forms of both the conformations along with their Erythromycin and Azithromycin bound states were used.

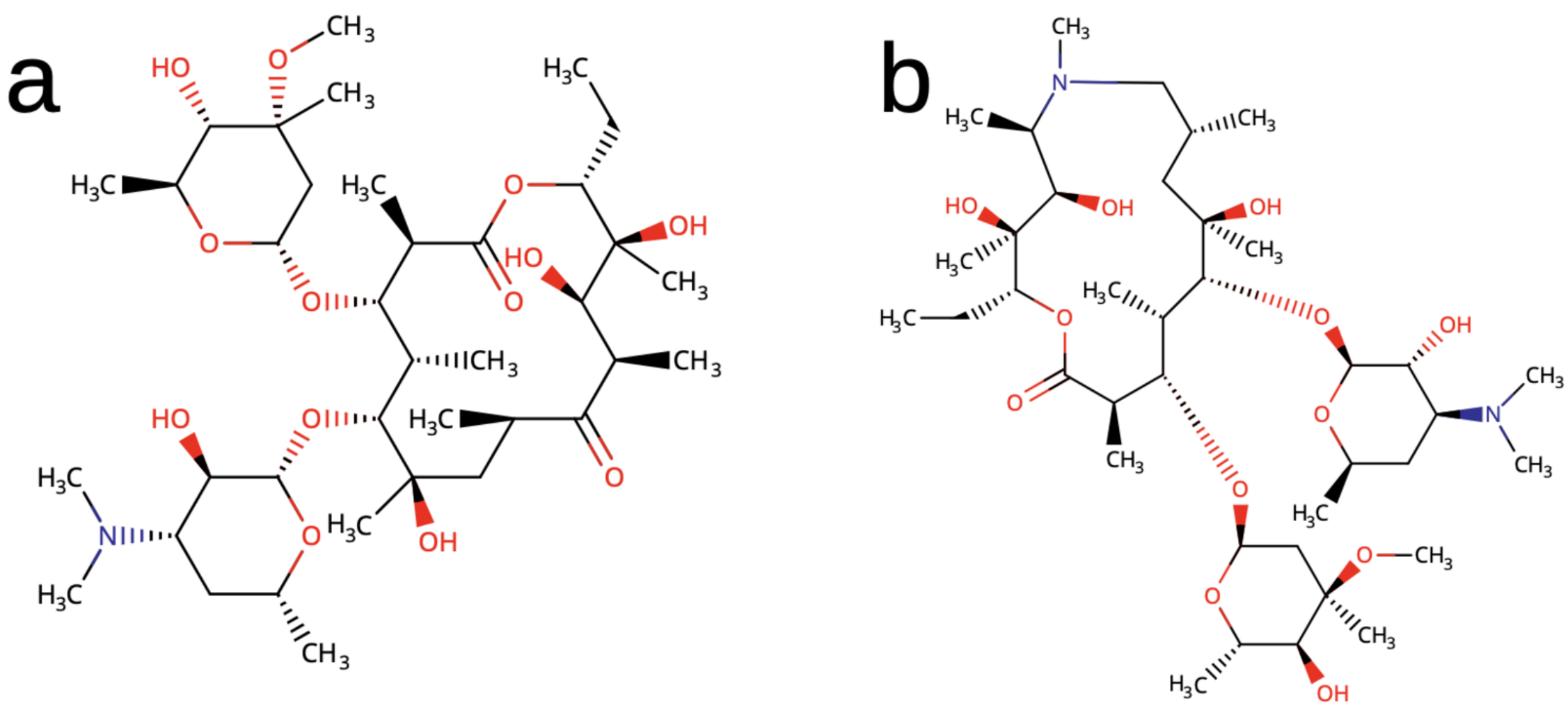

**Figure 28.** *Structures of the macrolide antibiotics (a) Erythromycin, (b) Azithromycin.*

### 3.2.2 Docking Studies

Docking of both the ligand molecules were performed on the Open and Closed conformers of EreC, which are obtained from RCSB. All the structures of molecules (ligands) were constructed using Avogadro 1.97 [65], following this a UFF optimization was done [66]. Vina-compatible receptor and ligand 'PDBQTs' were generated using Auto Dock Tools [61,62]. For the grid box setup, the dimensions were set as 18 Å × 18 Å × 18 Å with a grid spacing of 1.0, keeping the rest of the parameters as Vina default. The grid box was built beneath the flexible active loop, near the catalytic residues: His-50, Glu-78, His-289, Glu-47, Arg-261, and Asp-262. The purpose of these dockings was to get the ideal docking pose for further MD simulations. The selected docked poses are shown in Figure 27. These poses are so chosen that the ester group of the macrolide ring stays in close proximity to the catalytic residues and the tail stays above the pocket.

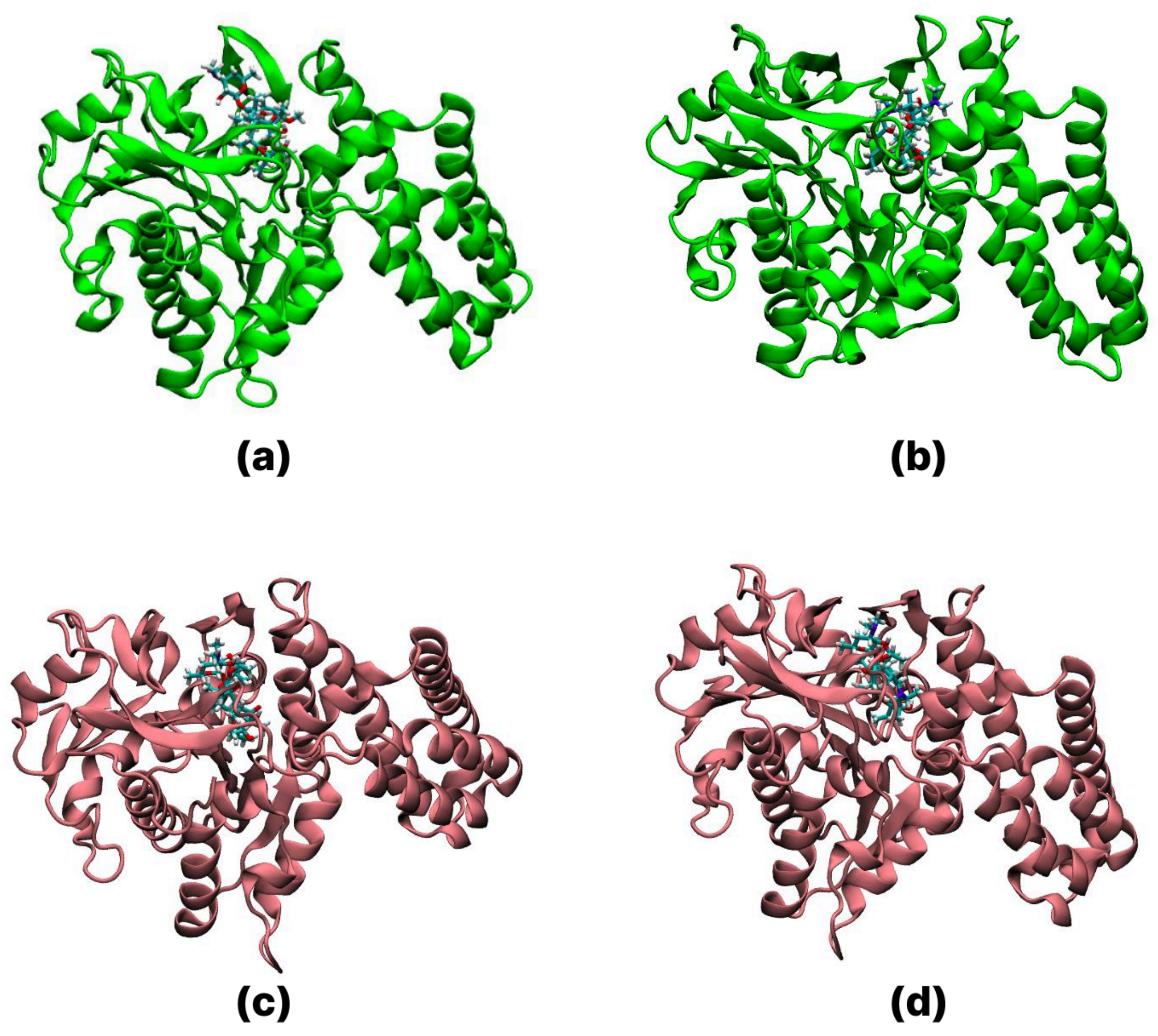

**Figure 29.** *Selected docked poses of the Ligands in EreC. (a) EreC$^{Open}$-Erythromycin, (b) EreC$^{Open}$-Azithromycin, (c) EreC$^{Closed}$-Erythromycin, and (d) EreC$^{Closed}$-Azithromycin.*

### 3.2.3 Classical MD Simulations

3.2.3.1 Preparation of Protein-Ligand Systems

The ligand structure and combined parameter/ topology files (prmtop) were generated in the Antechamber [88] module provided by AmberTools 18. All the molecules were optimized with the 6-311+G**/ B3LYP level in TeraChem [89-91], followed by calculation of restrained electrostatic potential charge (RESP) [92] using 6-31G/HF method. The ligand parameters were generated using GAFF [93] and the RESP charges are provided as partial atomic charges. The protein structures of EreC$^{Open}$ and EreC$^{Closed}$ were modeled using the Amber ff14SB force field [108]. The different protonation states of histidine residues were obtained from the H++ server [95] at pH 7.5 at the time of input generation. The ligand-bound EreC$^{Open}$ and EreC$^{Closed}$ structure and prmtop were prepared using the LEaP module. Each component was then assembled and solvated with the TIP3P water model [96] with 21 Å. The molarity and neutrality of each system was maintained at 150 mM by adding an appropriate number of Na+ and Cl− ions, 136 and 124 respectively. For comparative trajectory analysis, solvated apo-open, apo-close, and individual ligand systems were also prepared using the same procedure. Periodic boundary conditions were applied in each direction. A cutoff value of 10 Å was set for the calculation of van der Waals (vdW) and electrostatic interactions. Particle mesh Ewald (PME) [97] was used for electrostatics and SHAKE algorithm [98] for covalent interactions of heavy and hydrogen atoms. It has to be noted that the resids are 'original-10' throughout this study.

#### 3.2.3.2 Molecular Dynamics simulation

All the MD simulations were done using the AMBER 18 package [99]. Energy minimization was performed in a sequence of seven steps. In the first step, only the water molecules were minimized by imposing restraints into the complex. Then, gradually, restraints were decreased for the protein complex in the following five steps, and finally, the entire unrestrained system was energy-minimized. Successively, the heating was done from 0 to 310 K over a 600 ps period. followed by a total 6 ns simulation to equilibrate the system in three stages (restraints on the backbone and Ligand with the restraint wt 2.5, 1.0, and 0.5 kcal/Å-mol) in NPT ensemble at a constant temperature of 310 K and pressure of 1 atm. The production run was done for 400 ns, where the MD trajectory was produced in four 100 ns segments without imposing any restraint. A 2 fs integration step was set for the equilibration run and long simulations. During this NPT simulation, the temperature was maintained using a Langevin thermostat [100] with a collision frequency of 1.0 $ps^{-1}$ and 1 atm pressure was maintained using a Berendsen barostat [101].

#### 3.2.3.3 Structural analysis and Binding Free-Energy Estimation

From the last 400 ns trajectory data, the Root Mean Square Deviation (RMSD) was calculated using the CPPTRAJ module implemented in AmberTools 22, the Root Mean Square Fluctuations (RMSF), Radius of Gyration (RoG), Helix analysis, Contacts analysis, and H bonds analysis were done using MDAnalysis Python library [109, 110]. Snapshots of structures were visualized and generated in VMD [111]. The H-bond distance and angle cutoffs were set to 3.0 Å and 135°, respectively.

The binding energy of the molecules for both open and closed states was calculated using the well-established implicit solvation model, the Generalized Born surface area (QM/GBSA) method [112]. The gmx_MMPBSA module [113] implemented in GROMACS [114] was invoked to calculate the QM/GBSA binding free energy (Eq. 15). The QM region was defined within 6 Å of the ligand, the ResIDs of the residues in the QM region are: 40, 41, 97, 101, 127, 129, 131, 169, 170, 171, 231, 234, 235, 238, 239, 244, 248, 249, 251, 255, 276, 278, 279, 285, 286, 287, 292, 293, 322, 323, 325, 334, 378, 379, 380, and 381. The internal dielectric was set as $\varepsilon_{in}$=2 while the external dielectric was set as $\varepsilon_{out}$=80. The QM region was treated using the semiempirical QM Hamiltonian Parameterized Model number 3 (PM3); the QM/MM system was minimized using a self-consistent field (SCF) convergence with a threshold of 10-4. The GB model used was the $GB^{Neck2}$ (igb=8) [115], and the SASA was estimated using the LCPO algorithm. For these, a total of 250 frames were extracted from the last 20 ns of trajectories of each system. Entropy contribution to the overall free-energy change (ΔGb) was not calculated due to the small contribution for protein−small-molecule systems.

$$\Delta G_{b} \approx \Delta G_{MM} + \Delta G_{SCF} + \Delta G_{solv} - T\Delta S \quad (15)$$

## 3.3 Results and Discussion

### 3.3.1 MD Data Analyses

#### 3.3.1.1 Structural Changes to Binding Region

The RMSD data of the binding region consisting of the residues His-50, Glu-78, His-289, Glu-47, Arg-261, and Asp-262 over the last 400 ns trajectory were analysed. The average RMSD values of this region were calculated for each ligand-bound $\text{EreC}^{\text{Open}}$ and $\text{EreC}^{\text{Closed}}$ system to gain insights about the structural changes of the binding region (Figure 28). From the results, it was evident that the open structure experienced more structural changes than the closed structure after ligand binding. It can be seen that the RMSD of the erythromycin and azithromycin bound open system decreases significantly from that of the apo-open form, while the closed bound systems have a less noticeable decrease from the apo-closed form. This strongly suggests that the active loop tends to close, and the protein is likely to assume the closed structure after the binding of ligands into the pocket.

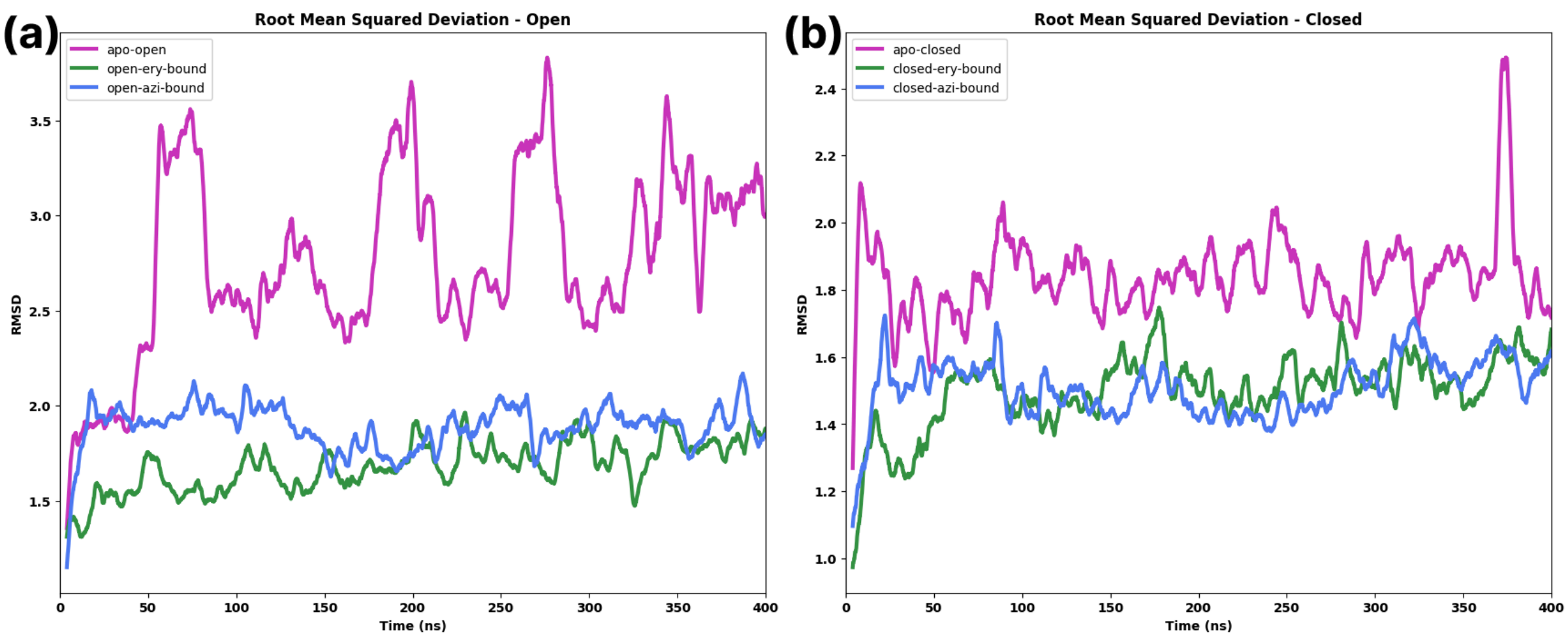

**Figure 30.** *Root Mean Squared Deviation of (a) $\text{EreC}^{\text{Open}}$ systems, (b) $\text{Erec}^{\text{Closed}}$ systems.*

The RoG of the region enclosed by the residues 238, 239, 286, 292, 324, 325, 380, and 381 also provided useful insights against structural changes. The decreased average RoG of this region from its apo state indicates the closure of the pocket upon ligand binding (Figure 29). This effect was found to be more prominent in the case of the Ery-bound open system. For the Azi-bound system, the RoG increases slightly at the beginning, then decreases with respect to the apo state; this shows that the pocket widens in the beginning because of the large size of azithromycin.

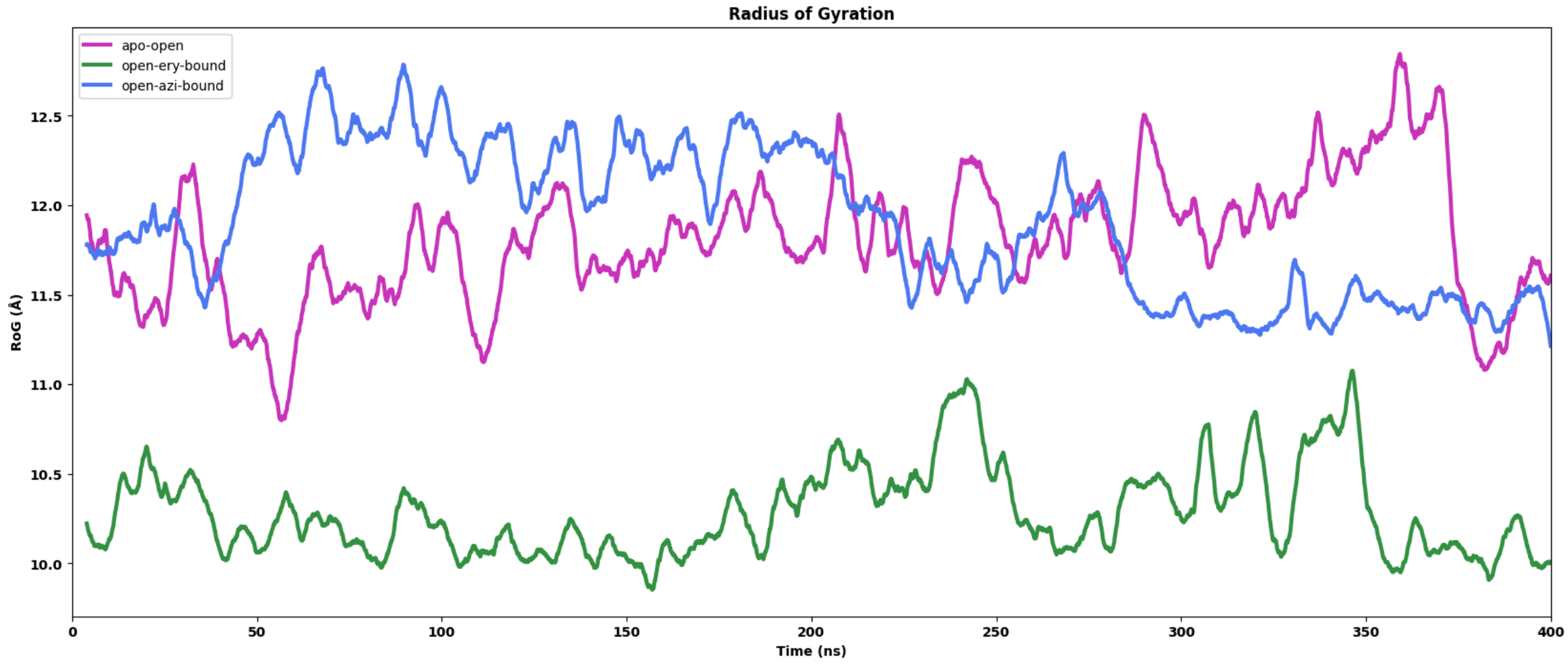

**Figure 31.** *Radius of Gyration of EreC$^{Open}$ systems.*

The helix analysis of the active loop (resid 282-295) shows a significant decrease in local twist angles, indicating a decrease in fluctuations. For the lateral β-sheet (resid 374-385) and opposite β-sheet (resid 321-340), this change is less noticeable (Figure 30).

**Figure 32.** *Helix analysis of (a) Active loop, (b) Lateral β-sheet, and (c) Opposite β-sheet.*

3.3.1.2 QM/GBSA and H-Bonding Analyses

Both erythromycin and azithromycin record a good binding affinity (ΔG) for both ligand-bound Open and Closed systems. From the data obtained from the QM/GBSA studies (Table 7), it can be clearly seen that the binding affinities of the ligands are much larger in the Closed form than in the Open form. This further solidifies the argument that the protein adapts the closed conformation after macrolide binding. In the case of $EreC^{Open}$, the ΔG values of erythromycin and azithromycin are -16.16 ± 4.57 kcal/mol and -24.38 ± 4.67 kcal/mol respectively, while in the case of $EreC^{Closed}$, they are -27.52 ± 3.41 kcal/mol and -33.86 ± 3.34 kcal/mol respectively. In both the conformations, azithromycin binds very tightly, as evidenced by its higher binding energy.

**Table 7.** *$\Delta G_{QMGBSA}$ values in kcal/mol shown for Erythromycin and Azithromycin bound Open and Closed structures of EreC.*

| **Energy Terms** | **$EreC^{Open}$** | | **$EreC^{Closed}$** | |
|---|---|---|---|---|
| | **Erythromycin** | **Azithromycin** | **Erythromycin** | **Azithromycin** |
| vdW | -6.86 ± 0.84 | -11.15 ± 1.60 | -13.64 ± 1.30 | -16.05 ± 1.03 |
| $E_{GB}$ | 15.41 ± 2.10 | 16.64 ± 1.83 | 22.99 ± 1.39 | 23.59 ± 1.38 |
| $E_{SURF}$ | -6.41 ± 0.57 | -7.20 ± 0.50 | -7.46 ± 0.29 | -7.64 ± 0.20 |
| $E_{SCF}$ | -18.30 ± 5.23 | -22.68 ± 4.16 | -29.44 ± 8.98 | -33.77 ± 3.63 |
| $\Delta G_{gas}$ | -25.16 ± 5.53 | -33.83 ± 5.22 | -43.05 ± 4.03 | -49.82 ± 3.85 |
| $\Delta G_{solv}$ | 9.00 ± 1.71 | 9.44 ± 1.60 | 15.53 ± 1.29 | 15.96 ± 1.33 |
| $\Delta G_{QMGBSA}$ | -16.16 ± 4.57 | -24.38 ± 4.67 | -27.52 ± 3.41 | -33.86 ± 3.34 |

From the Hydrogen bond analysis (Table 8), it can be clearly seen that the macrolide oxygen of both erythromycin and azithromycin form H-bond with Arg-261 (ARG_251 in the table, since resids are 'original-10' in this study as mentioned previously in methods section). This result is consistent with the proposed mechanism shown in Figure 25, in which the macrolide oxygen forms H-bond with the amine group of Arg-261. This validates that the catalytic residues His-50, Glu-78, His-289, Glu-47, Arg-261, and Asp-262 play very important roles in the macrolide degradation process.

**Table 8.** *Hydrogen Bonding information for (a) Erythromycin and (b) Azithromycin with EreC$^{Open}$.*

(a)

| Acceptor | Donor | #Frames | AvgDist | AvgAng |
|---|---|---|---|---|
| LIG_409@O7 | SER_249@OG | 2 | 2.9286 | 162.2420 |
| LIG_407@O10 | SER_378@OG | 347 | 2.8244 | 164.7192 |
| LIG_409@O3 | HIP_40@NE2 | 2 | 2.8155 | 165.5008 |
| LIG_409@O1 | HIP_40@NE2 | 1 | 2.8775 | 172.7548 |
| LIG_409@O7 | THR_248@OG1 | 40 | 2.8291 | 163.2128 |
| LIG_409@O3 | ARG_251@NH1 | 47 | 2.8382 | 158.3607 |
| LIG_409@O10 | GLN_379@NE2 | 71 | 2.8987 | 158.9385 |
| LIG_409@O10 | ASN_278@ND2 | 1 | 2.8315 | 152.7384 |

(b)

| Acceptor | Donor | #Frames | AvgDist | AvgAng |
|---|---|---|---|---|
| LIG_409@O1 | SER_169@OG | 175 | 2.7566 | 165.0037 |
| LIG_407@O9 | SER_288@OG | 57 | 2.7883 | 161.4835 |
| LIG_409@O6 | ASN_129@ND2 | 20 | 2.8792 | 161.9463 |
| LIG_409@O6 | ARG_251@NH1 | 4 | 2.8923 | 163.2286 |

#### 3.3.1.3 Ligand Binding and Active Loop Dynamics

The binding of macrolide is supposed to be followed by the closure of the active loop, as discussed above. This observation can be inferred from the RMSF data of the apo and ligand-bound states of Open conformation. Also, after careful visual inspection, it was found that the bound ligand is held near a β-sheet laterally to the ligand and another β-sheet opposite to it. Thus, the RMSF of these β-sheets may also give useful insights into the binding mechanism of macrolide. For this, the RMSF of the apo- and ligand-bound Open conformation were computed from their respective trajectories (Figure 31).

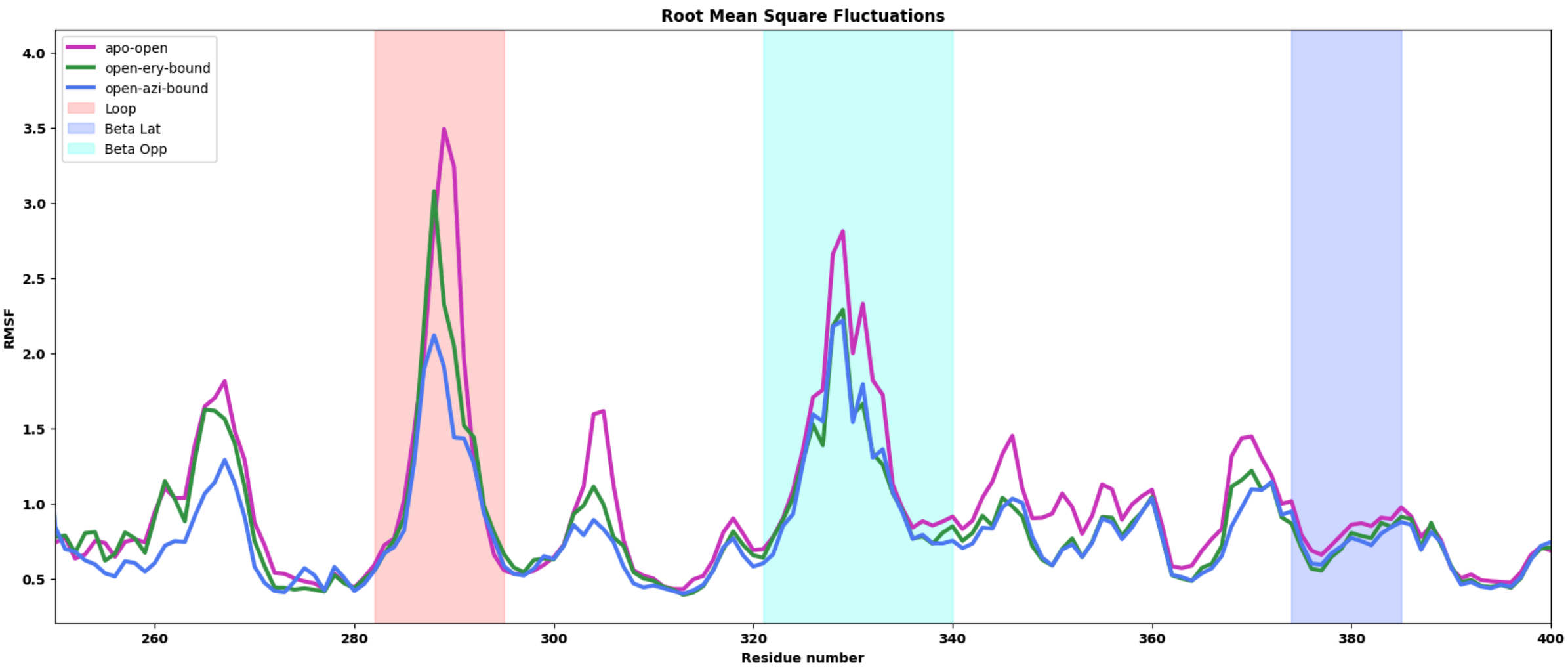

**Figure 33.** *Root Mean Square Fluctuations for apo-Open, Erythromycin-bound Open and Azithromycin-bound Open systems, highlighting the active loop, lateral β-sheet, and opposite β-sheet regions.*

From the results, it can be seen that the binding of Erythromycin or Azithromycin decreased the fluctuations of the active loop and the lateral β-sheet from the apo- form. In contrast, no significant change can be observed for the opposite β-sheet. The reduction in the fluctuation of the active loop is more pronounced in the case of azithromycin. These results indicate that after the binding of the ligand, the active loop stabilizes and tends to close above the ligand, ultimately going towards the closed conformation.

# 4. CONCLUSION

## 4.1 AcrAB-TolC and MexAB-OprM Efflux Pumps

The goal of this study was to identify potential EPI candidates for AcrB and MexB using various ML classification models trained upon the structural and pharmacokinetic properties of previously reported inhibitors. To generate targets in the binary classification approach, intrinsic MIC values for all 53 known molecules were utilized. The target was set to "0" for less potent and "1" for highly potent, as mentioned earlier in the Methods section. Thereafter, 53 molecules were divided into the "training (80%)" and "test (20%)" sets, followed by implementing 10 classification models onto these. SVM, Random Forest, AdaBoost, and LightGBM showed quite good performance in terms of prediction quality for our available data set, where LightGBM scored the highest in all comparison parameters. Hence, LightGBM along with the other three was applied onto an unknown data set of 5043 generated molecules, and from that, a total of 493 molecules were popped out as "highly potent". Lipinski's Ro5 and AcrB/MexB-ligand docking scores were employed sequentially to filter out the top eight hits, followed by MD simulations. From the simulation studies, Lig6 (mol.3488) showed an overall good in-silico activity for both AcrB and MexB. Also, Lig2 (mol.2977) for AcrB and Lig3 (mol.3032) showed high binding affinities. It is noteworthy that, of these top ligands, the "pyridopyrimidone and benzothiazole" moieties are found to be very important for ligand binding inside the DBP. Therefore, the three molecules Lig6, Lig2, and Lig3 might act as potential EPIs to cope with AcrB and MexB drug efflux by disrupting the functional rotating mechanism. In conclusion, our findings might help both the in-silico and in-vitro pharmacophore-based EPI modeling for Gram-negative bacteria in the near future.

## 4.2 Macrolide Degradation by EreC

The goal of this study was to gain insights into the mechanism of macrolide binding into the catalytic pocket using molecular dynamics simulations. The mechanism proposed in the literature states that an active loop between residues 295 and 303 closes upon ligand binding into the pocket beneath that loop. The protein thus changes its conformation from open to closed. Following this, the ligand aligns itself with the catalytic residues His-50, Glu-78, His-289, Glu-47, Arg-261, and Asp-262 and finally gets catalytically degraded. For this purpose, the crystal structures of $EreC^{Open}$ and $EreC^{Closed}$ conformations were taken and bound with erythromycin and azithromycin. Classical MD simulations were run using apo-, ery-bound, and azi-bound states of both the conformations. The analyses were done using the last 400ns of trajectory. The RMSD analysis suggests that the active loop tends to close, and the protein is likely to assume the closed structure after the binding of ligands into the pocket since the ligand-bound open systems have significant deviations from that of the apo-open form, while the ligand-bound closed systems this change is less noticeable. The decreased average RoG of the binding region after ligand binding from its apo state also indicates the closure of the pocket. The binding energy from the QM/GBSA showed that the binding affinities of the ligands are much larger in the closed form than in the open form, which further solidifies the argument that the protein adapts to closed conformation after macrolide binding. The RMSF analysis of the active loop showed that the binding of the ligand decreases the fluctuations; this indicates that after the binding of the ligand, the active loop stabilizes and tends to close above the ligand, ultimately going towards the closed conformation. The hydrogen bond analysis showed that the macrolide oxygen of both the ligands form an H-bond with Arg-261, which is consistent with the proposed mechanism and validates that the catalytic residues His-50, Glu-78, His-289, Glu-47, Arg-261, and Asp-262 play critical roles in the macrolide degradation process.

# 5. FUTURE SCOPE

In the prediction of AcrB and MexB efflux pump inhibitors we have used a handful of basic machine learning models along with some bagging and boosting techniques for classification of Highly Potent and Less Potent inhibitors. More robust deep learning and neural networks methods can be employed to get better predictions. In the study of macrolide degradation of the MD simulation results suggest that the active loop closes upon ligand binding into the catalytic pocket. To get more information about the catalytic bond breaking of macrolide ester, QM/MM simulations can be done.

# REFERENCES

(1) Bartholomew, J. W.; Mittwer, T. The Gram Stain. *Bacteriol. Rev.* 1952, *16* (1), 1–29. https://doi.org/10.1128/br.16.1.1-29.1952.

(2) O'Toole, G. A. Classic Spotlight: How the Gram Stain Works. *J. Bacteriol.* 2016, *198* (23), 3128–3128. https://doi.org/10.1128/jb.00726-16.

(3) Libenson, L.; McIlroy, A. P. On the Mechanism of the Gram Stain. *J. Infect. Dis.* 1955, *97* (1), 22–26. https://doi.org/10.1093/infdis/97.1.22.

(4) Shugar, D.; Baranowska, J. Studies on the Gram Stain; the Importance of Proteins in the Gram Reaction. *Acta Microbiol. Pol. (1952)* 1954, *3* (1), 11–20.

(5) Haslett, A. S. W. The Chemical Significance of the Gram Test for Bacteria. *Aust. J. Sci.* 1947, *9* (6), 211.

(6) Fleming, A. On the Antibacterial Action of Cultures of a Penicillium, with Special Reference to Their Use in the Isolation of B. Influenzae. *Rev. Infect. Dis.* 1980, *2* (1), 129–139.

(7) Pankey, G. A.; Sabath, L. D. Clinical Relevance of Bacteriostatic versus Bactericidal Mechanisms of Action in the Treatment of Gram‐positive Bacterial Infections. *Clin. Infect. Dis.* 2004, *38* (6), 864–870. https://doi.org/10.1086/381972.

(8) *Antimicrobials, Antibiotic Resistance, Antibiofilm Strategies and Activity Methods*; Kırmusaoglu, S., Ed.; IntechOpen: London, England, 2019. https://doi.org/10.5772/intechopen.78751.

(9) Henkin, T. M.; Peters, J. E. *Snyder and Champness Molecular Genetics of Bacteria*, 5th ed.; American Society for Microbiology: Washington, D.C., DC, 2020.

(10) Tortora, G. J.; Funke, B. R.; Case, C. L.; Weber, D.; Bair, W. *Microbiology: An Introduction*, 13th ed.; Pearson, 2018.

(11) Rice, L. B. Federal Funding for the Study of Antimicrobial Resistance in Nosocomial Pathogens: No ESKAPE. *J. Infect. Dis.* 2008, *197* (8), 1079–1081. https://doi.org/10.1086/533452.

(12) Catalano, A.; Iacopetta, D.; Ceramella, J.; Scumaci, D.; Giuzio, F.; Saturnino, C.; Aquaro, S.; Rosano, C.; Sinicropi, M. S. Multidrug Resistance (MDR): A Widespread Phenomenon in Pharmacological Therapies. *Molecules* 2022, *27* (3), 616. https://doi.org/10.3390/molecules27030616.

(13) Blair, J. M. A.; Webber, M. A.; Baylay, A. J.; Ogbolu, D. O.; Piddock, L. J. V. Molecular Mechanisms of Antibiotic Resistance. *Nat. Rev. Microbiol.* 2015, *13* (1), 42–51. https://doi.org/10.1038/nrmicro3380.

(14) Silhavy, T. J.; Kahne, D.; Walker, S. The Bacterial Cell Envelope. Cold Spring Harbor Perspect. *Biol* 2010, *2* (a000414).

(15) Blair, J. M. A.; Richmond, G. E.; Piddock, L. J. V. Multidrug Efflux Pumps in Gram-Negative Bacteria and Their Role in Antibiotic Resistance. *Future Microbiol.* 2014, *9* (10), 1165–1177. https://doi.org/10.2217/fmb.14.66.

(16) Alav, I.; Kobylka, J.; Kuth, M. S.; Pos, K. M.; Picard, M.; Blair, J. M. A.; Bavro, V. N. Structure, Assembly, and Function of Tripartite Efflux and Type 1 Secretion Systems in Gram-Negative Bacteria. *Chem. Rev.* 2021, *121* (9), 5479–5596. https://doi.org/10.1021/acs.chemrev.1c00055.

(17) Okada, U.; Yamashita, E.; Neuberger, A.; Morimoto, M.; van Veen, H. W.; Murakami, S. Crystal Structure of Tripartite-Type ABC Transporter MacB from Acinetobacter Baumannii. *Nat. Commun.* 2017, *8* (1), 1336. https://doi.org/10.1038/s41467-017-01399-2.

(18) Kuroda, T.; Tsuchiya, T. Multidrug Efflux Transporters in the MATE Family. *Biochim. Biophys. Acta* 2009, *1794* (5), 763–768. https://doi.org/10.1016/j.bbapap.2008.11.012.

(19) Morrison, E. A.; DeKoster, G. T.; Dutta, S.; Vafabakhsh, R.; Clarkson, M. W.; Bahl, A.; Kern, D.; Ha, T.; Henzler-Wildman, K. A. Antiparallel EmrE Exports Drugs by Exchanging between Asymmetric Structures. *Nature* 2011, *481* (7379), 45–50. https://doi.org/10.1038/nature10703.

(20) Hassan, K. A.; Liu, Q.; Henderson, P. J. F.; Paulsen, I. T. Homologs of the Acinetobacter Baumannii AceI Transporter Represent a New Family of Bacterial Multidrug Efflux Systems. *MBio* 2015, *6* (1). https://doi.org/10.1128/mBio.01982-14.

(21) Sulavik, M. C.; Houseweart, C.; Cramer, C.; Jiwani, N.; Murgolo, N.; Greene, J.; Didomenico, B.; Shaw, K. J.; Miller, G. H.; Hare, R.; Shimer, G. Antibiotic Susceptibility Profiles of Escherichiacoli Strains Lacking Multidrug Efflux Pump Genes. *Antimicrob. Agents Chemother* 2001, *45*, 1126–1136.

(22) Nikaido, H. RND Transporters in the Living World. *Res. Microbiol.* 2018, *169* (7–8), 363–371. https://doi.org/10.1016/j.resmic.2018.03.001.

(23) Tseng, T. T.; Gratwick, K. S.; Kollman, J.; Park, D.; Nies, D. H.; Goffeau, A.; Saier, M. H. Jr The RND Permease Superfamily: An Ancient, Ubiquitous and Diverse Family That Includes Human Disease and Development Proteins. *J. Mol. Microbiol. Biotechnol* 1999, *1*, 107–125.

(24) Zhang, B.; Li, J.; Yang, X.; Wu, L.; Zhang, J.; Yang, Y.; Zhao, Y.; Zhang, L.; Yang, X.; Yang, X.; Cheng, X.; Liu, Z.; Jiang, B.; Jiang, H.; Guddat, L. W.; Yang, H.; Rao, Z. Crystal Structures of Membrane Transporter MmpL3, an Anti-TB Drug Target. *Cell* 2019, *176* (3), 636-648.e13. https://doi.org/10.1016/j.cell.2019.01.003.

(25) Kumar, N.; Su, C.-C.; Chou, T.-H.; Radhakrishnan, A.; Delmar, J. A.; Rajashankar, K. R.; Yu, E. W. Crystal Structures of the Burkholderia Multivorans Hopanoid Transporter HpnN. *Proc. Natl. Acad. Sci. U. S. A.* 2017, *114* (25), 6557–6562. https://doi.org/10.1073/pnas.1619660114.

(26) Nies, D. H. Efflux-Mediated Heavy Metal Resistance in Prokaryotes. *FEMS Microbiol. Rev.* 2003, *27* (2–3), 313–339. https://doi.org/10.1016/s0168-6445(03)00048-2.

(27) Pak, J. E.; Ekendé, E. N.; Kifle, E. G.; O'Connell, J. D., 3rd; De Angelis, F.; Tessema, M. B.; Derfoufi, K.-M.; Robles-Colmenares, Y.; Robbins, R. A.; Goormaghtigh, E.; Vandenbussche, G.; Stroud, R. M. Structures of Intermediate Transport States of ZneA, a Zn(II)/Proton Antiporter. *Proc. Natl. Acad. Sci. U. S. A.* 2013, *110* (46), 18484–18489. https://doi.org/10.1073/pnas.1318705110.

(28) Su, C.-C.; Long, F.; Zimmermann, M. T.; Rajashankar, K. R.; Jernigan, R. L.; Yu, E. W. Crystal Structure of the CusBA Heavy-Metal Efflux Complex of Escherichia Coli. *Nature* 2011, *470* (7335), 558–562. https://doi.org/10.1038/nature09743.

(29) Chacón, K. N.; Mealman, T. D.; McEvoy, M. M.; Blackburn, N. J. Tracking Metal Ions through a Cu/Ag Efflux Pump Assigns the Functional Roles of the Periplasmic Proteins. *Proc. Natl. Acad. Sci. U. S. A.* 2014, *111* (43), 15373–15378. https://doi.org/10.1073/pnas.1411475111.

(30) Saier, M. H., Jr; Tam, R.; Reizer, A.; Reizer, J. Two Novel Families of Bacterial Membrane Proteins Concerned with Nodulation, Cell Division and Transport. *Mol. Microbiol.* 1994, *11* (5), 841–847. https://doi.org/10.1111/j.1365-2958.1994.tb00362.x.

(31) Ohara, K.; Kanda, T.; Kono, M. Structure of a Phosphorylated Derivative of Oleandomycin, Obtained by Reaction of Oleandomycin with an Extract of an Erythromycin-Resistant Strain of Escherichia Coli. *J. Antibiot* 1988, *41*, 823–827.

(32) Golkar, T.; Zieliński, M.; Berghuis, A. M. Look and Outlook on Enzyme-Mediated Macrolide Resistance. *Front. Microbiol.* 2018, *9*. https://doi.org/10.3389/fmicb.2018.01942.

(33) Barthélémy, P.; Autissier, D.; Gerbaud, G.; Courvalin, P. Enzymic Hydrolysis of Erythromycin by a Strain of Escherichia Coli. A New Mechanism of Resistance. *J. Antibiot. (Tokyo)* 1984, *37* (12), 1692–1696. https://doi.org/10.7164/antibiotics.37.1692.

(34) Zieliński, M.; Park, J.; Sleno, B.; Berghuis, A. M. Structural and Functional Insights into Esterase-Mediated Macrolide Resistance. *Nat. Commun.* 2021, *12* (1). https://doi.org/10.1038/s41467-021-22016-3.

(35) Cundliffe, E. Glycosylation of Macrolide Antibiotics in Extracts of Streptomyces Lividans. *Antimicrob. Agents Chemother.* 1992, *36* (2), 348–352. https://doi.org/10.1128/AAC.36.2.348.

(36) Vilches, C.; Hernandez, C.; Mendez, C.; Salas, J. A. Role of Glycosylation and Deglycosylation in Biosynthesis of and Resistance to Oleandomycin in the Producer Organism, Streptomyces Antibioticus. *J. Bacteriol.* 1992, *174* (1), 161–165. https://doi.org/10.1128/jb.174.1.161-165.1992.

(37) Ambler, R. P.; Coulson, A. F. W.; Frère, J. M.; Ghuysen, J. M.; Joris, B.; Forsman, M.; Levesque, R. C.; Tiraby, G.; Waley, S. G. A Standard Numbering Scheme for the Class A *β*-Lactamases. *Biochem. J.* 1991, *276* (1), 269–270. https://doi.org/10.1042/bj2760269.

(38) Bush, K.; Jacoby, G. A.; Medeiros, A. A. A Functional Classification Scheme for Beta-Lactamases and Its Correlation with Molecular Structure. *Antimicrob. Agents Chemother.* 1995, *39* (6), 1211–1233. https://doi.org/10.1128/aac.39.6.1211.

(39) Bush, K. Metallo-β-lactamases: A Class Apart. *Clin. Infect. Dis.* 1998, *27* (s1), S48–S53. https://doi.org/10.1086/514922.

(40) Kumar, A.; Schweizer, H. P. Bacterial Resistance to Antibiotics: Active Efflux and Reduced Uptake. *Adv. Drug Deliv. Rev.* 2005, *57* (10), 1486–1513. https://doi.org/10.1016/j.addr.2005.04.004.

(41) Nakamura, H.; Hachiya, N.; Tojo, T. Second Acriflavine Sensitivity Mutation,*acrB*, in*Escherichia Coli*K-12. *J. Bacteriol.* 1978, *134* (3), 1184–1187. https://doi.org/10.1128/jb.134.3.1184-1187.1978.

(42) Sharma, O.; Zakharov, S. D.; Zhalnina, M. V.; Yamashita, E.; Cramer, W. A. Colicins. In *Handbook of Biologically Active Peptides*; Elsevier, 2013; pp 93–100.

(43) Poole, K.; Krebes, K.; McNally, C.; Neshat, S. Multiple Antibiotic Resistance in Pseudomonas Aeruginosa: Evidence for Involvement of an Efflux Operon. *J. Bacteriol.* 1993, *175* (22), 7363–7372. https://doi.org/10.1128/jb.175.22.7363-7372.1993.

(44) Masuda, N.; Sakagawa, E.; Ohya, S. Outer Membrane Proteins Responsible for Multiple Drug Resistance in Pseudomonas Aeruginosa. *Antimicrob. Agents Chemother.* 1995, *39* (3), 645–649. https://doi.org/10.1128/aac.39.3.645.

(45) Murakami, S.; Nakashima, R.; Yamashita, E.; Yamaguchi, A. Crystal Structure of Bacterial Multidrug Efflux Transporter AcrB. *Nature* 2002, *419* (6907), 587–593. https://doi.org/10.1038/nature01050.

(46) Nakashima, R.; Sakurai, K.; Yamasaki, S.; Nishino, K.; Yamaguchi, A. Structures of the Multidrug Exporter AcrB Reveal a Proximal Multisite Drug-Binding Pocket. *Nature* 2011, *480* (7378), 565–569. https://doi.org/10.1038/nature10641.

(47) Eicher, T.; Cha, H.-J.; Seeger, M. A.; Brandstätter, L.; El-Delik, J.; Bohnert, J. A.; Kern, W. V.; Verrey, F.; Grütter, M. G.; Diederichs, K.; Pos, K. M. Transport of Drugs by the Multidrug Transporter AcrB Involves an Access and a Deep Binding Pocket That Are Separated by a Switch-Loop. *Proc. Natl. Acad. Sci. U. S. A.* 2012, *109* (15), 5687–5692. https://doi.org/10.1073/pnas.1114944109.

(48) Müller, R. T.; Travers, T.; Cha, H.-J.; Phillips, J. L.; Gnanakaran, S.; Pos, K. M. Switch Loop Flexibility Affects Substrate Transport of the AcrB Efflux Pump. *J. Mol. Biol.* 2017, *429* (24), 3863–3874. https://doi.org/10.1016/j.jmb.2017.09.018.

(49) Bera, A.; Roy, R. K.; Joshi, P.; Patra, N. Machine Learning-Guided Discovery of AcrB and MexB Efflux Pump Inhibitors. *J. Phys. Chem. B* 2024, *128* (3), 648–663. https://doi.org/10.1021/acs.jpcb.3c05845.

(50) Roy, R. K.; Patra, N. Probing the pH Sensitivity of OprM: Insights into Metastable States and Semi-Open Conformation. *J. Phys. Chem. B* 2024, *128* (3), 622–634. https://doi.org/10.1021/acs.jpcb.3c05384.

(51) Bolla, J.-M.; Alibert-Franco, S.; Handzlik, J.; Chevalier, J.; Mahamoud, A.; Boyer, G.; Kieć-Kononowicz, K.; Pagès, J.-M. Strategies for Bypassing the Membrane Barrier in Multidrug Resistant Gram-Negative Bacteria. *FEBS Lett.* 2011, *585* (11), 1682–1690. https://doi.org/10.1016/j.febslet.2011.04.054.

(52) Nakashima, R.; Sakurai, K.; Yamasaki, S.; Hayashi, K.; Nagata, C.; Hoshino, K.; Onodera, Y.; Nishino, K.; Yamaguchi, A. Structural Basis for the Inhibition of Bacterial Multidrug Exporters. *Nature* 2013, *500* (7460), 102–106. https://doi.org/10.1038/nature12300.

(53) Bohnert, J. A.; Kern, W. V. Selected Arylpiperazines Are Capable of Reversing Multidrug Resistance in Escherichia Coli Overexpressing RND Efflux Pumps. *Antimicrob. Agents Chemother.* 2005, *49* (2), 849–852. https://doi.org/10.1128/AAC.49.2.849-852.2005.

(54) Opperman, T. J.; Kwasny, S. M.; Kim, H.-S.; Nguyen, S. T.; Houseweart, C.; D'Souza, S.; Walker, G. C.; Peet, N. P.; Nikaido, H.; Bowlin, T. L. Characterization of a Novel Pyranopyridine Inhibitor of the AcrAB Efflux Pump of Escherichia Coli. *Antimicrob. Agents Chemother.* 2014, *58* (2), 722–733. https://doi.org/10.1128/AAC.01866-13.

(55) Yilmaz, S.; Altinkanat-Gelmez, G.; Bolelli, K.; Guneser-Merdan, D.; Ufuk Over-Hasdemir, M.; Aki-Yalcin, E. Yalcin, I. Binding Site Feature Description of 2-Substituted Benzothiazoles as Potential AcrAB-TolC Efflux Pump Inhibitors inE. Coli. *SAR QSAR Environ. Res* 2015, *26* (10), 853–871.

(56) Yoshida, K.-I.; Nakayama, K.; Yokomizo, Y.; Ohtsuka, M.; Takemura, M.; Hoshino, K.; Kanda, H.; Namba, K.; Nitanai, H.; Zhang, J. Z. MexAB-OprM Specific Efflux Pump Inhibitors in Pseudomonas Aeruginosa. Part 6: Exploration of Aromatic Sub-Stituents. *Bioorg. Med. Chem* 2006, No. 24, 8506–8518.

(57) Bishop, C. M. Model-Based Machine Learning. *Philos. Trans. A Math. Phys. Eng. Sci.* 2013, *371* (1984), 20120222. https://doi.org/10.1098/rsta.2012.0222.

(58) Berry, M. W.; Mohamed, A.; Yap, B. W. *Supervised and Unsupervised Learning for Data Science*; Springer Nature: Cham, Switzerland, 2019.

(59) Compagne, N.; Vieira Da Cruz, A.; Müller, R. T.; Hartkoorn, R. C.; Flipo, M.; Pos, K. M. Update on the Discovery of Efflux Pump Inhibitors against Critical Priority Gram-Negative Bacteria. *Antibiotics (Basel)* 2023, *12* (1), 180. https://doi.org/10.3390/antibiotics12010180.

(60) BIOVIA − Dassault systèmes®. 3ds.com. *BIOVIA Product Portfolio*; 2022.

(61) Eberhardt, J.; Santos-Martins, D.; Tillack, A. F.; Forli, S. AutoDock Vina 1.2.0: New Docking Methods, Expanded Force Field, and Python Bindings. *J. Chem. Inf. Model.* 2021, *61* (8), 3891–3898. https://doi.org/10.1021/acs.jcim.1c00203.

(62) Trott, O.; Olson, A. J. AutoDock Vina: Improving the Speed and Accuracy of Docking with a New Scoring Function, Efficient Optimization, and Multithreading. *J. Comput. Chem.* 2010, *31* (2), 455–461. https://doi.org/10.1002/jcc.21334.

(63) Pearson, K. VII. Note on Regression and Inheritance in the Case of Two Parents. *Proc. R. Soc. Lond.* 1895, *58* (347–352), 240–242. https://doi.org/10.1098/rspl.1895.0041.

(64) Pedregosa, F.; Varoquaux, G.; Gramfort, A.; Michel, V.; Thirion, B.; Grisel, O.; Blondel, M.; Müller, A.; Nothman, J.; Louppe, G.; Prettenhofer, P.; Weiss, R.; Dubourg, V.; Vanderplas, J.; Passos, A.; Cournapeau, D.; Brucher, M.; Perrot, M.; Duchesnay, É. Scikit-Learn: Machine Learning in Python. *arXiv [cs.LG]*, 2012. http://arxiv.org/abs/1201.0490.

(65) Hanwell, M. D.; Curtis, D. E.; Lonie, D. C.; Vandermeersch, T.; Zurek, E.; Hutchison, G. R. Avogadro: An Advanced Semantic Chemical Editor, Visualization, and Analysis Platform. *J. Cheminform.* 2012, *4* (1), 17. https://doi.org/10.1186/1758-2946-4-17.

(66) Rappe, A. K.; Casewit, C. J.; Colwell, K. S.; Goddard, W. A., III; Skiff, W. M. UFF, a Full Periodic Table Force Field for Molecular Mechanics and Molecular Dynamics Simulations. *J. Am. Chem. Soc.* 1992, *114* (25), 10024–10035. https://doi.org/10.1021/ja00051a040.

(67) Lloyd, S. Least Squares Quantization in PCM. *IEEE Trans. Inf. Theory* 1982, *28* (2), 129–137. https://doi.org/10.1109/tit.1982.1056489.

(68) Amendola, C.; Faugere, J.-C.; Sturmfels, B. Moment Varieties of Gaussian Mixtures. *J. Algebr. Stat.* 2016, *7* (1). https://doi.org/10.18409/jas.v7i1.42.

(69) Tolles, J.; Meurer, W. J. Logistic Regression: Relating Patient Characteristics to Outcomes: Relating Patient Characteristics to Outcomes. *JAMA* 2016, *316* (5), 533–534. https://doi.org/10.1001/jama.2016.7653.

(70) Jegorowa, A.; Górski, J.; Kurek, J.; Kruk, M. Use of Nearest Neighbors (k-NN) Algorithm in Tool Condition Identification in the Case of Drilling in Melamine Faced Particleboard. *Maderas Cienc. Tecnol.* 2020, No. ahead, 0–0. https://doi.org/10.4067/s0718-221x2020005000205.

(71) Cortes, C.; Vapnik, V. Support-Vector Networks. *Mach. Learn.* 1995, *20* (3), 273–297. https://doi.org/10.1007/bf00994018.

(72) Quinlan, J. R. Induction of Decision Trees. *Mach. Learn.* 1986, *1* (1), 81–106. https://doi.org/10.1007/bf00116251.

(73) Russell, S.; Norvig, P. Artificial Intelligence: A Modern Approach. *Artificial Intelligence: A Modern Approach; United States* 1999.

(74) Breiman, L. Bagging Predictors. *Mach. Learn.* 1996, *24* (2), 123–140. https://doi.org/10.1007/bf00058655.

(75) Breiman, L. Arcing Classifiers. *The Annals of Statistics* 1998, *26* (3), 801–824.

(76) Breiman, L. Random Forests. *Mach. Learn.* 2001, *45* (1), 5–32. https://doi.org/10.1023/a:1010933404324.

(77) Theoretic Generalization of On-Line Learning and an Application to Boosting. In *Lecture Notes in Computer Science*; Springer: Berlin Heidelberg; Berlin, Heidelberg, 1995; pp 23–37.

(78) Chen, T.; Guestrin, C. XGBoost: A Scalable Tree Boosting System. In *Proceedings of the 22nd ACM SIGKDD International Conference on Knowledge Discovery and Data Mining*; ACM: New York, NY, USA, 2016.

(79) Ke, G.; Meng, Q.; Finley, T.; Wang, T.; Chen, W.; Ma, W.; Ye, Q.; Liu, T.-Y. Lightgbm: A Highly Efficient Gradient Boosting Decision Tree. *Adv. Neural Inf. Process. Syst* 2017.

(80) Tankeo, S. B.; Kuete, V. African Plants Acting on Pseudomonas Aeruginosa: Cut-off Points for the Antipseudomonal Agents from Plants. In *Advances in Botanical Research*; Elsevier, 2022.

(81) Stehman, S. V. Selecting and Interpreting Measures of Thematic Classification Accuracy. Remote Sens. *Remote Sens. Environ* 1997, *62* (1), 77–89.

(82) McKinney, W. Data Structures for Statistical Computing in Python. In *Proceedings of the 9th Python in Science Conference*; SciPy, 2010.

(83) Fawcett, T. An Introduction to ROC Analysis. *Pattern Recognit. Lett.* 2006, *27* (8), 861–874. https://doi.org/10.1016/j.patrec.2005.10.010.

(84) Harris, C. R.; Millman, K. J.; van der Walt, S. J.; Gommers, R.; Virtanen, P.; Cournapeau, D.; Wieser, E.; Taylor, J.; Berg, S.; Smith, N. J.; Kern, R.; Picus, M.; Hoyer, S.; van Kerkwijk, M. H.; Brett, M.; Haldane, A.; Del Río, J. F.; Wiebe, M.; Peterson, P.; Gérard-Marchant, P.; Sheppard, K.; Reddy, T.; Weckesser, W.; Abbasi, H.; Gohlke, C.; Oliphant, T. E. Array Programming with NumPy. *Nature* 2020, *585* (7825), 357–362. https://doi.org/10.1038/s41586-020-2649-2.

(85) Chicco, D.; Warrens, M. J.; Jurman, G. The Matthews Correlation Coefficient (MCC) Is More Informative than Cohen's Kappa and Brier Score in Binary Classification Assessment. *IEEE Access* 2021, *9*, 78368–78381. https://doi.org/10.1109/access.2021.3084050.

(86) Taha, A. A.; Hanbury, A. Metrics for Evaluating 3D Medical Image Segmentation: Analysis, Selection, and Tool. *BMC Med. Imaging* 2015, *15* (1), 29. https://doi.org/10.1186/s12880-015-0068-x.

(87) Calders, T.; Jaroszewicz, S. Efficient AUC Optimization for Classification. In *Knowledge Discovery in Databases: PKDD 2007*; Springer Berlin Heidelberg: Berlin, Heidelberg, 2007; pp 42–53.

(88) Wang, J.; Wang, W.; Kollman, P. A.; Case, D. A. Automatic Atom Type and Bond Type Perception in Molecular Mechanical Calculations. *J. Mol. Graph. Model.* 2006, *25* (2), 247–260. https://doi.org/10.1016/j.jmgm.2005.12.005.

(89) Lee, C.; Yang, W.; Parr, R. G. Development of the Colle- Salvetti Correlation-Energy Formula into a Functional of the Electron Density. *Phys. Rev. B: Condens. Matter Mater. Phys* 1988, *37* (2), 785–789.

(90) Becke, A. D. Correlation Energy of an Inhomogeneous Electron Gas: A Coordinate-Space Model. *J. Chem. Phys.* 1988, *88* (2), 1053–1062. https://doi.org/10.1063/1.454274.

(91) Stephens, P. J.; Devlin, F. J.; Chabalowski, C. F.; Frisch, M. J. Ab Initio Calculation of Vibrational Absorption and Circular Dichroism Spectra Using Density Functional Force Fields. *J. Phys. Chem.* 1994, *98* (45), 11623–11627. https://doi.org/10.1021/j100096a001.

(92) Bayly, C. I.; Cieplak, P.; Cornell, W.; Kollman, P. A. A Well- Behaved Electrostatic Potential Based Method Using Charge Restraints for Deriving Atomic Charges: The RESP Model. *J. Phys. Chem* 1993, No. 40, 10269–10280.

(93) Wang, J.; Wolf, R. M.; Caldwell, J. W.; Kollman, P. A.; Case, D. A. Development and Testing of a General Amber Force Field. *J. Comput. Chem.* 2004, *25* (9), 1157–1174. https://doi.org/10.1002/jcc.20035.

(94) Tian, C.; Kasavajhala, K.; Belfon, K. A. A.; Raguette, L.; Huang, H.; Migues, A. N.; Bickel, J.; Wang, Y.; Pincay, J.; Wu, Q.; Simmerling, C. Ff19SB: Amino-Acid-Specific Protein Backbone Parameters Trained against Quantum Mechanics Energy Surfaces in Solution. *J. Chem. Theory Comput.* 2020, *16* (1), 528–552. https://doi.org/10.1021/acs.jctc.9b00591.

(95) Gordon, J. C.; Myers, J. B.; Folta, T.; Shoja, V.; Heath, L. S.; Onufriev, A. H++: A Server for Estimating pKas and Adding Missing Hydrogens to Macromolecules. *Nucleic Acids Res.* 2005, *33* (Web Server issue), W368-71. https://doi.org/10.1093/nar/gki464.

(96) Jorgensen, W. L.; Chandrasekhar, J.; Madura, J. D.; Impey, R. W.; Klein, M. L. Comparison of Simple Potential Functions for Simulating Liquid Water. *J. Chem. Phys.* 1983, *79* (2), 926–935. https://doi.org/10.1063/1.445869.

(97) Darden, T.; York, D.; Pedersen, L. Particle Mesh Ewald: An N• Log(N) Method for Ewald Sums in Large Systems. *J. Chem. Phys* 1993, No. 12, 10089–10092.

(98) Ryckaert, J.-P.; Ciccotti, G.; Berendsen, H. J. C. Numerical Integration of the Cartesian Equations of Motion of a System with Constraints: Molecular Dynamics of n-Alkanes. *J. Comput. Phys.* 1977, *23* (3), 327–341. https://doi.org/10.1016/0021-9991(77)90098-5.

(99) Case, D. A.; Ben-Shalom, I. Y.; Brozell, S. R.; Cerutti, D. S.; Cheatham, T. E., III; Cruzeiro, V. W. D.; Darden, T. A.; Duke, R. E.; Ghoreishi, D.; Gilson, M. K. *Amber 2018; University of California: San Francisco*; 2018.

(100) McPhie, M. G.; Daivis, P. J.; Snook, I. K.; Ennis, J.; Evans, D. J. Generalized Langevin Equation for Nonequilibrium Systems. *Physica A* 2001, *299* (3–4), 412–426. https://doi.org/10.1016/s0378-4371(01)00328-4.

(101) Berendsen, H. J. C.; Postma, J. P. M.; van Gunsteren, W. F.; DiNola, A.; Haak, J. R. Molecular Dynamics with Coupling to an External Bath. *J. Chem. Phys.* 1984, *81* (8), 3684–3690. https://doi.org/10.1063/1.448118.

(102) Laurent, B.; Chavent, M.; Cragnolini, T.; Dahl, A. C. E.; Pasquali, S.; Derreumaux, P.; Sansom, M. S. P.; Baaden, M. Epock: Rapid Analysis of Protein Pocket Dynamics. *Bioinformatics* 2015, *31* (9), 1478–1480. https://doi.org/10.1093/bioinformatics/btu822.

(103) Sharp, K. A.; Honig, B. Calculating Total Electrostatic Energies with the Nonlinear Poisson-Boltzmann Equation. *J. Phys. Chem.* 1990, *94* (19), 7684–7692. https://doi.org/10.1021/j100382a068.

(104) Miller, B. R.; Mcgee, T. D., Iii; Swails, J. M.; Homeyer, N.; Gohlke, H.; Roitberg, A. E. Py: An Efficient Program for End-State Free Energy Calculations. *J. Chem. Theory Comput* 2012, *8* (9), 3314–3321.

(105) Yoshida, K.-I.; Nakayama, K.; Kuru, N.; Kobayashi, S.; Ohtsuka, M.; Takemura, M.; Hoshino, K.; Kanda, H.; Zhang, J. Z.; Lee, V. J.; Watkins, W. J. MexAB-OprM Specific Efflux Pump Inhibitors in Pseudomonas Aeruginosa. Part 5: Carbon-Substituted Analogues at the C-2 Position. *Bioorg. Med. Chem.* 2006, *14* (6), 1993–2004. https://doi.org/10.1016/j.bmc.2005.10.043.

(106) Taubman, S. B.; Jones, N. R.; Young, F. E.; Corcoran, J. W. Sensitivity and Resistance to Erythromycin in Bacillus Subtilis 168: The Ribosomal Binding of Erythromycin and Chloramphenicol. *Biochim. Biophys. Acta* 1966, *123* (2), 438–440. https://doi.org/10.1016/0005-2787(66)90301-7.

(107) Svetlov, M. S.; Vázquez-Laslop, N.; Mankin, A. S. Kinetics of Drug–Ribosome Interactions Defines the Cidality of Macrolide Antibiotics. *Proc. Natl. Acad. Sci. U. S. A.* 2017, *114* (52), 13673–13678. https://doi.org/10.1073/pnas.1717168115.

(108) Maier, J. A.; Martinez, C.; Kasavajhala, K.; Wickstrom, L.; Hauser, K. E.; Simmerling, C. Ff14SB: Improving the Accuracy of Protein Side Chain and Backbone Parameters from ff99SB. *J. Chem. Theory Comput.* 2015, *11* (8), 3696–3713. https://doi.org/10.1021/acs.jctc.5b00255.

(109) Gowers, R.; Linke, M.; Barnoud, J.; Reddy, T.; Melo, M.; Seyler, S.; Domański, J.; Dotson, D.; Buchoux, S.; Kenney, I.; Beckstein, O. MDAnalysis: A Python Package for the Rapid Analysis of Molecular Dynamics Simulations. In *Proceedings of the Python in Science Conference*; SciPy, 2016.

(110) Michaud-Agrawal, N.; Denning, E. J.; Woolf, T. B.; Beckstein, O. MDAnalysis: A Toolkit for the Analysis of Molecular Dynamics Simulations. *J. Comput. Chem.* 2011, *32* (10), 2319–2327. https://doi.org/10.1002/jcc.21787.

(111) Humphrey, W.; Dalke, A.; Schulten, K. VMD - Visual Molecular Dynamics. *J. Molec. Graphics* 1996, *14*, 33–38.

(112) Valdés-Tresanco, M. S.; Valdés-Tresanco, M. E.; Valiente, P. A.; Moreno, E. Gmx_MMPBSA: A New Tool to Perform End-State Free Energy Calculations with GROMACS. *J. Chem. Theory Comput.* 2021, *17* (10), 6281–6291. https://doi.org/10.1021/acs.jctc.1c00645.

(113) Miller, B. R., III; McGee, T. D., Jr; Swails, J. M.; Homeyer, N.; Gohlke, H.; Roitberg, A. E. *MMPBSA.Py*: An Efficient Program for End-State Free Energy Calculations. *J. Chem. Theory Comput.* 2012, *8* (9), 3314–3321. https://doi.org/10.1021/ct300418h.

(114) Berendsen, H. J. C.; van der Spoel, D.; van Drunen, R. GROMACS: A Message-Passing Parallel Molecular Dynamics Implementation. *Comput. Phys. Commun.* 1995, *91* (1–3), 43–56. https://doi.org/10.1016/0010-4655(95)00042-e.

(115) Nguyen, H.; Roe, D. R.; Simmerling, C. Improved Generalized Born Solvent Model Parameters for Protein Simulations. *J. Chem. Theory Comput.* 2013, *9* (4), 2020–2034. https://doi.org/10.1021/ct3010485.